\documentclass{lmcs}
\usepackage[T1]{fontenc} 
\usepackage{lmodern}
\usepackage{hyperref}
\usepackage{graphicx}
\usepackage{subcaption}
\usepackage{color}

\usepackage{listings}
\lstdefinestyle{mystyle}{
    basicstyle=\ttfamily\footnotesize,
    breakatwhitespace=false,         
    breaklines=true,                 
    captionpos=b,                    
    keepspaces=true,                 
    numbers=left,                    
    numbersep=5pt,                  
    showspaces=false,                
    showstringspaces=false,
    showtabs=false,                  
    tabsize=2,
    frame=leftline
}
\usepackage{multicol}
\usepackage{algorithm}
\usepackage{algpseudocode}

\usepackage{mathtools}
\usepackage[disable]{todonotes}
\usepackage{ulem}
\usepackage{comment}
\usepackage{amsmath}

\usepackage{amsfonts}
\usepackage{amssymb}
\usepackage{mathtools}
\usepackage{stmaryrd}
\usepackage[capitalize]{cleveref}
\usepackage{tikz}
\usepackage{wrapfig}
\usepackage{thmtools, thm-restate}
\usepackage{multicol}
\usepackage{caption}
\usepackage{xspace}
\usetikzlibrary{arrows.meta}
\usetikzlibrary{automata,positioning}
\usetikzlibrary{shapes.geometric}

\makeatletter
\DeclareRobustCommand{\emph}[1]{%
  \ifmmode
    \nfss@text{\itshape #1}%
  \else
    \ifdim\fontdimen\@ne\font>\z@ 
      \textup{#1}%
    \else
      \textit{#1}%
    \fi
  \fi
}
\makeatother

\newcommand{\disjointunion}{\uplus}

\newcommand{\set}[1]{\{#1\}}
\newcommand{\nset}[2]{\llbracket #1, #2 \rrbracket}

\newcommand{\deff}{\coloneq}
\newcommand{\ffed}{\eqcolon} 
\newcommand{\nats}{\mathbb{N}}
\newcommand{\rels}{\mathbb{Z}}
\newcommand{\proglike}[1]{\textup{\texttt{#1}}}

\newcommand{\size}[1]{|#1|}

\newcommand{\tuple}[1]{\langle #1 \rangle}
\newcommand{\recdeff}{\Coloneqq}

\newcommand{\prog}{\mathcal{P}}

\newcommand{\threads}{\mathcal{T}}
\newcommand{\athread}{t}
\newcommand{\mutexes}{\mathcal{M}}
\newcommand{\amutex}{m}
\newcommand{\codeif}{\proglike{if} \xspace}
\newcommand{\codethen}{\proglike{then}}

\newcommand{\lock}[1]{\proglike{lock}(\instrstyle{#1})}
\newcommand{\unlock}[1]{\proglike{unlock}(\instrstyle{#1})}

\newcommand{\create}[1]{\proglike{create}(\instrstyle{#1})}
\newcommand{\join}[1]{\proglike{join}(\instrstyle{#1})}
\newcommand{\trueguard}[1]{\proglike{true}(\mathsf{#1})}
\newcommand{\falseguard}[1]{\proglike{false}(\mathsf{#1})}

\newcommand{\codefalse}{\proglike{false}}
\newcommand{\CFG}{\mathcal{C}}
\newcommand{\CFGof}[1]{\CFG(#1)}
\newcommand{\nodes}{\mathcal{N}}
\newcommand{\nodesof}[1]{\nodes_{#1}}
\newcommand{\startnode}{\mathsf{st}}
\newcommand{\startnodeof}[1]{\startnode_{#1}}
\newcommand{\transitions}{\mathrm{\Delta}} 
\newcommand{\transitionsof}[1]{\transitions_{#1}}
\newcommand{\instructions}{\mathsf{Instr}}

\newcommand{\aninstr}{\iota}

\newcommand{\anode}{n}
\newcommand{\atrans}{\delta}
\newcommand{\stateUncreated}{\mathsf{Uncreated}}
\newcommand{\stateActive}[1]{#1}
\newcommand{\stateJoined}{\mathsf{Joined}}
\newcommand{\maint}{\mathsf{main}}

\newcommand{\varsof}[1]{\mathcal{V}_{#1}}
\newcommand{\vars}{\mathcal{V}}
\newcommand{\instrstyle}[1]{\mathsf{#1}}

\newcommand{\states}{\mathsf{States}}
\newcommand{\values}{\nu}
\newcommand{\stateFunction}{\mathsf{St}}
\newcommand{\held}{\mathsf{h}}
\newcommand{\config}{\gamma}
\newcommand{\configs}{\mathrm{\Gamma}}
\newcommand{\step}[1]{\xrightarrow{#1}}
\newcommand{\initialconfig}{\config_0}

\newcommand{\exec}{\rho}
\newcommand{\apath}{\pi}
\newcommand{\uninit}{\#}
\newcommand{\threadof}[1]{\mathsf{th}(#1)}
\newcommand{\binop}{\mathrel{\mathsf{op}}}
\newcommand{\evalof}[2]{\mathsf{ev}_{#1}(#2)}
\newcommand{\PTGof}[2]{\mathsf{ptg}_{#2}(#1)}
\newcommand{\PTG}{\mathsf{G}}
\newcommand{\subptg}{\preccurlyeq}
\newcommand{\deref}[1]{\text{*}#1}
\newcommand{\dummyderef}[2]{\text{*}#1^{#2}}
\newcommand{\dummyvars}{\vars_*}

\newcommand{\extvars}{\vars_+}

\newcommand{\addr}[1]{\&#1}
\newcommand{\pto}{\mathsf{pto}}
\newcommand{\procpto}{\alpha}
\newcommand{\uncreatedsymb}{U}
\newcommand{\uncreated}[1]{\uncreatedsymb(#1)} 
\newcommand{\joined}[1]{J(#1)}
\newcommand{\lockset}[1]{L(#1)}
\newcommand{\termnode}{\mathsf{end}}
\newcommand{\termnodeof}[1]{\mathsf{end}_{#1}}
\newcommand{\MHP}{\mathrel{\mathsf{MHP}}}
\newcommand{\MHB}[1]{\mathsf{MHB}(#1)}
\newcommand{\veccomp}[2]{{#1}[#2]}

\newcommand{\predcreate}[1]{\mathsf{P}_{\mathsf{cr}}(#1)}
\newcommand{\prednocreate}[1]{\mathsf{P}_{\mathsf{o}}(#1)}
\newcommand{\predjoin}[1]{\mathsf{P}_j(#1)}
\newcommand{\prednojoin}[1]{\mathsf{P}_{o}(#1)}
\newcommand{\childsymb}{\mathsf{child}}
\newcommand{\parentsymb}{\mathsf{par}}
\newcommand{\grandparentsymb}{\mathsf{gpar}}
\newcommand{\parentnodes}{\nodes_{\parentsymb}}
\newcommand{\grandparentnodes}{\nodes_{\grandparentsymb}}

\newcommand{\auxstsymb}{\sigma}
\newcommand{\auxst}[1]{\auxstsymb(#1)}
\newcommand{\auxuncsymb}{\tilde{U}}
\newcommand{\auxunc}[1]{\auxuncsymb(#1)}
\newcommand{\auxnodes}{\tilde{\nodes}}
\newcommand{\vertices}{V}
\newcommand{\edges}{E}
\newcommand{\labelling}{\lambda}
\newcommand{\avertex}{v}
\newcommand{\vertexof}[1]{\mathsf{ver}(#1)}
\newcommand{\PTGunion}{\mathrel{\sqcup}}
\newcommand{\bigPTGunion}{\bigsqcup}
\newcommand{\DCR}[1]{\mathsf{DCR}(#1)}
\newcommand{\DCRbig}[1]{\mathsf{DCR}\left(#1\right)}
\newcommand{\pt}[1]{\rightarrow_{#1}}
\newcommand{\HB}[1]{\mathsf{HB}(#1)}
\newcommand{\newpt}[1]{\theta(#1)}

\newcommand{\readacc}[2]{\mathsf{R}_{#2}(#1)}
\newcommand{\writeacc}[2]{\mathsf{W}_{#2}(#1)}
\newcommand{\mayrace}[1]{\mathrel{\mathsf{MayRace}_{#1}}}
\newcommand{\mayread}[1]{\mathsf{MR}(#1)}
\newcommand{\maywrite}[1]{\mathsf{MW}(#1)}

\newcommand{\newmut}[1]{\mu_{#1}}

\newcommand{\newmutexes}{\mutexes_{\mathsf{new}}}

\newcommand{\criticalsection}{\mathsf{CS}}

\newcommand{\csfirst}[1]{\mathsf{fst}(#1)}
\newcommand{\cslast}[1]{\mathsf{lst}(#1)}
\newcommand{\lockordering}{\trianglelefteq}

\newcommand{\interferencetransitions}[2]{\mathsf{Interf}(#1,#2)}

\newcommand{\newprog}{\tilde{\prog}}
\newcommand{\newof}[1]{\tilde{#1}}
\newcommand{\newseqof}[1]{\mathsf{new}(#1)}
\newcommand{\oldnodeof}[1]{\mathsf{old}(#1)}
\newcommand{\oldnodeofsymb}{\mathsf{old}}

\newcommand{\lockseq}{\sigma_{\proglike{lock}}}
\newcommand{\unlockseq}{\sigma_{\proglike{unlock}}}

\newcommand{\locksymb}{\proglike{lock} \xspace}
\newcommand{\unlocksymb}{\proglike{unlock} \xspace}

\newcommand{\prottrans}[2]{T_{#1}^{#2}}
\newcommand{\protvars}[1]{V^{#1}}
\newcommand{\racevars}[1]{V_{\mathsf{drace}}^{#1}}
\newcommand{\interfvars}[1]{V_{\mathsf{intf}}^{#1}}
\newcommand{\homomorphism}{\phi}
\newcommand{\isomorphism}{\theta}
\newcommand{\isom}{\cong}
\begin{document}
\title[Lock Placement for Concurrent Programs with Pointers]{Sound Automatic Lock Placement for Concurrent Programs with Pointers}


\author[N.~Waldburger]{Nicolas Waldburger\lmcsorcid{0009-0002-7664-5828}}[a]
\author[F.~Faissole]{Florian faissole\lmcsorcid{0000-0001-5792-0658}}[a]
\author[R.~Okabe]{Ryo Okabe\lmcsorcid{0009-0007-4800-9011}}[a]
\author[D.~Cousineau]{Denis Cousineau\lmcsorcid{0000-0003-4078-3591}}
\address{Mitsubishi Electric R\&D Centre Europe, Rennes, France}
\email{n.waldburger@fr.merce.mee.com, f.faissole@fr.merce.mee.com, r.okabe@fr.merce.mee.com, d.cousineau@fr.merce.mee.com}

\specialissue{Special issue of FORTE 2026.}

\begin{abstract}
We propose a lightweight method for automatic lock placement in concurrent POSIX programs with explicit pointer manipulation, such as C programs. 
Given a program that lacks sufficient user-placed mutexes, or contains none at all, our method introduces new mutexes to prevent data races. 
The method is formally proven to guarantee the absence of data races and mainly consists of three steps. 
The first step performs lockset analysis and may-happen-in-parallel analysis. 
The second step conducts a concurrent points-to analysis derived from a Steensgaard-like procedural analysis, where per-thread points-to graphs are unified using a special completion rule. 
The final step places locks to prevent data races while respecting a lock ordering to avoid deadlocks. In addition, user-annotated critical sections are protected through insertion of dedicated locks.  
Since lock placement is a complex and error-prone task, our method aims at helping developers write correct concurrent programs.

\end{abstract}

\maketitle             
\section{Introduction}

Concurrent programs are notoriously cumbersome to implement and difficult to debug. Unlike sequential programs, their behavior does not only depend on the 
individual computations performed by each thread, but also on the order in which these computations are interleaved during execution. As a consequence, even 
a relatively simple concurrent program may exhibit a large number of possible execution behaviors. Concurrent programs are therefore prone to several categories
of subtle errors, including data races, deadlocks, atomicity violations, and more general race conditions. Such errors generally arise from unexpected 
interactions between concurrent threads accessing shared resources or from incorrect assumptions about the relative ordering of operations. Preventing them 
requires careful implementation of concurrency-control mechanisms, such as mutual exclusion, synchronization, and atomic operations, whose incorrect use may 
itself introduce additional issues, notably deadlocks or unintended serialization.

Detecting concurrency errors is particularly challenging because their occurrence may depend on a specific and often unlikely thread interleaving. Consequently,
an erroneous execution may only manifest under particular scheduling or timing conditions, making concurrency bugs notoriously difficult
to reproduce. Conventional testing is therefore insufficient to provide strong confidence in the absence of such errors, since it can explore only a small
fraction of all possible thread interleavings. This motivates the use of more systematic verification techniques capable of reasoning about multiple 
executions and explicitly accounting for the nondeterminism introduced by concurrency, such as model checking or static analysis, 
which aim to explore or reason about concurrent behaviors more exhaustively than conventional execution-based testing.

Among concurrency errors, shared-memory errors have a particular flavor: they are due to unexpected interferences among threads' executions, leading to unexpected 
values in memory. The simplest error of this category is a data race, which occurs when two conflicting accesses to a shared variable are executed without proper 
synchronization \cite{boehm_how_2011,lu_learning_2008}. At a higher level, an atomicity violation occurs when a critical portion of code suffers from unintended 
interference by another thread. Preventing data races and atomicity violations with mutexes is prone to errors such as missing locks and introduction of deadlocks.
Moreover, atomicity violations are inherently specification-dependent: determining whether a particular thread interleaving constitutes a violation requires 
knowledge of which operations are intended by the programmer to execute atomically. For both data races and atomicity violations, pointers increase the difficulty 
as they obscure which variables a statement accesses. 

\subsection{Related works}



Concurrency errors can be addressed at different levels, ranging from their elimination by construction to their detection in existing programs. 
A first line of defense consists in preventing certain classes of concurrency errors by construction. Languages such as 
Cyclone~\cite{gerakios_race-free_2010,grossman_type-safe_2003}, Rust~\cite{matsakis_rust_2014} 
and SPARK \cite{carre_sparkannotated_1990}, rely on language-level restrictions and type systems to rule out or 
restrict unsafe memory interactions. However, adopting such languages may require substantial changes to the programming model and 
existing code bases as well as adaptability from developers. Moreover, data-race freedom alone is not sufficient to guarantee the correctness of concurrent 
interactions: a program can be free of data races while still exhibiting incorrect interleavings at a higher semantic level.

Static analysis constitutes a second line of defense. In particular, abstract-interpretation-based
analyzers, like Frama-C Mthread~\cite{kirchner_frama-c_2015}, Astrée~\cite{mine_taking_2016} and Goblint~\cite{vojdani_goblint_2009},
have been developed specifically for concurrent programs and can provide sound analyses of shared-memory interactions,
notably for detecting data races and deadlocks. Such approaches alleviate part of the verification burden by automatically 
reasoning about potentially large sets of program executions. However, they primarily target the verification of an existing 
concurrent implementation: synchronization mechanisms must still be designed and introduced by the programmer. Shared-variable 
identification and points-to analysis are key to static analysis. Rugina et al.~\cite{rugina_pointer_1999,rugina_pointer_2003} 
proposed a flow-sensitive points-to analysis for structured multithreaded programs, which computes a points-to graph for each 
program point, using graphs summarizing inter-thread interference. Salcianu et al.~\cite{salcianu_pointer_2001} presented a 
flow-sensitive points-to analysis for parallel Java programs. Their approach of merging parent and child graphs to derive
thread interference is unsound for C, as global pointers allow more general inter-thread interference.
Sui et al.~\cite{sui_sparse_2016} developed a sparse flow-sensitive analysis for unstructured multithreaded C programs,
propagating points-to information sparsely along def-use chains for scalability.
Abstract interpreters compute shared memory locations and their contents, as well as abstract pointer values.
Mthread~\cite{kirchner_frama-c_2015} and Astrée~\cite{mine_taking_2016} iteratively re-analyze threads that read shared 
memory to propagate inter-thread interference until reaching a fixpoint. Their path-sensitive analysis is highly precise, 
but is expensive due to the combinatorial explosion of program paths and thread interleavings. 
Goblint~\cite{vojdani_goblint_2009} avoids this problem through thread-modular analysis, using global invariants summarizing inter-thread interference.

A promising middle ground between these approaches is to automate lock placement. This reduces development effort and makes 
concurrent programming more accessible to non-experts. Atomicity violations constitute a particularly natural target for automatic lock placement. 
As discussed above, atomicity is inherently specification-dependent: determining that an interleaving is erroneous requires knowing which sequence 
of operations is intended to execute atomically. Locks provide a direct implementation mechanism for such atomicity requirements. Protecting the 
relevant shared-memory accesses with a common mutex can prevent other threads from observing or modifying intermediate states of an atomic section.
 This establishes a close correspondence between the specification, expressed as a region that must be protected from interference, and its 
 implementation through mutual exclusion. Consequently, several approaches to automatic lock placement formulate their input in terms of 
 atomic sections provided by the programmer and synthesize lock acquisitions and releases that enforce their atomic execution.

Several works have proposed algorithms for automatic lock placement for atomic 
sections~\cite{mccloskey_autolocker_2006,hicks_lock_2006,cherem_inferring_2008,emmi_lock_2007}.
The developer specifies which code sections must be treated as atomic, \emph{i.e.}, must execute without any interference from other threads.
These tools insert appropriate mutex acquisitions and releases to enforce atomicity while avoiding deadlocks, using two-phase locking and a predetermined 
lock ordering. Autolocker~\cite{mccloskey_autolocker_2006} considers pessimistic atomic sections, in which atomic 
blocks are automatically translated into conventional lock-based synchronization. The motivation is to retain the simplicity of atomic sections while avoiding 
overhead of optimistic atomic sections. 

Static-analysis-based approaches infer the synchronization required to implement atomic sections specified by the programmer
by analyzing the shared-memory accesses performed by these regions. A central challenge is to determine which memory locations may be shared between threads
and therefore require protection. Hicks et al.~\cite{hicks_lock_2006} combine points-to analysis with continuation effects to infer thread sharing in a flow-sensitive manner. 
The inferred sharing information is subsequently used to determine the locks required by each atomic section. The precision of lock inference therefore closely
depends on the ability of the underlying static analysis to distinguish thread-local from shared memory locations.
Cherem et al.~\cite{cherem_inferring_2008} address a related problem using an interprocedural analysis combined with runtime support. Their approach notably considers lock granularity: 
fine-grained locks can be inferred when individual shared resources can be identified precisely, whereas coarser-grained locks can be used when the set of 
accessed locations cannot be represented conveniently using fine-grained locking. This provides a compromise between analysis precision, synchronization 
overhead, and contention. 

It is worth mentioning a posteriori approaches to detect atomicity violations. NIChecker \cite{zhang2025bounded} proposes a bounded verification method for atomicity violations in interrupt-driven
programs based on lazy sequentialization. It transforms the original program into a non-deterministic sequential one while preserving atomicity violations, 
enabling the use of existing sequential program verifiers based on model-checking. Velodrome~\cite{flanagan2008velodrome} and RegionTrack~\cite{ma2020regiontrack} 
are sound and complete trace-based atomicity checkers for explicitly identified atomic regions. Both exactly detect whether the observed 
execution is conflict-serializable, i.e., if its interleaved operations can be reordered into a serial execution of the atomic regions while preserving 
the order of all conflicting memory accesses, with no false positives or negatives for this criterion. RegionTrack~\cite{ma2020regiontrack} goes further
 by identifying which transactions suffer atomicity violations, rather than only establishing global trace non-serializability. Their guarantees 
 apply only to observed traces, not to unobserved executions.
A quite different approach is taken by Emmi et al.~\cite{emmi_lock_2007}, who formulate automatic synchronization as a lock-allocation and optimization problem. 
Given a multithreaded program annotated with atomic sections, their technique infers both an assignment of global variables to mutexes and the corresponding lock 
instrumentation. Under the assumption that all shared data is accessed within atomic sections, the generated program is guaranteed to preserve atomicity 
and deadlock freedom.



\subsection{Contributions.}

We propose an automatic lock placement method for concurrent POSIX programs with explicit pointer manipulation (\emph{e.g.}, C programs).
Given a program lacking sufficient user-placed mutexes (or containing none), our method operates on the source code by introducing new mutexes 
to eliminate data races and to guarantee atomicity of user-annotated critical sections. 

A first contribution is the introduction of a formal model for concurrent programs and of a preliminary analysis that are crucial for applying our 
methodology, namely lockset, joined threads and uncreated threads analyses. Although these analyses are relatively standard, we 
provide a novel formalization of them and rigorously prove their fundamental properties.

We build a concurrent points-to analysis derived from an inexpensive Steensgaard-like per-thread analysis, such as the one implemented in the 
\texttt{alias} plugin of Frama-C \cite{le_gall_alias_2024}. Overall, our approach is designed to run fast, sacrificing precision for efficiency. 
It targets use cases where rapid development is a priority, and contention in the obtained program is not a primary concern.
Moreover, our approach is formalized and proven, which strengthens the soundness claim and is, in our view, a contribution in itself.
The underlying points-to analysis is cumulative, does not rely on any value analysis and is path-insensitive. This makes the analysis both 
less expensive in terms of execution time and easier to extend to a concurrent setting. The points-to analysis of concurrent programs relies 
on a specific rule that we dub the diamond completion rule and that corresponds to a weakened notion of merging. Once again, we prove the correctness 
of the analysis. 

The last step is the proposal of a method that adds mutexes in a program in order to guarantee the absence of data races and atomicity violations 
according to the user's atomicity requirements. This method guarantees not to introduce resource deadlocks.

\subsection{Outline.}
The paper is organized as follows.
Section \ref{sec:model} defines our model for concurrent programs and its semantics.
Section \ref{sec:analysis} describes a preliminary analysis of concurrency between statements. Section \ref{sec:points-to-analysis} presents the points-to analysis. 
Section \ref{sec:automatic-placement} introduces the automatic lock placement technique, and Section \ref{sec:Conclusion} concludes the paper. 

\section{A Model for Concurrent Programs}\label{sec:model}
\subsection{Concurrent Programs}
\label{sec:concurrent-programs} 

A concurrent program is a tuple $\prog = \tuple{\threads, \vars, \mutexes, \CFG}$ where:
\begin{itemize}
\item $\threads$ is a finite set of \emph{threads} with a special thread $\maint$; 
\item $\vars$ is a finite set of \emph{variables};
\item $\mutexes$ is a finite set of \emph{mutexes};
\item a function $\CFG$ that maps each thread to its \emph{control-flow graph}.
\end{itemize} 
While the method is intended to work on source-level programs, we present it here on control-flow graphs; in practice, tools such as Frama-C \cite{CKKPSY12} allow to operate on such abstract representations of the program. 
A \emph{Control-Flow Graph} (CFG) is a tuple $(\nodes, \startnode, \termnode, \transitions)$ where $\nodes$ is a finite set of \emph{nodes} with \emph{start node} $\startnode$ and \emph{end node} $\termnode$, and $\transitions \subseteq \nodes \times \instructions \times \nodes$ is a finite set of \emph{transitions} labelled by \emph{instructions} from a set $\instructions$ described below.

Given $\atrans = (\anode, \aninstr, \anode') \in \transitions$, $\anode$ is the \emph{source} and $\anode'$ the \emph{target} of $\atrans$.
A \emph{path} is a sequence of transitions $\atrans_1, \dots, \atrans_k$ such that the target of $\atrans_i$ is the source of $\atrans_{i+1}$. 
Each thread $\athread$ is described by its CFG $\CFGof{\athread} = (\nodesof{\athread}, \startnodeof{\athread}, \termnodeof{\athread}, \transitionsof{\athread})$. 
Threads have disjoint sets of nodes: for all $\athread \ne \athread'$, $\nodesof{\athread} \cap \nodesof{\athread'} = \emptyset$. We let $\nodes \deff \bigcup_{\athread \in \threads} \nodesof{\athread}$ denote the set of all nodes of all threads, and $\transitions \deff \bigcup_{\athread \in \threads} \transitionsof{\athread}$ denote the set of all transitions of all threads. Each node $\anode \in \nodes$ has a single thread, denoted $\threadof{\anode}$, such that $\anode \in \nodesof{\threadof{\anode}}$. Similarly, each transition $\atrans$ has a single thread, denoted $\threadof{\atrans}$, such that $\atrans \in \transitionsof{\threadof{\atrans}}$.  We assume that, for every thread $\athread$, $\startnodeof{\athread}$ has no incoming transition ($\set{(\anode, \aninstr, \startnode) \mid \aninstr \in \instructions, \anode \in \nodes} \cap \transitions = \emptyset$), that $\termnodeof{\athread}$ has no outgoing transition ($\set{(\termnode, \aninstr, \anode) \mid \aninstr \in \instructions, \anode \in \nodes} \cap \transitions = \emptyset$) and that, for every node $\anode \in \nodesof{\athread}$, there is a path from $\startnodeof{\athread}$ to $\anode$. 

Our model includes pointers of arbitrary depth (\emph{e.g.}, in C, $\deref{\proglike{int}}$, $\deref{\deref{\proglike{int}}}$...). 
Each variable of the program has a fixed \emph{depth}; for each $k \geq 0$, $\varsof{k}$ denotes the set of variables of depth $k$. 
Variables of depth $0$ are integer-valued, variables of depth $1$ are pointers to integer variables, variables of depth $2$ are pointers to pointers to integer variables, and so on.  
We define the set $e_k$ of \emph{expressions} of depth $k$ as follows:
\begin{align*}
& e_0 \recdeff c \mid e_0 \binop e_0 \mid x_0 \mid \deref{x_1}
\\
\forall k\geq 1, \quad & e_{k} \recdeff x_{k} \mid \addr{x_{k-1}} \mid \deref{x_{k+1}} 
\end{align*}
where $x_\ell \in \varsof{\ell}$ is a variable of depth $\ell$.
In words, an expression $e_0$ of depth $0$ can be a constant $c \in \rels$, a (binary) operator (\emph{e.g.}, $+$, $-$ or $\times$) applied to two expressions of depth $0$, a variable of depth $0$ or the dereference of a variable of depth $1$. The exact set of operators allowed is irrelevant to our purposes.  
An expression $e_k$ of depth $k\geq 1$ is either a variable of depth $k$, the dereference of a variable of depth $k+1$ or the address of a variable of depth $k-1$. 

An instruction of $\instructions$ is of one of the following types:
\begin{description}
\item[\emph{Regular assignment}:] $\instrstyle{x_k \deff e_k}$ where $k \in \nats$, $x_k \in \varsof{k}$ and $e_k$ is an expression of depth $k$;
\item[\emph{Dereference assignment}:] $\instrstyle{\deref{x_{k+1}} \deff e_k}$ where $k \in \nats$, $x_{k+1} \in \varsof{k+1}$ and $e_k$ is an expression of depth $k$;
\item[\emph{Conditional guard}:] $\trueguard{e_0}, \falseguard{e_0}$ with $e_0$ an expression of depth $0$ (this expression is considered false when its value is $0$ and true otherwise);
\item[\emph{Locking and unlocking}:] $\lock{\amutex}, \unlock{\amutex}$ where $\amutex \in \mutexes$;
\item[\emph{Thread creation}:] $\create{\athread}$ where $\athread \in \threads \setminus \set{\maint}$; 
\item[\emph{Joining}:] $\join{\athread}$ where $\athread \in \threads \setminus \set{\maint}$.
\end{description}

\begin{figure}
\begin{center}
\begin{lstlisting}[language=C, basicstyle=\ttfamily\footnotesize, escapeinside={(*@}{@*)}, label=input, multicols = 2]
int** dbl_ptr;
int* p;
int* q;
int k;
int j;
int out;


main() {
  dbl_ptr = &p;
  q = &j;
  k=0, j=0;
  create(t1);
  create(t2); (*@\label{create-t2}@*)
  out = k; (*@\label{access-out}@*) 
} (*@\label{end-main}@*) 
t1() {
  lock(m); (*@\label{t1-line1}@*) 
  dbl_ptr = &q; (*@\label{t1-line2}@*) 
  p = &k; 
  unlock(m); 
  return;
}

t2() {
  lock(m);
  *dbl_ptr = p; (*@\label{dblptr-to-p}@*) 
  *q = 2; (*@\label{access-q}@*) 
  unlock(m); (*@\label{t2-unlock-m}@*)
  return;
}
\end{lstlisting}
\end{center}
\caption{An example of concurrent program, written in C-like syntax. 
There is a possible data race: if \texttt{t1} runs before \texttt{t2}, the instruction at line~\ref{dblptr-to-p} causes \texttt{q} to become aliased with \texttt{p} so that \texttt{q} points to \texttt{k}, making \texttt{t2} at line~\ref{access-q} race with \texttt{main} at line~\ref{access-out}.}
\label{fig:example-program}
\end{figure}

We introduce an example of concurrent program in \autoref{fig:example-program}. 
Here, the set of threads is $\threads = \set{\texttt{main}, \texttt{t1}, \texttt{t2}}$, the set of variables is $\vars = \set{\texttt{dbl\_ptr},\texttt{p},\texttt{q}, \texttt{k}, \texttt{j}, \texttt{out}}$ ($\texttt{dbl\_ptr}$ has depth $2$, $\texttt{p}$ and $\texttt{q}$ have depth $1$, $\texttt{j},\texttt{k}$ and $\texttt{out}$ have depth $0$) and the set of mutexes is $\mutexes = \set{\texttt{m}}$. 
The CFG of thread \texttt{t1} is represented in \autoref{cfg-example}. 

\begin{figure}
\centering
\begin{tikzpicture}[
  ->, >=Stealth,
  node distance=15mm,
  every node/.style={font=\small},
  block/.style={draw, rounded corners, align=center, minimum width=15mm, minimum height=7mm},
  term/.style={draw, ellipse, align=center, minimum width=10mm, minimum height=7mm}
]
  \node[term]  (start) {$\startnode$};
  \node[term] (1) [right=of start] {$\anode_1$};
  \node[term] (2) [right=of 1] {$\anode_2$};
  \node[term] (3)[right = of 2] {$\anode_3$};
  \node[term]  (end)   [right=of 3] {$\termnode$};

  \draw (start) -- (1) node[midway, below=5pt, font={\footnotesize}]{$\lock{m}$};
  \draw (1) -- (2) node[midway, below=5pt, font={\footnotesize}]{$\texttt{dbl\_ptr} = \addr{\texttt{q}}$};
  \draw (2) -- (3) node[midway, below=5pt, font={\footnotesize}]{$\texttt{p}= \addr{\texttt{k}}$};
  \draw (3) -- (end) node[midway, below=5pt, font={\footnotesize}]{$\unlock{m}$};
\end{tikzpicture}
\caption{The CFG of thread \texttt{t1}. In the code, nodes are program locations in between transitions; for example $\anode_1$ is between lines~\ref{t1-line1} and \ref{t1-line2}.}
\label{cfg-example}
\end{figure}

\vspace{-0.5cm}
\subsection{Semantics}

We define the set of configurations of our system.
The \emph{state} of a thread is an element of
$\states \deff \nodes \disjointunion \set{\stateUncreated, \stateJoined}$.
State $\stateUncreated$ is for a thread that has not yet been created. A thread $\athread$ whose state is a node $\anode \in \nodes$ (including $\anode = \termnodeof{\athread}$) is \emph{active}. A thread $\athread$ is in $\stateJoined$ once $\join{\athread}$ has been performed. A thread may only be created once, and a created thread can only be joined once. 

The main features of the semantics are the following. Threads are statically identified and not re-entrant: there can only be a single copy of every thread in $\threads$, and thread creation cannot be performed if the thread is already created. The memory model is simple, focusing on statically allocated integer and pointer variables. More sophisticated concepts such as pointer arithmetic or heap allocation are not captured; moreover, sequential consistency is assumed. Reading an uninitialized value is a blocking step, so that executions with uninitialized behaviors are not considered.    

A configuration is a tuple $\config = \tuple{\stateFunction, \values, \held}$ where:
\begin{itemize}
\item $\stateFunction: \threads \to \states$ maps each thread to its state,
\item $\values : \vars \to \rels \cup \vars \cup \set{\uninit}$ maps each variable to its value, which is either an integer (for variables in $\varsof{0}$), the variable that it points to (for variables in $\varsof{k}$, $k \geq 1$) or a special value $\uninit$ when the variable has not been initialized.
\item $\held : \mutexes \to \threads \cup \set{\bot}$ maps each mutex to the thread holding it, and $\bot$ if none.
\end{itemize}

The valuation $\values$ must be \emph{well-formed}: $\values(x_0) \in \rels \cup \set{\uninit}$ for all $x_0 \in \varsof{0}$ and $\values(x_{k+1}) \in \varsof{k} \cup \set{\uninit}$ for all $k \geq 0$ and $x_{k+1} \in \varsof{k+1}$. 
The initial configuration is $\initialconfig = \tuple{\stateFunction_0, \values_0, \held_0}$ where
$\stateFunction_0(\maint) = \stateActive{\startnodeof{\maint}}$ and $\stateFunction_0(\athread) = \stateUncreated$ for all $\athread \ne \maint$ (only $\maint$ is created),
 $\values_0(x) = \uninit$ for all $x \in \vars$ (variables are uninitialized) and
 $\held_0(\amutex) = \bot$ for all $\amutex \in \mutexes$ (no mutex is held).

We explain how expressions of depth $0$ (integer-valued expressions) are evaluated. Let $\values$ be a valuation, let $e$ be an expression of depth $0$.
The evaluation $\evalof{\values}{e} \in \rels \cup \set{\uninit}$ is defined recursively by:
\begin{itemize}
\item if $e = c \in \rels$ then $\evalof{\values}{e} = c$;
\item if $e = x_0$ with $x_0 \in \varsof{0}$ then $\evalof{\values}{e} = \values(x_0)$;
\item if $e = \deref{x_1}$ with $x_1 \in \varsof{1}$ then $\evalof{\values}{e} = \values(\values(x_1))$;
\item if $e = e_a \binop e_b$ then $\evalof{\values}{e} = (\evalof{\values}{e_a}) \binop (\evalof{\values}{e_b})$. 
\end{itemize}
We use conventions that $\values(\uninit) = \uninit$ , so that $\values(\values(x_1)) = \uninit$ if $\values(x_1) = \uninit$, and that $x \binop \uninit = \uninit \binop x = \uninit$ for all $x$. 

Let $\config = \tuple{\stateFunction, \values, \held}$, $\config' = \tuple{\stateFunction', \values', \held'}$ and $\atrans = (\anode, \aninstr, \anode') \in \transitions$. There is a \emph{step} $\config \step{\atrans} \config'$ when, by letting $\athread = \threadof{\atrans}$, we have $\stateFunction(\athread) = \stateActive{\anode}$ and the following are true. 
\begin{description}
\item [Assignments] If $\aninstr$ is an assignment, then we must have $\stateFunction'(\athread) = \stateActive{\anode'}$, $\stateFunction(\tilde\athread) = \stateFunction'(\tilde\athread)$ for all $\tilde\athread \ne \athread$ and $\held' = \held$. Moreover, the left-hand side of the assignment is either of the form $\instrstyle{x_k}$ or $\instrstyle{\deref{x_{k+1}}}$; in the latter case, we require that $\values(x_{k+1}) \ne \uninit$ (otherwise the step is not possible).  Let $x \in \varsof{k}$ be the variable equal to $x_k$ in the first case and to $\values(x_{k+1})$ in the second case ($\values(x_{k+1}) \in \varsof{k}$ as $\values$ is well-formed). We require that $\values(z) = \values'(z)$ for all $z \in \vars \setminus \set{x}$. The value accessed at the right-side must be correctly initialized, otherwise the step is not possible. We make a case disjunction on the right-hand side:
\begin{itemize}
\item if it is an expression $e$ of depth $0$ ($k=0$) then $\values'(x) = \evalof{\values}{e} \ne \uninit$ (if the expression evaluates to $\uninit$ then the step is not possible);
\item if it is a variable $y \in \varsof{k}$ then $\values'(x) = \values(y) \ne \uninit$;
\item if it is  a $\deref{y}$ (which implies $y \in \varsof{k+1}$) then $\values'(x) = \values(\values(y)) \ne \uninit$ (we have $\values(y) \in \varsof{k}$ by well-formedness);
\item if it is a $\addr{y}$ (which implies $k \geq 1$ and $y \in \varsof{k-1}$) then $\values'(x) = y$.
\end{itemize}
\item[Conditional guards.] If $\aninstr$ is $\trueguard{e_0}$ then $\evalof{\values}{e_0} \notin \set{0,\uninit}$  and if $\aninstr$ is $\falseguard{e_0}$ then $\evalof{\values}{e_0} = 0$; in both cases, $\values' = \values$, $\stateFunction'(\athread) = \stateActive{\anode'}$, $\stateFunction(\tilde\athread) = \stateFunction'(\tilde\athread)$ for all $\tilde\athread \ne \athread$ and $\held' = \held$.
\item[Mutex operations.] If $\aninstr$ is $\lock{\amutex}$ then $\held(\amutex) = \bot$ and $\held'(\amutex) = \athread$, and if $\aninstr$ is $\unlock{\amutex}$ then $\held(\amutex) = \athread$ and $\held'(\amutex) = \bot$; in both cases $\values' = \values$, $\stateFunction'(\athread) = \stateActive{\anode'}$ and $\stateFunction(\tilde\athread) = \stateFunction'(\tilde\athread)$ for all $\tilde\athread \ne \athread$.
\item[Create.] If $\aninstr$ is $\create{\athread_c}$ then $\stateFunction(\athread_c) = \stateUncreated$, $\stateFunction'(\athread_c) = \stateActive{\startnodeof{\athread_c}}$, $\stateFunction'(\athread) = \stateActive{\anode'}$, $\stateFunction(\tilde\athread) = \stateFunction'(\tilde\athread)$ for all $\tilde\athread \in \threads \setminus \set{\athread, \athread_c}$, $\values' = \values$ and $\held' = \held$.
\item[Join.] If $\aninstr$ is $\join{\athread_j}$ then $\stateFunction(\athread_j)  = \stateActive{\termnodeof{\athread_j}}$, $\stateFunction'(\athread_j) = \stateJoined$, $\stateFunction'(\athread) = \stateActive{\anode'}$, $\stateFunction(\tilde\athread)= \stateFunction'(\tilde\athread)$ for all $\tilde\athread \in \threads \setminus \set{\athread, \athread_j}$, $\values' = \values$ and $\held' = \held$.
\end{description}

The valuation $\values'$ is well-formed by construction. From now on, we implicitly assume that all considered valuations are well-formed.  
Given a sequence $s = \atrans_1, \dots, \atrans_k \in \transitions^*$, we let $\config_0 \step{s} \config_k$ when there are $\config_1, \dots, \config_{k-1}$ such that $\config_i \step{\atrans_i} \config_{i+1}$ for all $i \leq k-1$. We denote $\config_1 \step{*} \config_k$ the existence of such a sequence.   
A configuration $\config'$ is \emph{reachable} when $\initialconfig \step{*} \config'$. An \emph{execution} is an alternating sequence $\exec = \initialconfig, \atrans_1, \config_1, \dots, \atrans_k, \config_k$ such that $\config_{i} \step{\atrans_{i+1}} \config_{i+1}$ for all $i \leq k-1$.

\begin{exa}\label{example-reachable-config}
On the example of \autoref{fig:example-program}, by letting $\texttt{t1}$ return before $\texttt{t2}$ starts and assuming that $\texttt{main}$ stops after line~\ref{create-t2} and $\texttt{t2}$ stopping after line~\ref{dblptr-to-p}, we obtain a reachable configuration $\config = \tuple{\stateFunction, \values, \held}$ where:
\begin{itemize}
\item $\stateFunction(\texttt{main}) = \anode_{\ref{create-t2} \rightarrow \ref{access-out}}$, $\stateFunction(\texttt{t1}) = \termnodeof{\texttt{t1}}$ and $\stateFunction(\texttt{t2}) = \anode_{\ref{dblptr-to-p}\rightarrow\ref{access-q}}$ where $\anode_{\ell_1 \rightarrow \ell_2}$ denotes the node between lines $\ell_1$ and $\ell_2$;
\item $\values(\texttt{dbl\_ptr}) = \texttt{q}$, $\values(\texttt{p}) = \values(\texttt{q}) = \texttt{k}$, $\values(\texttt{k}) = \values(\texttt{j}) = 0$ and $\values(\texttt{out}) = \uninit$;
\item $\held(\texttt{m}) = \texttt{t2}$.
\end{itemize}
From $\config$, $\texttt{main}$ and $\texttt{t2}$ may perform concurrent accesses to $\texttt{k}$, \emph{i.e.}, a data race.
\end{exa}


\section{Concurrency Analysis}\label{sec:analysis}
We start with a preliminary concurrency analysis of the program. This analysis concerns when two statements may run in parallel as well as when locks are held. It does not perform any sort of value analysis (\emph{e.g.}, we do not analyze which branch of a \codeif is taken). Indeed, value analysis for concurrent programs tends to be expensive as it requires to take into account interferences between threads, whereas we aim for a fast analysis at the cost of reduced precision. 
We compute, for every node $\anode$, three sets: $\uncreated{\anode}$ (\emph{analysis of uncreated threads}), $\joined{\anode} \subseteq \threads$ (\emph{analysis of joined threads}) and $\lockset{\anode} \subseteq \mutexes$ (\emph{lockset analysis}). When the corresponding thread $\threadof{\anode}$ is in $\anode$, threads in $\uncreated{\anode}$ cannot be created yet, threads in $\joined{\anode}$ must already be joined, and mutexes in $\lockset{\anode}$ are held by $\threadof{\anode}$. 

\subsection{Lockset Analysis}

We start with the analysis of mutexes held. Trivially, $\lockset{\startnodeof{\athread}} = \emptyset$: a thread that was just created holds no mutex. For a node $\anode$, a mutex $\amutex$ is in $\lockset{\anode}$ when it is guaranteed to be held when $\threadof{\anode}$ is in $\anode$, \emph{i.e.}, when $\lockset{\anode}$ is held regardless of the path taken by $\threadof{\anode}$ to get to $\anode$. More precisely, $\amutex$ is guaranteed to be held by $\threadof{\anode}$ at $\anode$ when, for every $\anode'$ with a transition $(\anode',\aninstr, \anode) \in \transitions$, the following condition holds: $\aninstr$ is not $\unlock{\amutex}$ and, if $\aninstr$ is \emph{not} $\lock{\amutex}$, then $\amutex \in \lockset{\anode'}$. In words, no transition leading to $\anode$ should unlock $\amutex$ and, if a given transition does not lock $\amutex$, then $\amutex$ was already held before the transition.  
Overall, $(\lockset{\anode})_{\anode \in \nodes}$ is the greatest $\size{\nodes}$-uple of subsets of $\mutexes$ (for component-wise inclusion) that satisfies:

\vspace{-0.4cm}
\begin{gather*}
\forall \athread \in \threads, \,  \lockset{\startnodeof{\athread}} \deff \emptyset  \\
\forall \anode \in \nodes \setminus \set{\startnodeof{\athread} \mid \athread \in \threads},  \lockset{\anode} \deff \\
\bigcap_{\substack{\anode' \in \nodes, \aninstr \in \instructions  \\ \aninstr \ne \lock{\_} \\ \aninstr \ne \unlock{\_} \\ (\anode',\aninstr, \anode) \in \transitions}} \lockset{\anode'} \cap \bigcap_{\substack{\anode' \in \nodes, \amutex \in \threads \\ (\anode', \lock{\amutex}, \anode) \in \transitions}} (\lockset{\anode'} \cup \set{\amutex}) \cap \bigcap_{\substack{\anode' \in \nodes, \amutex \in \threads \\ (\anode', \unlock{\amutex}, \anode) \in \transitions}} (\lockset{\anode'} \setminus \set{\amutex})
\end{gather*}
We use convention $\bigcap_{x \in \emptyset} \_ = \mutexes$.
We call \emph{tuple of subsets} a $\size{\nodes}$-uple of subsets of $\mutexes$.
Given a tuple of subsets $\vec{x} \in (2^\mutexes)^\nodes$ and a node $\anode \in \nodes$, we let $\veccomp{\vec{x}}{\anode} \in \nodes$ denote the value of $\vec{x}$ at the index corresponding to $\anode$. Tuples of subsets are ordered using component-wise inclusion, so that $\vec{x}$ is smaller than $\vec{y}$ when, for all $\anode$, $\veccomp{\vec{x}}{\anode} \subseteq \veccomp{\vec{y}}{\anode}$. 

\begin{prop}
There is a greatest tuple of subsets satisfying the condition above.
\end{prop}
\begin{proof}
We prove it using the Knaster-Tarski theorem \cite{tarski}.
To do so, we define $\vec{L} = (\lockset{\anode})_{\anode \in \nodes} \in (2^\mutexes)^\nodes$ as the fixpoint of a function $\phi: (2^\mutexes)^\nodes \to (2^\mutexes)^\nodes$. 
The function $\phi$ is defined by letting, for every $\vec{\ell} \in (2^\mutexes)^\nodes$:
\begin{gather*}
\forall \athread \in \threads, \,  \veccomp{\phi(\vec{\ell})}{\startnodeof{\athread}} \deff \emptyset  \\
\forall \anode \in \nodes \setminus \set{\startnodeof{\athread} \mid \athread \in \threads},  \veccomp{\phi(\vec{\ell})}{\anode} \deff \\
\bigcap_{\substack{\anode' \in \nodes, \aninstr \in \instructions  \\ \aninstr \ne \lock{\_} \\ \aninstr \ne \unlock{\_} \\ (\anode',\aninstr, \anode) \in \transitions}} \veccomp{\vec{\ell}}{\anode'} \cap \bigcap_{\substack{\anode' \in \nodes, \amutex \in \threads \\ (\anode', \lock{\amutex}, \anode) \in \transitions}} (\veccomp{\vec{\ell}}{\anode'} \cup \set{\amutex}) \cap \bigcap_{\substack{\anode' \in \nodes, \amutex \in \threads \\ (\anode', \unlock{\amutex}, \anode) \in \transitions}} (\veccomp{\vec{\ell}}{\anode'} \setminus \set{\amutex})
\end{gather*}
Again, we use convention $\bigcap_{x \in \emptyset} \_ = \mutexes$. 
It is easy to see that $\phi$ is order-preserving for component-wise inclusion in the complete lattice $(2^{\mutexes})^{\nodes}$: if we have $\vec{\ell}, \vec{\ell'}$ such that $\veccomp{\vec{\ell}}{\anode} \subseteq \veccomp{\vec{\ell'}}{\anode}$ for all $\anode$, then $\veccomp{\phi(\vec{\ell})}{\anode} \subseteq \veccomp{\phi(\vec{\ell'})}{\anode}$ for all $\anode$.  By the Knaster-Tarski theorem, it admits a unique greatest fixpoint $\vec{L} = (\lockset{\anode})_{\anode \in \nodes}$.
\end{proof}

We consider the greatest fixpoint, because a larger $(\lockset{\anode})_{\anode \in \nodes}$ means a more precise analysis. Considering the greatest fixpoint is important in the case of loops, \emph{i.e.}, cycles in the CFG. In that case, setting $\lockset{\anode} = \emptyset$ for every $\anode$ in the cycle would satisfy the constraints, even if the cycle can be accessed by a single transition which locks a mutex. 

In practice, this fixpoint analysis is easy to implement by exploring the control flow of each thread; performing an actual fixpoint computation is needed only when the CFG has a cycle. This analysis is sound:

\begin{prop}
\label{soundness-lockset}
For every $\anode \in \nodesof{\athread}$,  for every $\config = \tuple{\stateFunction, \values, \held}$ such that $\initialconfig \step{*} \config$ and $\stateFunction(\threadof{\anode}) = \stateActive{\anode}$, for every $\amutex \in \lockset{\anode}$, we have $\held(\amutex) = \threadof{\anode}$. 
\end{prop}
\begin{proof}
We proceed by induction on the length of the considered execution $\exec: \initialconfig \step{*} \config$.
If $\exec$ is of length $0$, then $\config = \initialconfig$ so that $\athread = \maint$ and $\anode = \startnodeof{\maint}$; in this case, $\lockset{\anode} = \emptyset$ so the property trivially holds. 
Suppose that the property holds for $k$, and let $\exec: \initialconfig \step{*} \config$ be an execution of length $k+1$. We decompose $\exec$ into an execution $\exec_k: \initialconfig \step{*} \config_1 = \tuple{\stateFunction_1, \values_1, \held_1}$ of length $k$ and a final step $\config_1 \step{\atrans} \config_2 = \tuple{\stateFunction_2, \values_2, \held_2}$.  Let $(\anode_1, \aninstr, \anode_2) \deff \atrans$; let $\athread := \threadof{{\anode_1}}$ denote the thread performing $\atrans$. 
We perform a case disjunction on $\aninstr$.
\begin{itemize}
\item If $\aninstr$ is not a mutex locking or unlocking nor a create instruction then $\held_1 = \held_2$: all threads hold the same mutexes in $\config_1$ and in $\config_2$. By induction hypothesis, all threads except $\athread$ are in the same node in $\config_1$ and $\config_2$. Moreover, by definition of $\lockset{\anode_2}$, $\lockset{\anode_2} \subseteq \lockset{\anode_1}$ and $\config_2$ satisfies the property.
\item If $\aninstr = \create{\athread_c}$ then the same argument works except for $\athread_c$. However, the property trivially holds for $\athread_c$ since $\lockset{\startnodeof{\athread_c}} = \emptyset$.
\item If $\aninstr = \unlock{\amutex}$, all threads except $\athread$ are in the same node and hold the same mutexes in $\config_1$ and $\config_2$. Also, $\lockset{\anode_2} \subseteq \lockset{\anode_1} \setminus \set{\amutex}$ by definition of $\lockset{\anode_2}$. Moreover, $\held_2(\amutex') = \held_1(\amutex')$ for all $\amutex' \ne \amutex$, so that the property holds.
\item If $\aninstr = \lock{\amutex}$, all threads except $\athread$ are in the same node and hold the same mutexes in $\config_1$ and $\config_2$. Also, we have $\lockset{\anode_2} \subseteq \lockset{\anode_1} \cup \set{\amutex}$. Thanks to the semantics of $\lock{\amutex}$, $\held_2(\amutex) = \athread$ and the property holds. 
\end{itemize}
\end{proof}

\subsection{Analysis of Joined Threads}

We continue with the analysis of joined threads. Trivially, $\joined{\startnodeof{\maint}} = \emptyset$: there can be no joined thread while $\maint$ has not yet started executing. When a thread $\athread \ne \maint$ has been created by a transition $(\anode_1, \create{\athread}, \anode_2)$, threads in $\joined{\anode_2}$ can be considered joined at the start of $\athread$. If there are several transitions creating $\athread$, we must conservatively intersect over all such transitions, giving constraint 
\(\joined{\startnodeof{\athread}} \subseteq \bigcap_{(\anode_1, \create{\athread}, \anode_2) \in \transitions} \joined{\anode_2}.\)
For $\anode \ne \startnodeof{\threadof{\anode}}$, a thread $\athread$ is considered joined if the following holds for every $(\anode', \aninstr, \anode) \in \transitions$. If $\aninstr$ is not a join, then we need $\athread \in \joined{\anode'}$. If $\aninstr$ is a $\join{\athread}$, there is no condition. Finally, if $\aninstr$ is a $\join{\athread_j}$ with $\athread_j \ne \athread$, then $\athread$ must be in $\joined{\termnodeof{\athread_j}}$: if $\athread$ must be joined at the end of $\athread_j$, then once $\athread_j$ is joined, $\athread$ must also be joined. Overall, $(\joined{\anode})_{\anode \in \nodes}$ is the greatest $\size{\nodes}$-uple of subsets of threads such that:

\vspace{-0.4cm}
\begin{align*}
& \joined{\startnodeof{\maint}} \deff \emptyset  \\
& \forall \athread \in \threads \setminus \set{\maint}, \,   \joined{\startnodeof{\athread}} \deff 
\bigcap_{\substack{\anode_1, \anode_2 \in \nodes \\ (\anode_1, \create{\athread}, \anode_2) \in \transitions}} \joined{\anode_2} \\ &
\forall \anode \in \nodes \setminus \set{\startnodeof{\athread} \mid \athread \in \threads}, \\ & 
\joined{\anode} \deff \bigcap_{\substack{\anode_p \in \prednojoin{\anode}}} \joined{\anode_p} \cap \bigcap_{\substack{(\anode_p, \athread_j) \in \predjoin{\anode}}} \left( \joined{\anode_p} \cup \set{\athread_j} \cup \joined{\termnodeof{\athread_j}} \right)
\end{align*}
where we use the convention $\bigcap_{x \in \emptyset} \_ = \threads$ as before, and where:

\begin{align*}
\predjoin{\anode} & \deff \set{(\anode_p, \athread_j) \in \nodes \times \threads \mid (\anode_p, \join{\athread_j}, \anode) \in \transitions} \\
\prednojoin{\anode} & \deff \set{\anode_p \in \nodes \mid \exists \, \aninstr \text{ s.t. } (\anode_p, \aninstr, \anode) \in \transitions \text{ and } \aninstr \notin \set{\join{\athread} \mid \athread \in \threads}}. 
\end{align*}
In words, $\predjoin{\anode}$ contains pairs $(\anode_p, \athread_j)$ such that there is a transition from $\anode_p$ to $\anode$ with instruction $\join{\athread_j}$, while $\prednojoin{\anode}$ contains all nodes $\anode_p$ from which there is a transition to $\anode$ whose instruction is not a join.  

Proving the existence of $(\joined{\anode})_{\anode \in \nodes}$ is once again done using the Knaster-Tarski theorem, by noting that the function defined above is order-preserving for component-wise inclusion. 
This analysis is sound:
\begin{prop}
\label{soundness-joined}
For every $\anode \in \nodes$, for every $\config = \tuple{\stateFunction, \values, \held}$ such that $\initialconfig \step{*} \config$ and $\stateFunction(\threadof{\anode}) = \stateActive{\anode}$, for every $\athread_j \in \joined{\anode}$, we have $\stateFunction(\athread_j) = \stateJoined$. 
\end{prop} 
\begin{proof}
We proceed by induction on the length of the considered execution $\exec: \initialconfig \step{*} \config$.
If $\exec$ is of length $0$, then $\config = \initialconfig$ so that $\maint$ is the only active thread and no thread is joined, therefore the property holds since $\joined{\startnodeof{\maint}} = \emptyset$.
Suppose that the property holds for $k$, and let $\exec: \initialconfig \step{*} \config$ be an execution of length $k+1$. We decompose $\exec$ into an execution $\exec_k: \initialconfig \step{*} \config_1 = \tuple{\stateFunction_1, \values_1, \held_1}$ of length $k$ and a final step $\config_1 \step{\atrans} \config_2 = \tuple{\stateFunction_2, \values_2, \held_2}$.  Let $(\anode_1, \aninstr, \anode_2) \deff \atrans$; let $\athread := \threadof{{\anode_1}}$ denote the thread performing $\atrans$. 
We perform a case disjunction on $\aninstr$.
\begin{itemize}
\item If $\aninstr$ is not a join nor a create instruction, joined threads are the same in $\config_1$ and $\config_2$. Moreover, $\anode_1 \in \prednojoin{\anode_2}$ hence, by definition of $\joined{\anode_2}$, $\joined{\anode_2} \subseteq \joined{\anode_1}$. By hypothesis, every thread in $\joined{\anode_1}$ is joined in $\config_1$ hence also in $\config_2$. 
\item If $\aninstr = \join{\athread_j}$, then $\joined{\anode_2} \subseteq \joined{\anode_1} \cup \set{\athread_j} \cup \joined{\termnodeof{\athread_j}}$; we prove that each thread $\athread' \in  \joined{\anode_1} \cup \set{\athread_j} \cup \joined{\termnodeof{\athread_j}}$ is joined in $\config_2$. By induction hypothesis, every thread in $\joined{\anode_1}$ is joined in $\config_1$, therefore also in $\config_2$. Because $\config_1 \step{\atrans} \config_2$ and thanks to the semantics of $\join{\athread_j}$, $\athread_j$ is in $\stateJoined$ in $\config_2$. Finally, since $\athread_j$ is in $\termnodeof{\athread_j}$ in $\config_1$, by induction hypothesis, all threads in $\joined{\termnodeof{\athread_j}}$ are joined in $\config_1$, therefore also in $\config_2$. 
\item If $\aninstr = \create{\athread_c}$ then it corresponds to the first case above except that $\stateFunction'(\athread_c) = \startnodeof{\athread_c}$. By definition, $\joined{\startnodeof{\athread_c}} \subseteq \joined{\anode_1}$ since $\atrans$ appears in the intersection in the definition of $\joined{\startnodeof{\athread_c}}$. All threads in $\joined{\anode_1}$ are joined in $\config_1$ by induction hypothesis, therefore also in $\config_2$. This proves that all threads $\joined{\startnodeof{\athread_c}}$ are joined in $\config_2$.       
\end{itemize}  
\end{proof}

\subsection{Analysis of Uncreated Threads}

Finally, we present the analysis of uncreated threads. It relies on the same principles but is more complex. 
There are two main challenges: how a creation transition $(\anode', \create{\athread_c}, \anode)$ impacts $\uncreated{\anode}$, and what lies in the set $\uncreated{\startnodeof{\athread}}$ of the start node of a thread. 

\begin{wrapfigure}{r}{2.0cm}
\tikz[yscale = 0.7]{\node at (0,0) (t1) {$\athread_1$}; \node at (-0.5,-1) (t2){$\athread_2$}; \node at (0.5,-1) (t3) {$\athread_3$}; \node at (-0.5,-2) (t4) {$\athread_4$}; \draw (t1) -- (t2); \draw (t2) -- (t4); \draw (t1) -- (t3);}
\end{wrapfigure}
To explain our solution, we illustrate it on the following running example. Consider a program where executions behave as follows. A thread $\athread_1 = \maint$ creates a thread $\athread_2$, then joins $\athread_2$, then creates another thread $\athread_3$; $\athread_2$ itself creates a thread $\athread_4$. See opposite for a depiction of the creation graph.

Consider the transition $(\anode', \create{\athread_2}, \anode) \in \transitionsof{\athread_1}$ creating $\athread_2$. We must have $\athread_2 \notin \uncreated{\anode}$. Moreover, $\athread_2$ may in turn create thread $\athread_4$, so that, at $\anode$, $\athread_4$ is potentially created. We express this with constraint
\[ 
\uncreated{\anode} \subseteq (\threads \setminus \set{\athread_2}) \cap \bigcap_{\anode_2 \in \nodesof{\athread_2}} \uncreated{\anode_2}.
\]

We now explain which threads are considered uncreated at a start node $\startnodeof{\athread}$. 
First, consider the situation at the start node $\startnodeof{\athread_2}$ of $\athread_2$. Obviously, we must have $\uncreated{\startnodeof{\athread_2}} \subseteq \threads \setminus \set{\athread_2}$. 
Since $(\anode', \create{\athread_2}, \anode) \in \transitionsof{\athread_1}$ is the transition creating $\athread_2$, at the start of $\athread_2$, threads in $\threads \setminus \uncreated{\anode}$ are possibly created. 
Moreover, $\athread_1$ keeps running in parallel with $\athread_2$: subsequent nodes of $\athread_1$ are also taken into account, but only until $\athread_2$ is joined. In particular, $\athread_3$ is never created before $\athread_2$ is joined, so it is uncreated at every node in $\athread_2$. 
This leads us to the constraint 
\[\uncreated{\startnodeof{\athread_2}} \subseteq  \bigcap_{\anode \in \nodesof{\athread_1}, \athread_2 \notin \uncreated{\anode}, \athread_2 \notin \joined{\anode}} \uncreated{\anode}.\]

Consider now the case of the start node of $\athread_4$. While $\athread_2$ cannot run in parallel with $\athread_3$,  its child $\athread_4$ can: $\athread_4$ remains active after $\athread_2$ is joined. If $\athread_4$ obtained information from its parent's nodes only, it would wrongly deduce that $\athread_3$ is uncreated. For this reason, there is a transfer of information from the \emph{grandparents} to the \emph{grandchildren}, here from $\athread_1$ to $\athread_4$. This here gives $\uncreated{\startnodeof{\athread_4}} \subseteq \bigcap_{\anode \in \nodesof{\athread_1}} \uncreated{\anode}$. 

We define $(\uncreated{\anode})_{\anode \in \nodes}$ as the greatest $\size{\nodes}$-uple of subsets of $\threads$ such that:

\begin{align*}
& \uncreated{\startnodeof{\maint}}  \deff \threads \setminus \set{\maint}  \\
& \forall \athread \in \threads \setminus \set{\maint}, \,   \uncreated{\startnodeof{\athread}} \deff (\threads \setminus \set{\athread}) \cap \bigcap_{\substack{\anode \in \parentnodes(\athread) \\ \athread \notin \joined{\anode}, \athread \notin \uncreated{\anode}}} \uncreated{\anode} \cap \bigcap_{\substack{\anode \in \grandparentnodes(\athread)}} \uncreated{\anode} \\ &
\forall \anode \in \nodes \setminus \set{\startnodeof{\athread} \mid \athread \in \threads}, \\ &
\uncreated{\anode} \deff \bigcap_{\substack{\anode_p \in \prednocreate{\anode}}} \uncreated{\anode_p} \cap \bigcap_{\substack{(\anode_p, \athread_c) \in \predcreate{\anode}}} \left( (\uncreated{\anode_p} \setminus \set{\athread_c}) \cap \bigcap_{\anode_c \in \nodesof{\athread_c}} \uncreated{\anode_c} \right)
\end{align*}
where we use convention $\bigcap_{x \in \emptyset} \_ = \threads$ and where:
\begin{align*} 
\predcreate{\anode} & \deff \set{(\anode_p, \athread_\childsymb) \in \nodes \times \threads \mid (\anode_p, \create{\athread_\childsymb}, \anode) \in \transitions} \\
\prednocreate{\anode} & \deff \set{\anode_p \in \nodes \mid \exists \, \aninstr \text{ s.t. } (\anode_p, \aninstr, \anode) \in \transitions \text{ and } \aninstr \notin \set{\create{\athread} \mid \athread \in \threads}} \\
\nodes_{\mathsf{par}}(\athread) & \deff \set{\anode \in \nodes \mid \exists \athread_\parentsymb \in \threads: \, \anode \in \nodesof{\athread_\parentsymb} \land (\exists \anode_1, \anode_2: \, (\anode_1, \create{\athread}, \anode_2) \in \transitionsof{\athread_\parentsymb})} \\
\grandparentnodes(\athread) & \deff \set{\anode \in \nodes \mid \exists \athread_\parentsymb, \athread_{\grandparentsymb} \in \threads: \, \anode \in \nodesof{\athread_{\grandparentsymb}} \land (\exists \anode_1, \anode_2: \, \\  &(\anode_1, \create{\athread_\parentsymb}, \anode_2) \in \transitionsof{\athread_{\grandparentsymb}}) \land (\exists \anode_1, \anode_2: \, (\anode_1, \create{\athread_{\grandparentsymb}}, \anode_2) \in \transitionsof{\athread_\parentsymb})} 
\end{align*}
In words, $\predcreate{\anode}$ contains nodes and threads of creation transitions leading to $\anode$, $\prednocreate{\anode}$ contains other predecessors, $\nodes_{\mathsf{par}}(\athread)$ contains the nodes of (possible) parent threads of $\athread$, and $\grandparentnodes(\athread)$ contains nodes of grandparent threads of $\athread$. 

Like for the two previous cases, the existence of $(\uncreated{\anode})_{\anode \in \nodes}$ is proven using the Knaster-Tarski theorem. However, there is a subtlety concerning the order preservation in this case, which is the condition $\athread \notin \veccomp{\vec{u}}{\anode}$ in the first big intersection of the definition of $\uncreated{\startnode{\athread}}$. However, the set of terms in the intersection becomes smaller when $\uncreated{\anode}$ becomes bigger; because intersecting less terms gives a larger resulting set, this is not an obstacle to order preservation. 
The proof of soundness is significantly more complex than the previous two proofs. We introduce it in a separate section.

\subsection{Soundness for Uncreated Threads}

Our goal is to prove soundness of the analysis of uncreated threads, expressed as follows.
\begin{thm}
\label{soundness-uncreated}
For every $\anode \in \nodes$, for every $\config = \tuple{\stateFunction, \values, \held}$ such that $\initialconfig \step{*} \config$ and $\stateFunction(\threadof{\anode}) = \stateActive{\anode}$, we have $\stateFunction(\athread) = \stateUncreated$ for each $\athread \in \uncreated{\anode}$.
\end{thm}

The proof proceeds in two steps. First, we show that information about possibly created threads propagates upward in the creation tree (\autoref{uncreated-firstlemma}). Second, we show that information also propagates downward from ancestors to descendants (\autoref{uncreated-secondlemma}). This second point is the most difficult one, and in particular it requires additional notations.

Fix an execution $\exec: \initialconfig \step{*} \config$ with $\config \ffed \tuple{\stateFunction, \values, \held}$. Let $G$ be the graph whose vertices are the threads created in $\config$, with an edge from $\athread$ to $\athread'$ whenever $\athread$ is the thread that performs the transition creating $\athread'$ in $\exec$. Because threads cannot be created several times, $G$ is a tree whose root is $\maint$. 

\begin{wrapfigure}{r}{2.0cm}
\tikz[yscale = 0.7]{\node at (0,0) (t1) {$\athread_1$}; \node at (-0.5,-1) (t2){$\athread_2$}; \node at (0.5,-1) (t3) {$\athread_3$}; \node at (-0.5,-2) (t4) {$\athread_4$}; \draw (t1) -- (t2); \draw (t2) -- (t4); \draw (t1) -- (t3);}
\end{wrapfigure}
Consider an execution whose creation graph is the following, because the following events occured. A thread $\athread_1 = \maint$ creates a thread $\athread_2$, then joins $\athread_2$, then creates another thread $\athread_3$; $\athread_2$ itself creates a thread $\athread_4$.

We start with the following observation: whenever there is a path from $\anode_1$ to $\anode_2$, $\uncreated{\anode_2} \subseteq \uncreated{\anode_1}$. This is a direct induction on the length of the path. In particular, $\athread \notin \uncreated{\anode}$ for each $\anode \in \nodesof{\athread}$, because $\athread \notin \uncreated{\startnodeof{\athread}}$ by definition. 

The following lemma expresses that the information flows correctly from child to parent, \emph{i.e.}, that, after creating a child, the thread considers the child and all its possible descendants as possibly created. In the example above, this would imply that $\athread_1$ is aware that $\athread_2$ and $\athread_4$ are possibly created.  
\begin{lem}
\label{uncreated-firstlemma}
Let $\athread_1 \to \dots \to \athread_k$ be a path in $G$. For every $i \leq k-1$, let $(\anode_i, \create{\athread_{i+1}}, \anode_{i}')$ denote the transition performed by $\athread_i$ to create $\athread_{i+1}$ in $\exec$. For each $i \in \nset{2}{k}$, we have \newline $\set{\athread_i, \athread_{i+1}, \dots, \athread_k} \cap \uncreated{\anode_{i-1}'} = \emptyset$.
\end{lem}
\begin{proof}
We proceed by induction on $k$. The property trivially holds for $k=1$; let $k \geq 2$.
Let $\athread_1 \to \dots \to \athread_{k} \to \athread_{k+1}$, and let $i \in \nset{2}{k}$. By induction hypothesis applied to $\athread_{i+1} \to \dots \to \athread_{k+1}$, $\set{\athread_{i+1}, \dots, \athread_{k+1}} \cap  \uncreated{\anode_i'} = \emptyset$. 
From the definition of $\uncreated{\anode_{i-1}'}$, we have:
\[ \uncreated{\anode_{i-1}'} \subseteq \bigcap_{\substack{(\anode_p, \athread_c) \in \predcreate{\anode_{i-1}'}}} \left( (\uncreated{\anode_p} \setminus \set{\athread_c}) \cap \bigcap_{\anode_c \in \nodesof{\athread_c}} \uncreated{\anode_c} \right).\] 
Because there is a transition from $\anode_{i-1}$ to $\anode_{i-1}'$ creating $\athread_i$, we have $(\anode_{i-1}, \athread_{i}) \in \predcreate{\anode_{i-1}'}$. 
This implies that $\uncreated{\anode_{i-1}} \setminus \set{\athread_i}$ appears in the intersection above, therefore $\uncreated{\anode_{i-1}'} \subseteq (\uncreated{\anode_{i-1}} \setminus \set{\athread_i})$ and $\athread_i \notin \uncreated{\anode_{i-1}'}$. 
It remains to prove that $\set{\athread_{i+1}, \dots, \athread_{k+1}} \cap \uncreated{\anode_{i-1}'} = \emptyset$. 
Recall that, by induction hypothesis, $\set{\athread_{i+1}, \dots, \athread_{k+1}} \cap  \uncreated{\anode_i'} = \emptyset$; we must prove this same property for $\uncreated{\anode_{i-1}'}$. 
To do so, it suffices to observe that, since $(\anode_{i-1}, \athread_{i}) \in \predcreate{\anode_{i-1}'}$ and since $\anode_i' \in \nodesof{\athread_i}$, $\uncreated{\anode_i'}$ appears in the intersection above. 
From this and from the induction hypothesis, we deduce $\set{\athread_{i+1}, \dots, \athread_k} \cap \uncreated{\anode_{i-1}'} = \emptyset$. 
Combined with the observation above that $\athread_i \notin \uncreated{\anode_{i-1}'}$, we obtain $\set{\athread_i, \athread_{i+1},\dots, \athread_{k+1}} \cap \uncreated{\anode_{i-1}'} = \emptyset$. 
 \end{proof}

This proves that the information flows ``upward'' in the creation graph; it remains to prove that the information flows ``downward'' in that tree, in the sense that a thread has gathered enough information from its ancestors. Not only must the thread consider its ancestors as created, but also all descendants of its ancestors in other branches. In the running example, $\athread_4$ must for instance be aware that $\athread_3$ is possibly created. 
Towards this objective, we introduce a few definitions that will allow to treat nodes in $\nodes$ and the joined states symmetrically. To do so, we define an enriched set of nodes which includes joined states and such that each thread has its own, separate joined state. 
This will largely simplify the statement and proof of the second lemma below. 

\begin{defi}
\label{def-auxnodes}
Let $\auxnodes \deff \nodes  \cup (\set{\stateJoined} \times\threads)$ and, for each thread $\athread \in \threads$, $\auxnodes_{\athread} \deff \nodesof{\athread} \cup \set{(\stateJoined, \athread)}$.
We let $\auxuncsymb: \auxnodes \to 2^\threads$ with, for every $\anode \in \auxnodes$: \[ \auxunc{\anode} \deff \begin{cases} \uncreated{\anode} \text{ when } \anode \in \nodes \\
    \uncreated{\termnodeof{\athread}} \cap \bigcap_{\substack{\anode_{\parentsymb} \in \parentnodes(\athread) \\ \athread \notin \uncreated{\anode_{\parentsymb}}}} \uncreated{\anode_{\parentsymb}} \text{ when } \anode = (\stateJoined, \athread).
\end{cases}\]
Finally, we define $\auxstsymb: \threads \to \auxnodes$ by letting, for every $\athread \in \threads$ such that $\stateFunction(\athread) \ne \stateUncreated$:
\[ 
\auxst{\athread} \deff \begin{cases} \stateFunction(\athread) \text{ when }\stateFunction(\athread) \ne \stateJoined \\  (\stateJoined, \athread) \in \auxnodes \text{ when } \stateFunction(\athread) = \stateJoined \end{cases}
\]
\end{defi}
We now explain this definition. Recall that \[\nodes_{\mathsf{par}}(\athread) \deff \set{\anode \in \nodes \mid \exists \athread_\parentsymb \in \threads: \, \anode \in \nodesof{\athread_\parentsymb} \land (\exists \anode_1, \anode_2: \, (\anode_1, \create{\athread}, \anode_2) \in \transitionsof{\athread_\parentsymb})}\] is the set of nodes of threads that could be parents of $\athread$. 
The set $\auxnodes$ denotes the set of nodes enriched with the joined state of each thread. 
Elements of $\auxnodes$ describe the states of created threads; once thread $\athread$ has been joined, it is in state $(\stateJoined,\athread)$. A crucial feature of $\auxnodes$ is that the sets of elements of $\auxnodes$ that two distinct threads may access are disjoint; while different threads can be in state $\stateJoined$, only thread $\athread$ can be in $(\stateJoined, \athread)$. 

Regarding $\auxuncsymb$, it is an extension of $\uncreatedsymb$ that also contains information relevant after $\athread$ is joined. For this reason, it includes constraints related to nodes of creating threads that were not considered in $\veccomp{\vec{u}}{\startnodeof{\athread}}$ because of the condition $\athread \notin \joined{\anode}$. For example, in the example above, we could have $\athread_3 \in \uncreated{\termnodeof{\athread_2}}$ because $\athread_2$ is joined before $\athread_3$ is created, but we would still have $\athread_3 \notin \auxunc{(\stateJoined, \athread_2)}$. This will allow for a more convenient induction hypothesis in the proof of \autoref{uncreated-secondlemma} below. 
Observe that we have $\auxunc{\anode} = \uncreated{\anode}$ for all $\anode \in \nodes$, so that the previous lemma holds with $\auxunc{\anode}$ instead of $\uncreated{\anode}$. 

Regarding $\auxstsymb: \threads \to \auxnodes$, it is the adaptation of $\stateFunction$ that returns an element in $\auxnodes$. Recall that $\stateFunction$ is the state function of the configuration $\config$, which was fixed at the beginning of the proof. For every $\athread \in \threads$, because there is a path from $\startnodeof{\athread}$ to $\termnodeof{\athread}$, the observation extends to $\auxnodes$:  $\auxunc{\anode} \subseteq \uncreated{\startnodeof{\athread}}$ for every $\anode \in \auxnodes_{\athread}$. 

We prove the following lemma:
\begin{lem}
\label{uncreated-secondlemma}
For every edge $\athread \to \athread'$ in $G$ (\emph{i.e.}, $\athread$ creates $\athread'$ in $\exec$),  $\auxunc{\auxst{\athread'}} \subseteq \auxunc{\auxst{\athread}}$. 
\end{lem}
\begin{proof}
Note that $\auxstsymb$ can be applied to $\athread$ and $\athread'$, since they are created in $\config$ by hypothesis.
We distinguish three cases, depending on whether $\athread$ is in joined state and, if not, depending on whether $\athread'$ is in $\joined{\auxst{\athread}}$.
\begin{description}
\item[$\auxst{\athread} \in \nodes \text{ and } \athread' \notin \joined{\auxst{\athread}}$.] This is the case where $\athread$ is not yet joined and where $\athread'$ is not considered joined at the current node of $\athread$.  
In this case, $\auxst{\athread} \in \parentnodes(\athread')$ (recall that $\parentnodes(.)$ is defined in the definition of $\phi$ above). By considering the transition $(\anode_1, \create{\athread'},\anode_2) \in \transitionsof{\athread}$ creating $\athread'$ in $\exec$, there is a path from $\anode_2$ to $\auxst{\athread}$ therefore $\athread' \notin \uncreated{\auxst{\athread}}$. We have proven that $\auxst{\athread}$ appears in the first big intersection in the definition of $\uncreated{\startnodeof{\athread'}}$, \emph{i.e.}, the one with condition that $\anode \in \parentnodes(\athread')$, that $\athread' \notin \joined{\auxst{\athread}}$ (true by hypothesis) and that $\athread' \notin \uncreated{\auxst{\athread}}$ (proven above).
Therefore, $\uncreated{\startnodeof{\athread'}} \subseteq \uncreated{\auxst{\athread}} = \auxunc{\auxst{\athread}}$. Since $\auxunc{\auxst{\athread'}} \subseteq \uncreated{\startnodeof{\athread'}}$, this gives $\auxunc{\auxst{\athread'}} \subseteq \auxunc{\auxst{\athread}}$, concluding this case. 
\item[$\auxst{\athread} \in \nodes \text{ and } \athread' \in \joined{\auxst{\athread}}$.] This is the case where $\athread$ is not joined, and where $\athread'$ is considered joined at the current node of $\athread$. 
By \autoref{soundness-joined}, this implies that $\stateFunction(\athread') = \stateJoined$. Therefore, $\auxunc{\auxst{\athread'}} = \auxunc{(\stateJoined, \athread')} = \uncreated{\termnodeof{\athread'}} \cap \bigcap_{\substack{\anode_{\parentsymb} \in \parentnodes(\athread') \\ \athread' \notin \uncreated{\anode_{\parentsymb}}}} \uncreated{\anode_{\parentsymb}}$. In particular, $\auxst{\athread}$ appears in this intersection, because $\auxst{\athread} \in \parentnodes(\athread')$ since $\athread$ creates $\athread'$ and, by \autoref{uncreated-firstlemma}, $\athread' \notin \uncreated{\auxst{\athread}}$. We obtain that $\auxunc{\auxst{\athread'}} \subseteq \uncreated{\auxst{\athread}} = \auxunc{\auxst{\athread}}$. Note that this case justifies proving the lemma with $\auxuncsymb$ instead of $\uncreatedsymb$. Indeed, it would not suffice to consider $\auxunc{\termnodeof{\athread'}}$, because it does not always hold that $\uncreated{\termnodeof{\athread'}} \subseteq \uncreated{\auxst{\athread}}$.
\item[$\auxst{\athread} = (\stateJoined,\athread)$.] This is the case where $\athread$ is joined.  
In this case, $\auxst{\athread} = (\stateJoined, \athread)$ and $\auxunc{\auxst{\athread}} = \uncreated{\termnodeof{\athread}} \cap \bigcap_{\substack{\anode_{\parentsymb} \in \parentnodes(\athread) \\ \athread \notin \uncreated{\anode_{\parentsymb}}}} \uncreated{\anode_{\parentsymb}}$. The previous two cases show that $\auxunc{\auxst{\athread'}} \subseteq \uncreated{\termnodeof{\athread}}$. Because $\athread$ has a transition creating $\athread'$, $\parentnodes(\athread) \subseteq \grandparentnodes(\athread')$. Therefore, \( \bigcap_{\substack{\anode_{\grandparentsymb} \in \grandparentnodes(\athread')}} \uncreated{\anode_{\grandparentsymb}} \subseteq \bigcap_{\substack{\anode_{\parentsymb} \in \parentnodes(\athread) \\ \athread \notin \uncreated{\anode_{\parentsymb}}}} \uncreated{\anode_{\parentsymb}}.\). This proves that $\uncreated{\startnodeof{\athread'}}$ and therefore $\auxunc{\auxst{\athread'}}$ are included in the second intersection. Overall, this proves that $\auxunc{\auxst{\athread'}} \subseteq \auxunc{\auxst{\athread}}$.
\end{description} 
\end{proof}

We now prove \autoref{soundness-uncreated}. If $\stateFunction(\maint) \in \nodes$ then, by \autoref{uncreated-firstlemma}, as $\maint$ is at the root of $G$, none of the threads created in $\exec$ appear in $\auxunc{\auxst{\maint}}$. Indeed, for every $(\anode_1, \create{\athread}, \anode_2) \in \transitionsof{\maint}$ performed in $\exec$, there is a path from $\anode_2$ to $\stateFunction(\maint)$. If $\stateFunction(\maint) = \stateJoined$ then \autoref{uncreated-firstlemma} gives that none of threads created in $\exec$ appear in $\uncreated{\termnodeof{\maint}}$; this implies that the same holds for $\auxunc{(\stateJoined,\maint)}$ since, by definition, $\auxunc{(\stateJoined,\maint)} \subseteq \uncreated{\termnodeof{\maint}}$. Overall, we have proven that all threads in $\auxunc{\auxst{\maint}}$ are uncreated in $\config$. By the second lemma and a direct induction along each branch, for every thread $\athread$ created in $\exec$, $\auxunc{\auxst{\athread}} \subseteq \auxunc{\auxst{\maint}}$. This proves that, for each thread $\athread$ created, $\auxunc{\auxst{\athread}}$ contains only threads that are uncreated in $\config$. Because $\uncreated{\stateFunction(\athread)} = \auxunc{\auxst{\athread}}$ whenever $\stateFunction(\athread) \in \nodes$, this concludes the proof of \autoref{soundness-uncreated}. \qed  

\section{Points-to analysis}
\label{sec:points-to-analysis}
In this section, we introduce a points-to analysis of the concurrent program. 
Such an analysis allows to understand which variables can be accessed by a given instruction, which in particular is needed for data race detection. For example, if $\values(x) = y$ then an access to $\deref{x}$ is an access to $y$, which may race with a concurrent access to $y$. 
We compute an overapproximation of the points-to graph at each node $\anode$, \emph{i.e.}, a relation $\pt{\pto(\anode)}$ on $\vars$ such that, for every reachable $\config$ where $\threadof{\anode}$ is in $\anode$, if $x$ points to $y$ in $\config$ then $x \pt{\pto(\anode)} y$. 

\subsection{Points-to graphs}

A \emph{points-to graph} (PTG) represents, given a fixed set of variables, a family of possible valuations of the pointer variables. It consists in a graph where vertices are labelled by sets of variables, and where edges represent the points-to relation. If a vertex labelled $\set{x_2, y_2}$ has a single outgoing edge, which is to a vertex labelled $\set{x_1, y_1}$, the PTG represents valuations where $\values(x_2), \values(y_2) \in \set{x_1, y_1}$. 

It shall later be convenient to artificially extend the set of variables considered in points-to graphs, \emph{i.e.}, consider points-to graphs whose set of variables allows for dummy variables that are not in the program. For this reason, we consider below points-to graphs defined on an abstract set of variables $\extvars$. We will instantiate this set in \autoref{subsec:proc-pta}.  

\begin{defi}
\label{def-ptg}    
A points-to graph is a labelled (directed) graph $\PTG = (\vertices,\edges, \labelling)$ with $\labelling: \vertices \to 2^{\extvars} \setminus \set{\emptyset}$ (vertices are labelled by non-empty sets of variables) such that, for every $x \in \extvars$, there is a unique vertex $\vertexof{x} \in \vertices$ for which $x \in \labelling(\vertexof{x})$. Moreover, each vertex $\avertex \in \vertices$ has a depth $k \in \nats$ and $\labelling(\avertex)$ may only contain variables of depth $k$; children of $\avertex$ must have depth $k-1$\footnote{As a consequence, points-to graphs are acyclic. More precisely, points-to graphs are layered DAGs (Directed Acyclic Graphs).}. 
\end{defi}

We shall consider points-to graphs up to isomorphism. An isomorphism between $\PTG = (\vertices, \edges, \labelling)$ and $\PTG'=(\vertices', \edges', \labelling')$ is a bijective function $\isomorphism: \vertices \rightarrow \vertices'$ such that, for every $\avertex \in \vertices$, $\labelling(\avertex) = \labelling'(\isomorphism(\avertex))$  and, for every $\avertex_1, \avertex_2 \in \vertices$, $(\avertex_1,\avertex_2) \in \edges$ if and only if $(\isomorphism(\avertex_1), \isomorphism(\avertex_2)) \in \edges'$. Intuitively, two PTGs are isomorphic if they are identical up to vertex renaming. We let $\PTG \isom \PTG'$ denote the fact that $\PTG$ and $\PTG'$ are isomorphic. Below, we always consider points-to graphs up to isomorphism.

Given $x, y \in \vars$ and a PTG $\PTG = (\vertices, \edges, \labelling)$, $x$ \emph{points to} $y$ in $\PTG$, denoted $x \pt{\PTG} y$ (or simply $x \pt{} y$ when $\PTG$ is clear from context), when there is an edge from $\vertexof{x}$ to $\vertexof{y}$ in $\PTG$. 
We define a \emph{merge} operation of two vertices of same depth: the two vertices are merged into one, with their edges combined and, as labelling set for the merged vertex, the union of the labelling sets of the two vertices.
\begin{defi}
\label{def-merge}
Given $\PTG = (\vertices, \edges, \labelling)$ and $\avertex_1, \avertex_2 \in \vertices$ of same depth, the points-to graph obtained from $\PTG$ by \emph{merging} $\avertex_1$ and $\avertex_2$ is the PTG $(\vertices', \edges',\labelling')$ s.t.:
\begin{itemize}
\item $\vertices' \deff (\vertices \setminus \set{\avertex_1, \avertex_2}) \cup \set{\avertex}$ with $\avertex \notin \vertices$ a fresh vertex;
\item all edges not touching $\avertex_1$ or $\avertex_2$ remain the same; the set of parents (resp. children) of $\avertex$ is the union of the set of parents (resp. children) of $\avertex_1$ and $\avertex_2$, formally \begin{multline*} \edges' \deff \set{(u, w) \in E \mid u, w \notin \set{\avertex_1, \avertex_2}} \quad \cup \quad \set{(u, \avertex) \mid (u, \avertex_1) \in \edges \vee (u, \avertex_2) \in \edges} \\ \cup \quad \set{(\avertex, w) \mid (\avertex_1, w) \in \edges \vee (\avertex_2, w) \in \edges} \end{multline*}
\item $\labelling'(\avertex) = \labelling(\avertex_1) \cup \labelling(\avertex_2)$ and $\labelling'(\avertex') = \labelling(\avertex')$ for all $\avertex' \in \vertices' \setminus \set{\avertex}$.
\end{itemize}   
\end{defi}
Note that if $\avertex_1= \avertex_2$ then merging $\avertex_1$ and $\avertex_2$ has no effect. 
We now define an order over points-to graphs, which expresses a sufficient condition under which a points-to graph induces a superset of the points-to relations between variables induced by the other.

\begin{defi}
\label{def-subptg}
Given $\PTG = (\vertices, \edges, \labelling)$ and $\PTG' = (\vertices', \edges', \labelling')$, we let $\PTG \subptg \PTG'$ when there is a homomorphism $\homomorphism: \vertices \rightarrow \vertices'$ such that:
\begin{itemize}
\item for every $\avertex \in \vertices$, $\labelling(\avertex) \subseteq \labelling'(\homomorphism(\avertex))$;
\item for every $\avertex_1, \avertex_2 \in \vertices$, if $(\avertex_1, \avertex_2) \in \edges$ then $(\homomorphism(\avertex_1), \homomorphism(\avertex_2)) \in \edges'$. 
\end{itemize}
\end{defi}
In other words, $\PTG \subptg \PTG'$ when $\PTG'$ can be obtained (up to isomorphism) from $\PTG$ by merging vertices and adding edges. 
It is easy to see that $\PTG \subptg \PTG'$ implies that  $\PTG'$ gives more points-to pairs than $\PTG$:
\begin{prop}
\label{sub-less-pt-pairs}
If $\PTG \subptg \PTG'$ then, for every $x,y \in \extvars$, if $x \pt{\PTG} y$ then $x \pt{\PTG'} y$.
\end{prop}
\begin{proof}
Let $\PTG = (\vertices, \edges, \labelling)$ and $\PTG' = (\vertices', \edges', \labelling')$. Let $\homomorphism: \vertices \rightarrow \vertices'$ be a witness that $\PTG \subptg \PTG'$. Let $x,y$ such that $x \pt{\PTG} y$; let $v_x$ be the vertex of $x$ and $v_y$ be the vertex of $y$ in $\PTG$. By hypothesis, $(v_x,v_y) \in \edges$. By hypothesis on $\homomorphism$, $x \in \labelling'(\homomorphism(v_x))$, $y \in \homomorphism'(\homomorphism(v_y))$ and $(\homomorphism(v_x), \homomorphism(v_y)) \in \edges'$. Therefore, $x \pt{\PTG'} y$.  
\end{proof}

It is easy to see that $\subptg$ is a quasi-order: reflexivity is immediate with $\homomorphism = \mathsf{id}$, and transitivity can be proven by composing the homomorphisms. 
Given $\PTG = (\vertices, \edges, \labelling)$ and $\PTG' = (\vertices', \edges', \labelling')$, we let $\PTG \isom \PTG'$ when $\PTG$ and $\PTG'$ are \emph{isomorphic}, \emph{i.e.}, when there is $\isomorphism: \vertices \to \vertices'$ bijective such that: for all $\avertex_1, \avertex_2 \in \vertices$, $(\avertex_1, \avertex_2) \in \edges$ if and only if $(\isomorphism(\avertex_1), \isomorphism(\avertex_2)) \in \edges'$; and, for all $\avertex \in \vertices$, $\labelling(\avertex) = \labelling'(\isomorphism(\avertex))$. In fact, equivalence classes for $\subptg$ are isomorphism classes:

\begin{prop}
\label{isom-subptg}
For every $\PTG = (\vertices, \edges, \labelling)$ and $\PTG' = (\vertices', \edges', \labelling')$, if $\PTG \subptg \PTG'$ and $\PTG' \subptg \PTG$ then $\PTG \isom \PTG'$.
\end{prop}
\begin{proof}
Let $\homomorphism_a: \vertices \to \vertices'$ and $\homomorphism_b: \vertices' \to \vertices$ obtained by applying the definition of $\PTG \subptg \PTG'$ and $\PTG' \subptg \PTG$. We prove that $\homomorphism_a$ and $\homomorphism_b$ are inverses, \emph{i.e.}, $\homomorphism_b \circ \homomorphism_a = \mathsf{id}_{\vertices}$ and $\homomorphism_a \circ \homomorphism_b = \mathsf{id}_{\vertices'}$. Both statements are symmetric, it suffices to prove the first one. Let $\avertex \in \vertices$, and let $\avertex' \deff \homomorphism_a(\avertex)$. By hypothesis, $\labelling(\avertex) \ne \emptyset$ so there is $x \in \labelling(\avertex)$. By hypothesis on $\homomorphism_a$, $x \in \labelling'(\avertex')$. By hypothesis on $\homomorphism_b$, $x \in \labelling(\homomorphism_b(\avertex'))$. Since $x$ is in the label of a single vertex in $\PTG$, we obtain $\homomorphism_b(\homomorphism_a(\avertex)) = \homomorphism_b(\avertex') = \avertex$. 

We have proven that $\homomorphism_b = \homomorphism_a^{-1}$. We now prove that $\homomorphism_a$ is a valid isomorphism between $\PTG$ and $\PTG'$. Let $\avertex_1, \avertex_2 \in \vertices$. By definition of $\homomorphism_a$, $(\avertex_1, \avertex_2) \in \edges$ implies $(\homomorphism_a(\avertex_1), \homomorphism_a(\avertex_2)) \in \edges'$. Conversely, by applying the definition of $\homomorphism_b$ on vertices $\homomorphism_a(\avertex_1)$ and $\homomorphism_a(\avertex_2)$, we obtain that $(\homomorphism_a(\avertex_1), \homomorphism_a(\avertex_2)) \in \edges'$ implies $(\homomorphism_b(\homomorphism_a(\avertex_1)), \homomorphism_b(\homomorphism_a(\avertex_2))) \in \edges$ therefore $(\avertex_1, \avertex_2) \in \edges$. For the labelling function, we obtain $\labelling(\avertex) \subseteq \labelling'(\homomorphism_a(\avertex))$ thanks to $\homomorphism_a$ and $\labelling'(\homomorphism_a(\avertex)) \subseteq \labelling(\avertex)$ thanks to $\homomorphism_b$, so overall $\labelling(\avertex) = \labelling'(\homomorphism_a(\avertex))$. This concludes the proof that $\PTG \isom \PTG'$. 
\end{proof} 

We end this sequence of definitions by introducing the union operator on points-to graphs.

\begin{defi} 
\label{def-ptg-union}
Let $\PTG = (\vertices, \edges, \labelling)$ and $\PTG' = (\vertices', \edges', \labelling')$ be two points-to graphs. Let $\bowtie$ be the relation on $\vertices \disjointunion \vertices'$ defined by letting, for each $\avertex \in \vertices$ and $\avertex' \in \vertices'$, $\avertex \bowtie \avertex'$ whenever $\labelling(\avertex) \cap \labelling'(\avertex') \ne \emptyset$. Let $\sim$ be the reflexive and transitive closure of $\bowtie$. We define the \emph{union points-to graph} $\PTG \PTGunion \PTG' = (\vertices_{\PTGunion}, \edges_{\PTGunion}, \labelling_{\PTGunion})$ where:
\begin{itemize}
\item $\vertices_{\PTGunion} \subseteq 2^{\vertices \cup \vertices'}$ is the set of equivalence classes of $\sim$; 
\item given $W_1, W_2 \in \vertices_{\PTGunion}$, there is an edge $(W_1, W_2) \in \edges_{\PTGunion}$ if and only if there are $\avertex_1 \in W_1$, $\avertex_2 \in W_2$ such that $(\avertex_1, \avertex_2) \in \edges \cup \edges'$;
\item for each $W \in \vertices_{\PTGunion}$, $\labelling_{\PTGunion}(W) \deff \bigcup_{\avertex \in W \cap \vertices} \labelling(\avertex) \cup \bigcup_{\avertex \in W \cap \vertices'} \labelling'(\avertex)$.
\end{itemize}
\end{defi}
Informally, $\PTG \PTGunion \PTG'$ is obtained by iteratively merging vertices of $\PTG$ and $\PTG'$ that share a variable in either $\PTG$ or $\PTG'$. By transitivity, if $x$ is in the same vertex as $z$ in $\PTG$ and $y$ is in the same vertex as $z$ in $\PTG'$, then $x$ and $y$ are in the same vertex in $\PTG \PTGunion \PTG'$.  

\begin{prop}
\label{ptg-union-properties}
Let $\PTG_0, \PTG_1, \PTG_2$ be three points-to graphs. The operator $\PTGunion$ satisfies the following properties.
\begin{enumerate}
    \item \label{union-assos} Associativity: $\PTG_0 \PTGunion (\PTG_1 \PTGunion \PTG_2) \isom (\PTG_0 \PTGunion \PTG_1) \PTGunion \PTG_2$.
    \item \label{union-commu} Commutativity: $\PTG_0 \PTGunion \PTG_1 \isom \PTG_1 \PTGunion \PTG_0$.
    \item \label{union-idempot} Idempotence: $\PTG_0 \PTGunion \PTG_0 \isom \PTG_0$.
    \item \label{union-nondec} Non-decreasivity: if $\PTG_0 \subptg \PTG_1$ then $\PTG_0 \PTGunion \PTG_2 \subptg \PTG_1 \PTGunion \PTG_2$.
    \item \label{union-extensivity} Extensivity:  $\PTG_0 \subptg \PTG_0 \PTGunion \PTG_1$.
    \item \label{union-absorption} Absorption: if $\PTG_0 \subptg \PTG_1$ then $\PTG_1 \isom \PTG_0 \PTGunion \PTG_1$.
\end{enumerate}
\end{prop}
\begin{proof} Below, we let $\PTG_0 = (\vertices_0, \edges_0, \labelling_0), \PTG_1 = (\vertices_1, \edges_1, \labelling_1)$ and $\PTG_2 = (\vertices_2, \edges_2, \labelling_2)$. We let $\labelling: \vertices_0 \disjointunion \vertices_1 \disjointunion \vertices_2 \rightarrow 2^\vars$ be the function such that $\labelling(\avertex) = \labelling_i(\avertex)$ if $\avertex \in \vertices_i$ for each $i \in \set{0,1,2}$. For each $i, j \in \set{0,1,2}$, we denote by $\sim_{ij}$ (resp. $\bowtie_{ij}$, $\edges_{ij}$, $\labelling_{ij}$) the equivalence relation (resp. base relation, set of edges, labelling) constructed to build the union points-to graph $\PTG_i \PTGunion \PTG_j$ in \autoref{def-ptg-union}. 

For (\ref{union-assos}), it suffices to observe that, both for  $\PTG_0 \PTGunion (\PTG_1 \PTGunion \PTG_2)$ and for $(\PTG_0 \PTGunion \PTG_1) \PTGunion \PTG_2$, the resulting points-to graph corresponds to the construction from \autoref{def-ptg-union} generalized to three points-to graphs, \emph{i.e.}, where $\sim$ is extended as the equivalence relation over $\vertices_0 \disjointunion \vertices_1 \disjointunion \vertices_2$. 
Proving (\ref{union-commu}) follows directly from the symmetry of \autoref{def-ptg-union} with respect to $\PTG_0$ and $\PTG_1$. 

For (\ref{union-idempot}), we prove that, when computing the equivalence relation $\sim_{00}$ for $\PTG_0 \PTGunion \PTG_0$, the equivalence class of a vertex contains precisely the two copies of the vertex. Given a vertex $v$ of $\PTG_0$, we let $v_\ell$ (resp. $v_r$) denote its vertex in the copy of $\PTG_0$ appearing on the left-hand side (resp. right-hand side) of the union. Suppose that there are two distinct vertices $v,w$ of $\vertices_0$ such that $v_\ell \bowtie_{00} w_r$; this implies $\labelling_0(v) \cap \labelling_0(w) \ne \emptyset$, contradicting \autoref{def-ptg} which requires that each variable appears in exactly one vertex of points-to graph $\PTG_0$. This proves that equivalence classes for $\sim_{00}$ are of the form $\set{v_\ell, v_r}$ for $v \in \vertices_0$. We let $\isomorphism$ be the bijective function that maps each such equivalence class $\set{v_\ell, v_r}$ to its corresponding vertex $v$ in $\PTG_0$. It is easy to see that edges and labels are preserved by $\isomorphism$, proving that $\PTG_0 \PTGunion \PTG_0 \isom \PTG_0$. 

For (\ref{union-nondec}), 
suppose that $\PTG_0 \subptg \PTG_1$ and let $\homomorphism: \vertices_0 \rightarrow \vertices_1$ be a homomorphism witnessing that. 
We define a homomorphism $\psi$ witnessing that $\PTG_0 \PTGunion \PTG_2 \subptg \PTG_1 \PTGunion \PTG_2$; observe that $\psi$ takes an input an equivalence class for $\sim_{02}$ (represented as a subset of $\vertices_0 \cup \vertices_2$) and gives as output an equivalence class for $\sim_{12}$ (represented as a subset of $\vertices_1 \cup \vertices_2$). We let $\psi$ be function that, given an equivalence class $W$ for $\sim_{02}$, maps the equivalence class $\psi(W)$ for $\sim_{12}$ that contains all vertices in $\homomorphism(W \cap \vertices_0)$ and all vertices in $W \cap \vertices_2$. The fact that all such vertices lie in the same equivalence class for $\sim_{12}$ is not obvious, so we start by proving it. Let $W$ be an equivalence class for $\sim_{02}$, and let $f: \vertices_0 \cup \vertices_2 \rightarrow \vertices_1 \cup \vertices_2$ such that $f(\avertex) = \homomorphism(\avertex)$ when $\avertex \in \vertices_0$ and $f(\avertex) = \avertex$ when $\avertex \in \vertices_2$. With this definition, it suffices to prove that all vertices in $f(W)$ are in the same equivalence class for $\sim_{12}$. We have $\labelling(\avertex) \subseteq \labelling(f(\avertex))$ for all $\avertex \in W$: indeed, it is trivial when $\avertex \in \vertices_2$ because then $f(\avertex) = \avertex$, and when $\avertex \in \vertices_0$ this holds because $\homomorphism$ is a witness that $\PTG_0 \subptg \PTG_1$. Therefore, for every $\avertex, \avertex' \in \vertices_0 \cup \vertices_2$, if $\labelling(\avertex) \cap \labelling(\avertex') \ne \emptyset$ then $\labelling(f(\avertex)) \cap \labelling(f(\avertex')) \ne \emptyset$; in other words, if $\avertex \bowtie_{02} \avertex'$ then $f(\avertex) \bowtie_{12} f(\avertex')$. Since $\sim_{02}$ (resp. $\sim_{12}$) is the reflexive and transitive closure of $\bowtie_{02}$ (resp. $\bowtie_{12}$), this proves that $\avertex \sim_{02} \avertex'$ implies $f(\avertex) \sim_{12} f(\avertex')$. This implies the existence, for every equivalence class $W$ for $\sim_{02}$, of an equivalence class $\psi(W)$ for $\sim_{12}$ that contains all vertices in $f(W)$, \emph{i.e.}, all vertices in $\homomorphism(W \cap \vertices_0)$ and in $W \cap \vertices_2$. Note that uniqueness of $\psi(W)$ is obvious since $W \ne \emptyset$ and a given vertex lies in a single equivalence class. 

We now claim that, for every equivalence class $W$ for $\sim_{02}$, for every $x \in \labelling_{02}(W)$, we have $x \in \labelling_{12}(\psi(W))$. Let $W$ be an equivalence for $\sim_{02}$ and let $x \in \labelling_{02}(W)$. There exists $\avertex \in W$ such that $x \in \labelling_{02}(W)$. If $\avertex \in \vertices_2$, then trivially $\avertex \in \psi(W)$ hence $x \in \labelling_{12}(\psi(W))$. If  $\avertex \in \vertices_0$ then, since $\homomorphism$ is a witness that $\PTG_0 \subptg \PTG_1$, $x \in \labelling(\homomorphism(\avertex))$ and $\homomorphism(\avertex) \in \psi(W)$ therefore $x \in \labelling_{12}(\psi(W))$. 
Overall, this proves that, for every equivalence class $W$ for $\sim_{02}$, $\labelling_{02}(W) \subseteq  \labelling_{12}(\psi(W))$. Next, $\psi$ preserves edges, \emph{i.e.}, that for each edge $(W,W') \in \edges_{02}$, there is an edge $(\psi(W), \psi(W')) \in \edges_{12}$. Let $W,W'$ such that $(W,W') \in \edges_{02}$; by definition of $\PTG_0 \PTGunion \PTG_2$, there are $\avertex \in W, \avertex' \in W'$ such that $(\avertex, \avertex') \in \edges_0 \cup \edges_2$. If $(\avertex, \avertex') \in \edges_2$, then $\avertex \in \psi(W)$ and $\avertex' \in \psi(W')$  therefore there is an edge $(\psi(W), \psi(W')) \in \edges_{12}$. If $\avertex, \avertex' \in \vertices_0$, then by definition of $\homomorphism$ we have $(\homomorphism(\avertex), \homomorphism(\avertex')) \in \edges_1$; by definition of $\psi$ we have $\homomorphism(\avertex) \in \psi(W)$ and $\homomorphism(\avertex') \in \psi(W')$ therefore $(\psi(W), \psi(W')) \in \edges_{12}$. This proves that $\psi$ is a valid witness that $\PTG_0 \PTGunion \PTG_2 \subptg \PTG_1 \PTGunion \PTG_2$. 

For (\ref{union-extensivity}), let $\homomorphism$ that maps to each vertex $\avertex \in \vertices_0$ its equivalence class for $\sim_{01}$. Trivially, for each $\avertex \in \vertices_0$, since $\homomorphism(\avertex)$ is its equivalence class for $\sim_{01}$, $\labelling(\avertex) \subseteq \labelling_{01}(W)$ by definition of $\labelling_{01}$. Moreover, if $(\avertex,\avertex') \in \edges_{01}$, then $(\homomorphism(\avertex), \homomorphism(\avertex')) \in \edges_{01}$ by definition of $\edges_{01}$.

For (\ref{union-absorption}), (\ref{union-extensivity}) directly gives $\PTG_1 \subptg \PTG_0 \PTGunion \PTG_1$ while (\ref{union-nondec}) and (\ref{union-idempot}) together give that $\PTG_0 \PTGunion \PTG_1 \subptg \PTG_1 \PTGunion \PTG_1 = \PTG_1$. \autoref{isom-subptg} then gives that $\PTG_1 \isom \PTG_0 \PTGunion \PTG_1$. 
\end{proof}

From \autoref{ptg-union-properties}, we can derive an arguably more intuitive property, namely that $\PTGunion$ is the least upper bound with respect to $\subptg$.

\begin{cor}
  \label{union-lub}
For every points-to graphs $\PTG,\PTG'$, $\PTGunion$ is the least upper bound of $\PTG$ and $\PTG'$ for $\subptg$, \emph{i.e.}:
\begin{itemize}
\item $\PTG \subptg \PTG \PTGunion \PTG'$ and $\PTG' \subptg \PTG \PTGunion \PTG'$, and
\item for every $U$ such that $\PTG \subptg U$ and $\PTG' \subptg U$, we have $\PTG \PTGunion \PTG' \subptg U$. 
\end{itemize}
\end{cor}
\begin{proof}
The first claim directly corresponds to extensivity (\ref{union-extensivity} of \autoref{ptg-union-properties}). For the second item, consider $U$ such that $\PTG \subptg U$ and $\PTG' \subptg U$. Since $\PTG \subptg U$, by non-decreasivity (\ref{union-nondec} of \autoref{ptg-union-properties}) we have $\PTG \PTGunion \PTG' \subptg \PTG' \PTGunion U$. Since $\PTG' \subptg U$, we have $\PTG' \PTGunion U \subptg U$ by absorption (\ref{union-absorption} of \autoref{ptg-union-properties}). By transitivity of $\subptg$, we obtain $\PTG \PTGunion \PTG' \subptg U$. 
\end{proof}  

\subsection{Procedural points-to analysis}
\label{subsec:proc-pta}
Our concurrent points-to analysis is built from a \emph{procedural} points-to analysis, \emph{i.e.}, a points-to analysis that analyzes each thread in isolation. This procedural points-to analysis corresponds to the \texttt{alias} plugin of Frama-C \cite{le_gall_alias_2024}. It corresponds to a lightweight, Steensgaard-style \cite{Steensgaard96} points-to analysis, with some limited form of flow-sensitivity as an instruction only affects the analysis at future nodes. 

This procedural points-to analysis must sometimes encode in the points-to graph information such as "these two vertices must have a common child" or "this vertex must have this other vertex as grandchild". Encoding this using dummy vertex with empty set of variables is possible, but points-to graphs do not allow for empty labels and do not generalize well to the presence of vertices with empty labels (for example, \autoref{isom-subptg} would no longer hold). Instead, we introduce dummy variables for this purpose: more precisely, we introduce, for every variable $x$ of depth at least $2$, a dummy variable $\deref{x}$ with an initial edge from the vertex of $x$ to the vertex of $\deref{x}$. However, we will later combine procedural points-to graphs from different threads using points-to graph union $\PTGunion$, and we do not want to merge vertices because they share a $\deref{x}$ variable (otherwise, all children of $x$ would be merged). For this reason, we introduce distinct dummy variables for different threads. 
We let 
\[
\dummyvars \deff \set{\dummyderef{x}{\athread} \mid \athread \in \threads, x \in \varsof{k}, k \geq 2}. 
\]
For a variable $x \in \varsof{k}$, the depth of a variable $\dummyderef{x}{\athread}$ is $k-1$. 
We let $\extvars \deff \vars \cup \dummyvars$ be the \emph{extended} set of variable. Points-to graph therefore use the set $\extvars$ as set of variables.


Given a valuation $\values$ and a thread $\athread$, one can build the canonical points-to graph  $\PTGof{\values}{\athread} \deff (\extvars, \edges, \labelling)$ as follows. Its set of vertices is the extended set of variables $\extvars$. For every $x,y \in \extvars$, its set of edges $\edges$ is defined by
\[ \edges \deff \set{(x,\values(x)) \mid x \in \vars, \values(x) \in \vars} \cup \set{(x, \dummyderef{x}{\athread}) \mid x \in \varsof{k}, k \geq 2}.
\]
Finally, its labelling is simply $\labelling: x \in \extvars \mapsto \set{x}$. Formally, all dummy variables $\dummyderef{x}{\athread'}$ for all threads $\athread'$ are present in $\PTGof{\values}{\athread}$, but only the dummy variables of $\athread$ have edges; the dummy variables  $\dummyderef{x}{\athread'}$ for $\athread' \ne \athread$ remain isolated and irrelevant for the procedural points-to analysis of $\athread$. This is mostly for convenience: when we will combine points-to graphs for different threads, all points-to graphs will have the same underlying set of vertices.

Let $\athread \in \threads$ and consider a path $\apath$ from $\startnodeof{\athread}$ to a node $\anode \in \nodesof{\athread}$. We define the points-to graph $\PTG(\apath)$ generated by $\apath$ by induction as follows.
If $\apath$ is the empty path, then $\PTG(\apath) = \PTGof{\values_0}{\athread}$; $\PTGof{\values_0}{\athread}$ is the graph with one vertex per variable, and where the only edges are $(x, \dummyderef{x}{\athread})$ for variables $x \in \varsof{k}$ of depth $k \geq 2$. Let $\apath = \apath_p \cdot \atrans$.  If the instruction $\aninstr$ of $\atrans$ is an assignment at depth $0$ (\emph{i.e.}, integer assignment) or not an assignment, then $\PTG(\apath) = \PTG(\apath_p)$. If the assignment is at depth $k \geq 1$ (\emph{i.e.}, pointer assignment), then $\PTG(\apath)$ is obtained from $\PTG \deff \PTG(\apath_p)$ with the following transformation (see \autoref{alias_algo_fig}). 
\begin{itemize}
\item If $\aninstr$ is $\instrstyle{x \deff y}$, then we merge $\vertexof{x}$ and $\vertexof{y}$.
\item If $\aninstr$ is $\instrstyle{x \deff \deref{y}}$ then we add an edge from $\vertexof{y}$ to $\vertexof{x}$ (if it does not yet exist).
\item If $\aninstr$ is $\instrstyle{\deref{x} \deff y}$ then we add an edge from $\vertexof{x}$ to $\vertexof{y}$.
\item If $\aninstr$ is $\instrstyle{x \deff \addr{y}}$ then we add an edge from $\vertexof{x}$ to $\vertexof{y}$.
\item If $\aninstr$ is $\instrstyle{\deref{x} \deff \addr{y}}$, we add an edge from the vertex $\vertexof{\dummyderef{x}{\athread}}$ of the dummy variable $\dummyderef{x}{\athread}$ to $\vertexof{y}$.
\item If $\aninstr$ is $\instrstyle{\deref{x} \deff \deref{y}}$ then $x$ and $y$ have depth at least $2$ (otherwise this would be an assignment at depth $0$, treated above). We merge the vertices $\vertexof{\dummyderef{x}{\athread}}$ and $\vertexof{\dummyderef{y}{\athread}}$, so that $\vertexof{x}$ and $\vertexof{y}$ have a common child. 
\end{itemize} 
To obtain $\PTG(\apath)$, after the transformation above, we iteratively merge vertices with a common parent. For example, for instruction $\instrstyle{x \deff \addr{y}}$, $\vertexof{y}$ is merged with the child of $\vertexof{x}$. 
\begin{figure}
\begin{tikzpicture}[auto, yscale = 0.7]
\tikzstyle{vertex} = [draw,circle,minimum size = 0.6cm]
\tikzstyle{new} = [color=red!60!black, font=\scriptsize]
\begin{scope}
\node at (1,1) {$\instrstyle{x \deff y}$};
\node[vertex] at (0,-0.3) (x) {$x$};
\node[vertex] at (2,-0.3) (y) {$y$};
\draw[-,dashed,new] (x) edge node[above,new] {merge}  (y);
\end{scope}
\begin{scope}[xshift = 4cm]
\node at (1,1) {$\instrstyle{x \deff \deref{y}}$};
\node[vertex] at (0,-0.3) (x) {$x$};
\node[vertex] at (2,-0.3) (y) {$y$};
\draw[-latex,dashed,new] (y) edge node[above,new] {add edge} (x);
\end{scope}
\begin{scope}[xshift= 8cm]
\node at (1,1) {$\instrstyle{\deref{x} \deff y}$};
\node[vertex] at (0,-0.3) (x) {$x$};
\node[vertex] at (2,-0.3) (y) {$y$};
\draw[-latex,dashed,new] (x) edge node[above,new] {add edge} (y);
\end{scope}
\begin{scope}[xshift = 12cm]
\node at (1,1) {$\instrstyle{x \deff \addr{y}}$};
\node[vertex] at (0,-0.3) (x) {$x$};
\node[vertex] at (2,-0.3) (y) {$y$};
\draw[-latex,dashed,new] (x) edge node[above,new] {add edge} (y);
\end{scope}
\tikzstyle{vertex} = [draw,circle,minimum size = 0.8cm]
\begin{scope}[xshift = 3cm, yshift = -3.5cm]
\node at (1.1,1.5) {$\instrstyle{\deref{x} \deff \addr{y}}$};
\node[vertex] at (-0.6,0) (x) {$x$};
\node[vertex] at (3,0) (y) {$y$};
\node[new,vertex, inner sep = 1pt] at (1.2, 0) (aux) {$\dummyderef{x}{\athread}$};
\draw[-latex,dashed,new] (x) edge (aux);
\draw[-latex,dashed,new] (aux) edge (y);
\node[new, align = center] at (1.2,-1.3) {add grandchild relation \\ using dummy vertex $\dummyderef{x}{\athread}$};
\end{scope}
\begin{scope}[xshift = 9cm, yshift = -3.5cm]
\node at (1.1,1.5) {$\instrstyle{\deref{x} \deff \deref{y}}$};
\node[vertex] at (-0.6,0) (x) {$x$};
\node[vertex] at (3,0) (y) {$y$};
\node[new,vertex, inner sep = 1pt] at (1.2, 0) (aux) {$\dummyderef{x}{\athread}, \dummyderef{y}{\athread}$};
\draw[-latex,dashed,new] (x) edge (aux);
\draw[-latex,dashed,new] (y) edge (aux);
\node[new, align = center] at (1.2,-1.3) {merge children of $x$ and $y$ \\ (if $x$ and $y$ have depth $\geq 2$)};
\end{scope}
\draw[color = black!30!white,thick] (-0.5,-1.4) -- (14.5,-1.4);
\draw[color = black!30!white,thick] (3,1.2) -- (3, -1.4);
\draw[color = black!30!white,thick] (7,1.2) -- (7,-1.4);
\draw[color = black!30!white,thick] (11,1.2) -- (11,-1.4);
\draw[color = black!30!white,thick] (7,-1.4) -- (7, -5);
\end{tikzpicture}
\caption{A depiction of how the procedural points-to analysis takes assignments of depth $\geq 1$ into account. Here, vertex with label $x$ is the vertex $\vertexof{x}$; it may also be labelled by other variables that are not represented. If $x$ and $y$ are in the same vertex, there is nothing to do. In the second step of the transformation (not represented), for every vertex with several children, all children are merged. In particular, this implies that, when an edge from $x$ to $y$ is added, $y$ is merged with the vertex labelled by $\dummyderef{x}{\athread}$ if $x$ has depth two or more.}
\label{alias_algo_fig}
\end{figure}

The procedural points-to overapproximation $\procpto(\anode)$ at node $\anode$ of thread $\athread = \threadof{\anode}$ is computed as the union of all $\PTG(\apath)$ for all paths $\apath$ from $\startnodeof{\athread}$ to $\anode$. In presence of cycles, this set of paths may be infinite. However, by considering every case in \autoref{alias_algo_fig}, we observe that $\PTG(\apath \cdot \atrans)=\PTG(\apath)$ when $\atrans$ already appears in $\apath$. This implies that $\procpto(\anode)$ can be defined even if the set of paths from $\startnodeof{\athread}$ to $\anode$ is infinite. 

This analysis is \emph{path-insensitive}, since all paths are taken into account and conditional guards are not considered (for example, if we have $\codeif(\codefalse) \; \codethen \;  x:=y$, the vertices of $x$ and $y$ are merged even though this instruction can never be reached). While this absence of path-sensitivity may seem like a limitation, it makes it much easier to extend this points-to analysis to a concurrent points-to analysis. Indeed, a branch that is unreachable when the thread is considered in isolation is possibly reachable when considering interference from other threads. For example, suppose that there is an $\codeif(x==1)$ guard in thread $\athread$, but no instruction in $\athread$ sets $x$ to $1$, so that the $\codethen$ branch is unreachable in isolation. It is possible that another thread $\athread'$ has instruction $x \deff 1$, so that the $\codethen$ branch can in fact be taken. A (naive) path-sensitive points-to analysis could dismiss the $\codethen$ branch and give an unsound result. This observation justifies our choice to consider a path-insensitive analysis. 

Moreover, the analysis only accumulates points-to pairs along a path, never losing any: this is called \emph{cumulativity}, and is also a convenient feature towards concurrent points-to analysis. For example, when analyzing a sequence of instructions $x \deff \addr{y}; \deref{x} \deff z;$, one should not remove previous points-to edges from $y$, since in a concurrent setting the value of $x$ may be changed in between the two instructions. 

Implementation-wise, the \texttt{alias} plugin \cite{le_gall_alias_2024} does not explicitly build the points-to graph at each node, but stores it using a persistent union-find data structure \cite{CF07} which records changes along a path. 
We illustrate the procedural points-to analysis on thread \texttt{t1} from \autoref{sec:concurrent-programs}, using notations from \autoref{cfg-example}. At $\startnode$ and after the $\lock{\texttt{m}}$ instruction, the points-to graph has a single edge, from $\texttt{dbl\_ptr}$ to dummy variable $\dummyderef{\texttt{dbl\_ptr}}{\texttt{t1}}$. 
After instruction $\texttt{dbl\_ptr = \addr{p}}$, there is an edge from $\texttt{dbl\_ptr}$ to a vertex containing both variables $\texttt{q}$ and $\dummyderef{\texttt{dbl\_ptr}}{\texttt{t1}}$. After $\texttt{p = \addr{k}}$, there is an edge from $\texttt{p}$ to $\texttt{k}$.

\subsection{Concurrent points-to analysis}

We now introduce how the procedural points-to analysis from above is turned into a concurrent points-to analysis.

Let $\procpto$ be the result of the procedural points-to analysis. Recall that $\procpto$ maps each node $\anode$ to a points-to graph $\procpto(\anode)$. 
In words, the (concurrent) points-to overapproximation at node $\anode$ makes the union of its procedural points-to analysis $\procpto(\anode)$ with all procedural points-to analysis $\procpto(\anode')$ of nodes $\anode'$ that may have been visited before $\anode$ in the execution.
Therefore, we need, for every node $\anode$, an overapproximation of the set of nodes that may have been visited before. Nodes of the same thread $\threadof{\anode}$ need not be considered because $\procpto$ is cumulative, so $\procpto(\anode)$ already subsumes all $\procpto(\anode')$ with $\anode'$ a node of $\threadof{\anode}$ such that there is a path from $\anode'$ to $\anode$. We let 
\[ \MHB{\anode} \deff \set{\anode' \mid \threadof{\anode'} \ne \threadof{\anode} \land \threadof{\anode} \notin \joined{\anode'} \land \threadof{\anode'} \notin \uncreated{\anode}}.\]
MHB stand for ``May Happen Before''. Thanks to \autoref{soundness-joined} and \autoref{soundness-uncreated}, $\MHB{\anode}$ overapproximates the set of nodes that other threads may have visited before $\threadof{\anode}$ is in $\anode$.

\begin{wrapfigure}{r}{1.8cm}
\begin{tikzpicture}
\node at (0,0) (v1) {$\avertex_1$} ;
\node at (-0.75,-1) (v2) {$\avertex_2$};
\node at (0.75,-1) (v2p) {$\avertex_2'$};
\node at (0,-2) (v3) {$\avertex_3$};
\draw[-latex] (v1) -- (v2);
\draw[-latex] (v1)--(v2p);
\draw[-latex] (v2) -- (v3);
\draw[-latex,color=red!60!black,dashed,thick] (v2p) -- (v3);
\end{tikzpicture}
\end{wrapfigure}
One additional ingredient is needed to combine procedural points-to analysis of different threads. We do so using a new technique, which we dub diamond completion rule. It is a weakened notion of merging where two vertices are forced to have the same children, but not the same parents. It is applied whenever two nodes share a parent.  
Given a 
PTG $\PTG$, we let $\DCR{\PTG}$ denote the points-to graph obtained from $\PTG$ by applying, until saturation, the \emph{diamond completion rule} (DCR). The DCR can be applied to $(\avertex_1, \avertex_2, \avertex_2', \avertex_3) \in \vertices^4$ when there are edges $(\avertex_1,\avertex_2)$, $(\avertex_1,\avertex_2')$, $(\avertex_2, \avertex_3) \in \edges$; it adds an edge from $\avertex_2'$ to $\avertex_3$.  In $\DCR{\PTG}$, all nodes sharing a parent have the same set of children. 
See the figure opposite for an illustration. 

The function $G \mapsto \DCR{G}$ is non-decreasing for $\subptg$:
\begin{lem}
  \label{DCR-nondec-extensive}
Let $G$, $H$ be two points-to graphs. We have the following:
\begin{itemize}
  \item $G \subptg \DCR{G}$;
  \item if $G \subptg H$ then $\DCR{G} \subptg \DCR{H}$.
\end{itemize}
\end{lem}
\begin{proof}
Let $G = (\vertices_G, \edges_G, \labelling_G$ and $H = (\vertices_H, \edges_H, \labelling_H)$. To prove the first point, we observe $G$ and $\DCR{G}$ share the same set of vertices $\vertices_G$, the same labelling function $\labelling_G$ and that any edge of $G$ is an edge of $\DCR{G}$, so it suffices to let $\homomorphism$ be the identity function on $\vertices_G$. 

We now prove the second assertion. Suppose that $G \subptg H$, let $\homomorphism: \vertices_G \rightarrow \vertices_H$ be a witness of that. It suffices to prove that $D(G) \subptg \DCR{H}$ where $D(G)$ is obtained from $G$ by applying DCR \emph{once}. Let $(v_1,v_2, v_2',v_3) \in \vertices_G^4$ be the vertices on which DCR is applied; we have $(v_1,v_2), (v_1,v_2'), (v_2,v_3) \in \edges_G$. We claim that $\homomorphism$ is also a witness that $D(G) \subptg \DCR{H}$. Indeed, let $(v,w)$ be an edge in $D(G)$. If $(v,w) \ne (v_2',v_3)$ then $(v,w) \in \edges_G$ therefore, by definition of $\homomorphism$, $(\homomorphism(v), \homomorphism(w)) \in \edges_H$ so this edge also exists in $\DCR{H}$. Regarding edge $(v_2',v_3)$,  DCR can be applied in $H$ to $(\homomorphism(v_1), \homomorphism(v_2), \homomorphism(v_2'), \homomorphism(v_3))$ because, by definition of $\homomorphism$, $(\homomorphism(v_1), \homomorphism(v_2)) \in \edges_H$, $(\homomorphism(v_1), \homomorphism(v_2')) \in \edges_H$ and $(\homomorphism(v_2), \homomorphism(v_3)) \in \edges_H$. This proves the presence of an edge $(\homomorphism(v_2'), \homomorphism(v_3))$ in $\DCR{H}$.  
\end{proof}
We define the concurrent points-to analysis as follows.

\begin{defi}   
\label{def-pto}
For every $\anode \in \nodes$, 
$\pto(\anode) \deff \DCRbig{\procpto(\anode) \PTGunion \bigPTGunion\limits_{\anode' \in \MHB{\anode}} \procpto(\anode')}$. 
\end{defi}

In particular, note that dummy variables $\dummyderef{x}{\athread}$ are introduced solely to encode constraints in the procedural points-to analysis. Because dummy variables belonging to different threads are distinct, they do not cause any additional vertex merging when procedural points-to graphs of different threads are merged. 

\begin{exa}
In \autoref{fig:dcr-needed}, we illustrate this analysis on the example of \autoref{fig:example-program}. We present how $\pto(\anode_m)$ is computed, where $\anode_m$ is the node before the access to \texttt{out} in $\texttt{main}$.  Since no vertex has been merged in any of the analyses $\procpto$ of the tasks, we only have one variable per vertex. The DCR rule is needed for soundness, because there is indeed an execution where $\texttt{q}$ points to $\texttt{k}$, as highlighted in \autoref{example-reachable-config}. Without DCR, the analysis would miss the potential data race on $k$.  
\end{exa}

\begin{figure}[h]
\centering

\begin{tikzpicture}[yscale = 1.5]
\node at (0,0) (d1) {$\texttt{dbl\_ptr}$};
\node at (-1,-0.5) (p1) {\texttt{p}};
\node at (1,-0.5) (q1) {\texttt{q}};
\node at (-1,-1) (k1) {\texttt{k}};
\node at (1,-1) (j1) {\texttt{j}};
\draw[-latex] (d1) -- (p1);
\draw[-latex] (q1) -- (j1);
\node at (0,-1.3) [align = center, font = {\footnotesize}] {$\procpto$ at $\anode_{m}$ in \texttt{main}};
\begin{scope}[xshift = 3cm]
\draw[color = black!30!white,thick] (-1.5,0.2) -- (-1.5,-1.2);
\node at (0,0) (d2) {$\texttt{dbl\_ptr}$};
\node at (-1,-0.5) (p2) {\texttt{p}};
\node at (1,-0.5) (q2) {\texttt{q}};
\node at (-1,-1) (k2) {\texttt{k}};
\node at (1,-1) (j2) {\texttt{j}};
\draw[-latex] (d2) -- (q2);
\draw[-latex] (p2) -- (k2);
\node at (0,-1.3) {$\procpto$ at end of \texttt{t1}};
\end{scope}
\begin{scope}[xshift = 6cm]
\draw[color = black!30!white,thick] (-1.5,0.2) -- (-1.5,-1.2);
\node at (0,0) (d3) {$\texttt{dbl\_ptr}$};
\node at (-1,-0.5) (p3) {\texttt{p}};
\node at (1,-0.5) (q3) {\texttt{q}};
\node at (-1,-1) (k3) {\texttt{k}};
\node at (1,-1) (j3) {\texttt{j}};
\draw[-latex] (d3) -- (p3);
\node at (0,-1.3) {$\procpto$ at end of \texttt{t2}};
\end{scope}
\begin{scope}[xshift = 9cm]
\draw[color = black!30!white,thick] (-1.5,0.2) -- (-1.5,-1.2);
\node at (0,0) (d4) {$\texttt{dbl\_ptr}$};
\node at (-1,-0.5) (p4) {\texttt{p}};
\node at (1,-0.5) (q4) {\texttt{q}};
\node at (-1,-1) (k4) {\texttt{k}};
\node at (1,-1) (j4) {\texttt{j}};
\draw[-latex] (d4) -- (p4);
\draw[-latex] (d4) -- (q4);
\draw[-latex] (p4) -- (k4);
\draw[-latex] (q4) -- (j4);
\draw[-latex, dashed, color = red!50!black] (p4) -- (j4);
\draw[-latex, dashed, color = red!50!black] (q4) -- (k4);
\node[font={\footnotesize}] at (0,-1.3) {$\pto$ at $\anode_{m}$ in \texttt{main}.};
\end{scope}
\end{tikzpicture}
\caption{The computation of the concurrent points-to analysis on the program of \autoref{fig:example-program} at the node $\anode_{m}$ between lines~\ref{create-t2} and \ref{access-out} in \texttt{main}. 
Variable $\texttt{out}$ (which plays no role here) and dummy variables
 $\dummyderef{\texttt{dbl\_ptr}}{\athread}$ 
 (which are irrelevant here) are omitted from the picture. Dashed red edges are those added by DCR.}
\label{fig:dcr-needed}
\end{figure}

In practice, computing $\pto(\anode)$ does not require to iterate over \emph{all} nodes in $\MHB{\anode}$. Since the procedural analysis $\procpto$ is cumulative, it suffices to consider all $\anode' \in \MHB{\anode}$ that have no successor in $\MHB{\anode}$, \emph{i.e.}, who have no successor at all or whose successors $\anode''$ are all such that $\threadof{\anode} \in \joined{\anode''}$.

Soundness of the concurrent points-to analysis $\pto$ is stated as follows: 

\begin{thm}
\label{concurrent-alias-soundness}
For every $\athread \in \threads$, for every $\anode \in \nodesof{\athread}$, for every reachable $\config = \tuple{\stateFunction, \values, \held}$ such that $\stateFunction(\athread) = \stateActive{\anode}$, for every $x, y \in \vars$, if $\values(x) = y$ then $x \pt{\pto(n)} y$. 
\end{thm}
The proof of \autoref{concurrent-alias-soundness} is the subject of the next subsection.

\begin{rem}
\label{rem:dcr-better-than-merging}
The procedural points-to analysis $\procpto$ systematically merges children $\avertex_2$ and $\avertex_2'$ of a vertex $\avertex_1$. The concurrent points-to analysis $\pto$ does not: it uses DCR instead so that $\avertex_2$ and $\avertex_2'$ must have the same children, but not the same parents. This distinction is the reason why the points-to graph of each thread has its own copy of each $\dummyderef{x}{\athread}$; otherwise, vertices that are children of the same node in different threads would always be merged as they would share a variable $\deref{x}$. Within a given thread, the procedural points-to analysis always merges children, so this is not an issue. An alternative approach would be to make the procedural points-to analysis $\procpto$ also rely on DCR instead of merging children; this would give better precision but worse performance (merging allows to keep the number of vertices low). For our purposes, the procedural points-to analysis intends to reflect the implementation from the \texttt{alias} plugin of Frama-C, which systematically merges children. The technique introduced here therefore allows to lift the procedural points-to analysis of the \texttt{alias} plugin to a concurrent points-to analysis while limiting precision loss.  
\end{rem}

\subsection{Proof of Soundness for Concurrent Points-to Analysis}
We present in full details the proof of soundness for the concurrent points-to-analysis, \emph{i.e.}, the proof of \autoref{concurrent-alias-soundness}. The proof follows an induction on execution length. The presence of thread creations is a challenge for this proof, because a newly-created thread must also satisfy the induction hypothesis. We first sketch the natural induction argument (\ref{proof-pta-soundness-simplified}), then explain where it fails in presence of thread creations (\ref{proof-pta-soundness-subtlety}). We therefore present an alternative induction hypothesis (\ref{proof-pta-soundness-induction-hypothesis}), allowing us to provide a rigorous proof (\ref{proof-pta-soundness-wrapping-up}). 

\subsubsection{An intuitive but partial proof.} \label{proof-pta-soundness-simplified} The goal is to proceed by induction. The main interest of the proof consists in considering all possible cases of instructions, and explaining how the rules from \autoref{alias_algo_fig} along with \autoref{def-pto} suffice to capture all these cases. For this reason, we detail this case here in a simplified manner; we will later explain why the induction hypothesis must be enriched to obtain a full proof. 

Consider an instruction step $\config_1 = \tuple{\stateFunction_1, \values_1, \held_1} \step{\atrans} \config_2 = \tuple{\stateFunction_2, \values_2, \held_2}$ where $\atrans = (\anode_1, \aninstr, \anode_2)$ with $\aninstr$ an assignment of depth $k \geq 1$ (\emph{i.e.}, a pointer assignment). Let $\athread \deff \threadof{\atrans}$ be the thread performing this transition. Suppose that we know, by induction hypothesis, that 
 $\pto(\anode_1)$ has all points-to pairs from $\values_1$. By cumulativity, $\pto(\anode_1) \subptg \pto(\anode_2)$ so $\pto(\anode_2)$ also has every points-to pair in $\values_1$. We prove that $\pto(\anode_2)$ has any new points-to pair corresponding to the effect of the assignment. We proceed by case disjunction on $\aninstr$. Since only pointer assignments modify the points-to relation, we only consider such instructions.

\textbf{$\instrstyle{x \deff y}$:} \quad In the computation of $\procpto(\anode_2)$, the instruction $\instrstyle{x \deff y}$ yield a merge of the vertices of $x$ and $y$. Therefore, in $\procpto(\anode_2)$, $x$ and $y$ are in the same vertex, and thanks to $\pto(\anode_1)$, $y \pt{} \values_1(y)$; therefore $x \pt{} \values_1(y)$ in $\pto(\anode_2)$.

\textbf{$\instrstyle{x \deff \deref{y}}$:} \quad We want to apply DCR on $(\vertexof{y},\vertexof{\values_1(y)},\vertexof{x},\vertexof{\values_1(\values_1(y))})$. Thanks to $\pto(\anode_1)$, $y \pt{} \values_1(y)$ and thanks to $\procpto(\anode_2)$, $y \pt{} x$; this gives the two upper edges for DCR. Moreover, again thanks to $\pto(\anode_1)$, $\values_1(y) \pt{} \values_1(\values_1(y))$, which gives the lower edge for DCR. DCR completes the diamond: since $\vertexof{\values_1(y)}$ and $\vertexof{x}$ have a common parent, namely $\vertexof{\values_1(y)}$), any child of $\vertexof{\values_1(y)}$ becomes a child of $\vertexof{x}$. This is the case for $\vertexof{\values_1(\values_1(y))}$, hence giving $x \pt{} \values_1(\values_1(y))$ in $\pto(\anode_2)$. 

\textbf{$\instrstyle{\deref{x} \deff y}$:} \quad Let $p_x \deff \values_1(x)$ and $p_y \deff \values_1(y)$. We have $\values_2(p_x) = p_y$. Thanks to $\procpto(\anode_2)$, $x \pt{} y$. Thanks to $\pto(\anode_1)$, $x \pt{} p_x$ and $y \pt{} p_y$. By DCR on $(\vertexof{x}, \vertexof{y}, \vertexof{p_x}, \vertexof{p_y})$, $\vertexof{p_y}$ being a child of $\vertexof{y}$ implies it also being a child of $\vertexof{p_x}$, therefore $p_x \pt{} p_y$.

\textbf{$\instrstyle{x \deff \addr{y}}$:} \quad We have $\values_2(x) = y$. Thanks to $\procpto(\anode_2)$, $x \pt{} y$.

\textbf{$\instrstyle{\deref{x} \deff \addr{y}}$:} \quad Let $p_x \deff \values_1(x)$; we have $\values_2(p_x) = y$. Thanks to $\pto(\anode_2)$, $x \pt{} p_x$. Thanks to $\procpto(\anode_2)$, there is an edge from $\vertexof{x}$ to $\vertexof{\dummyderef{x}{\athread}}$ and an edge from $\vertexof{\dummyderef{x}{\athread}}$ to $\vertexof{y}$. By DCR applied to $(\vertexof{x}, \vertexof{\dummyderef{x}{\athread}}, \vertexof{p_x}, \vertexof{y})$, $p_x \pt{} y$.  See \autoref{soundness-deref-address} for an illustration.

\textbf{$\instrstyle{\deref{x} \deff \deref{y}}$:} \quad 
Let $p_x \deff \values_1(x)$ and $p_y \deff \values_1(y)$; we have $\values_2(p_x) = \values_1(p_y)$. Thanks to $\pto(\anode_1)$, $x \pt{} p_x$, $y \pt{} p_y$ and $p_y \pt{} \values_1(p_y)$. Thanks to $\procpto(\anode_2)$, $\vertexof{x}$ and $\vertexof{y}$ share a child $\avertex$ in $\pto(\anode_2)$: the vertex of dummy variables $\dummyderef{x}{\athread}$ and of $\dummyderef{y}{\athread}$. By DCR applied to $(\vertexof{y}, \vertexof{p_y}, \avertex, \vertexof{\values_1(p_y)})$, $\avertex$ has an edge to $\vertexof{\values_1(p_y)}$. By DCR applied to $(\vertexof{x}, \avertex, \vertexof{p_x}, \vertexof{\values_1(p_y)})$, $\vertexof{p_x}$ has an edge to $\vertexof{\values_1(p_y)}$. This proves that $p_x \pt{} \values_1(p_y)$. See \autoref{soundness-double-deref} for an illustration.

\begin{figure}
  \centering
  \begin{subfigure}[b]{0.45\linewidth}
    \centering
  \begin{tikzpicture}[yscale = 1, xscale = 1]
  \tikzstyle{vertex} = [draw,circle,minimum size = 0.8cm]
  \node[vertex] at (0,0) (x) {$x$};
  \node[vertex, inner sep = 1pt] at (-1,-2) (v) {$\dummyderef{x}{\athread}$};
  \node[vertex] at (1,-2) (px) {$p_x$};
  \node[vertex] at (0,-4) (y) {$y$};
  \draw[-latex] (x) -- (px);
  \draw[-latex] (x)--(v);
  \draw[-latex] (v)--(y);
  \draw[-latex,color=red!60!black,dashed,thick] (px) -- (y);
  \end{tikzpicture}
  \label{soundness-deref-address}
  \caption{Illustration of the case $\deref{x}=\addr{y}$ of the soundness proof.}
  \label{soundness-deref-address}
  \end{subfigure}
  \hfill
  \begin{subfigure}[b]{0.45\linewidth}
    \centering
  \begin{tikzpicture}[yscale = 1, xscale = 1]
    \tikzstyle{vertex} = [draw,circle,minimum size = 0.8cm] 
    \node[vertex] at (-1.5,0) (x) {$x$} ;
    \node[vertex] at (1.5,0) (y) {$y$} ;
    \node[vertex] at (-2.5,-2) (px) {$p_x$} ;
    \node[vertex, inner sep = 1pt] at (0,-2) (v) {$\dummyderef{x}{\athread}, \dummyderef{y}{\athread}$} ;
    \node[vertex] at (2.5,-2) (py) {$p_y$} ;
    \node[vertex, inner sep = 1pt] at (0,-4) (ppy) {$\values_1(p_y)$};
    \draw[-latex] (x) -- (px);
    \draw[-latex] (y)--(py);
    \draw[-latex] (x)--(v);
    \draw[-latex] (y)--(v);
    \draw[-latex] (py)--(ppy);
    \draw[-latex,color=red!60!black,dashed,thick] (v) -- (ppy);
    \draw[-latex,color=red!60!black,dashed,thick] (px) -- (ppy);
  \end{tikzpicture}
  \caption{Illustration of the case $\deref{x}=\deref{y}$ of the soundness proof. DCR must be applied twice, first to obtain the middle edge and then to obtain the left one.}
  \label{soundness-double-deref}
  \end{subfigure}
  \caption{Illustration of the last two cases of the soundness proof. Dashed red edges are obtained by DCR.}
\end{figure}

\subsubsection{A subtlety with thread creation}
\label{proof-pta-soundness-subtlety}
A technicality was hidden above that complicates the proof. It is related to the case (intentionally omitted above) where the instruction $\aninstr$ is a thread creation. 
Suppose that we naively perform an induction over the length of the execution and that the prefix execution ends at configuration $\config = \tuple{\stateFunction, \values, \held}$ from where a step is performed that creates a thread $\athread$. We must guarantee that $\pto(\startnodeof{\athread})$ contains all points-to pairs in $\values$. Recall that, by definition, \[ \pto(\startnodeof{\athread}) = \DCRbig{\procpto(\startnodeof{\athread}) \PTGunion \bigPTGunion\limits_{\anode \in \MHB{\startnodeof{\athread}}} \procpto(\anode)}. \]
The induction hypothesis gives information about $\pto(\anode)$, but $\pto(\startnodeof{\athread})$ is defined through $\procpto(\anode)$. In other words, the induction hypothesis is at the wrong abstract level, and it does not explain thanks to which $\procpto(\anode)$ the induction hypothesis holds. Consequently, when a thread is created, the hypothesis cannot be directly transferred to $\pto(\startnodeof{\athread})$. Proving that $\pto(\startnodeof{\athread})$ contains all points-to pairs requires to prove that all $\procpto(\anode)$ allowing the induction hypothesis to hold also appear in the definition of $\pto(\startnodeof{\athread})$. More precisely, this requires to prove that the union of all $\MHB{\anode}$ for all $\anode$ in $\config$ form a subset of $\MHB{\startnodeof{\athread}}$. However, this does not hold in general...
We therefore strengthen the induction hypothesis so that it specifies a set of $\procpto(\anode')$ that suffice, when combined, to obtain a sound points-to graph. 

\subsubsection{A new induction hypothesis}
\label{proof-pta-soundness-induction-hypothesis}
In fact, we will prove that, for soundness, it is enough to combine all $\procpto(\anode')$ where $\anode'$ has been visited before in the execution.
For every execution $\exec: \config_0 \step{*} \config = \tuple{\stateFunction,\values,\held}$ , we let $\HB{\exec} \subseteq \nodes$ denote the set of nodes that have been visited in $\exec$. We let $\newpt{\exec}$ be the points-to graph aggregating information $\procpto(\anode)$ from every node visited in $\exec$: \[\newpt{\exec} \deff \DCRbig{\bigPTGunion_{\anode' \in \HB{\exec}} \procpto(\anode')}.\] 
Unlike $\pto(\anode)$, $\newpt{\exec}$ depends explicitly on all nodes visited so far and is therefore easier to propagate when a new thread is created.
We now want to prove that $\newpt{\exec}$ contains less points-to pairs than $\pto(\anode)$ but still contains all points-to pairs at the end of $\exec$. 
To prove that $\newpt{\exec}$ contains less points-to pairs than $\pto(\anode)$, we need to prove non-decreasivity of DCR:

\begin{lem}
\label{lemma-newpt-smaller-pto}
For every $\exec: \initialconfig  \step{*} \config = \tuple{\stateFunction, \values, \held}$, for every $\anode$ such that $\stateFunction(\threadof{\anode}) = \anode$, $\newpt{\exec} \subptg \pto(\anode)$. 
\end{lem}
\begin{proof}
Let $\exec: \initialconfig  \step{*} \config = \tuple{\stateFunction, \values, \held}$ and let $\anode$ such that $\stateFunction(\threadof{\anode}) = \anode$. The assertion to prove expands to:
\[\DCRbig{\bigPTGunion_{\anode_{\mathsf{HB}} \in \HB{\exec}} \procpto(\anode_{\mathsf{HB}})} \subptg \DCRbig{\procpto(\anode) \PTGunion \bigPTGunion_{\anode' \in \MHB{\anode}} \procpto(\anode')}.\]
By non-decreasivity of $\subptg$ (\autoref{DCR-nondec-extensive}), it suffices to prove that \[\bigPTGunion_{\anode_{\mathsf{HB}} \in \HB{\exec}} \procpto(\anode_{\mathsf{HB}}) \subptg \procpto(\anode) \PTGunion \bigPTGunion_{\anode' \in \MHB{\anode}} \procpto(\anode').\]
Let $\nodes_R \deff \set{\anode} \cup \MHB{\anode}$ be the set of nodes appearing in the right-hand side. 
By extensivity of $\PTGunion$ (item \ref{union-extensivity} of \autoref{ptg-union-properties}), it suffices to prove that,
 for every $\anode_{\mathsf{HB}} \in \HB{\exec}$, there exists $\anode_R \in \nodes_R$ such that $\procpto(\anode_{\mathsf{HB}}) \subptg \procpto(\anode_R)$.
If $\threadof{\anode_{\mathsf{HB}}} \ne \threadof{\anode}$ then $\anode_{\mathsf{HB}} \in \MHB{\anode}$ by \autoref{soundness-joined} and \autoref{soundness-uncreated}, so that $\anode_{\mathsf{HB}} \in \nodes_R$ and it suffices to let $\anode_R \deff \anode_{\mathsf{HB}}$. Suppose now that $\threadof{\anode_{\mathsf{HB}}} = \threadof{\anode}$. This implies that $\anode_\mathsf{HB}$ has been visited by $\threadof{\anode}$ earlier in $\exec$. Therefore, following the trajectory of $\threadof{\anode}$ in $\exec$ witnesses the existence of a path from $\anode_{\mathsf{HB}}$ to $\anode$. This implies that $\procpto(\anode_{\mathsf{HB}}) \subptg \procpto(\anode)$ by cumulativity of $\procpto$, so that it suffices to let $\anode_R \deff \anode$. 
 \end{proof}

 \subsubsection{Wrapping up the proof}
 \label{proof-pta-soundness-wrapping-up}
We have proven that $\newpt{\exec}$ contains less points-to pairs than $\pto(\anode)$ for every $\anode$ populated in the last configuration $\config = \tuple{\stateFunction,\values,\held}$ of $\exec$. Formally, since $\newpt{\exec} \subptg \pto(\anode)$, we have that $x \pt{\newpt{\exec}} y$ implies $x \pt{\pto(\anode)} y$ for every $\anode$ such that $\stateFunction(\threadof{\anode}) = \anode$.
We now prove that $\newpt{\exec}$ contains all points-to pairs in the last configuration of $\config$. This allows for a stronger induction hypothesis than the one used in \autoref{proof-pta-soundness-simplified}: 

\begin{lem} 
\label{lemma-newpt-larger-values}
For every $\exec: \config_0 \step{*} \config = \tuple{\stateFunction, \values, \held}$, for every $x,y \in \vars$ such that $\values(x) = y$, $x \pt{\newpt{\exec}} y$. 
\end{lem}
\begin{proof}
We proceed by induction on the length of the execution. 
By definition, $\procpto(\startnodeof{\maint}) = \PTGof{\values_0}{\maint}$ so the property holds for the execution of length $0$. Suppose that the property is true for length $m$, and let $\exec_1: \initialconfig \step{*} \config_1 = \tuple{\stateFunction_1, \values_1, \held_1}$ of length $m$ and $\config_2 = \tuple{\stateFunction_2, \values_2, \held_2}$, $\atrans = (\anode_1, \aninstr, \anode_2) \in \transitions$ such that $\config_1 \step{\atrans} \config_2$. Let $\exec_2$ denote the execution obtained by appending $\config_1 \step{\atrans} \config_2$ at the end of $\exec_1$. We have $\HB{\exec_1} \subseteq \HB{\exec_2}$; since $\PTGunion$ is extensive by \autoref{ptg-union-properties} and $\PTG \mapsto \DCR{G}$ is non-decreasing by \autoref{DCR-nondec-extensive}, this gives $\newpt{\exec_1} \subptg \newpt{\exec_2}$. Moreover, since $\anode_2 \in \HB{\exec_2}$, $\procpto(\anode_2)$ appears in $\bigPTGunion_{\anode' \in \HB{\exec_2}} \procpto(\anode')$ therefore, by extensivity of $\PTGunion$ and because $\PTG \subptg \DCR{\PTG}$ by \autoref{DCR-nondec-extensive},  $\procpto(\anode_2) \subptg \newpt{\exec_2}$. Let $\athread \deff \threadof{\atrans}$.

Observe that, for every $x,y$ such that $\values_1(x) = y$, by induction hypothesis $x \pt{\newpt{\exec_1}} y$, therefore $x \pt{\newpt{\exec_2}} y$ thanks to $\newpt{\exec_1} \subptg \newpt{\exec_2}$. It only remains to prove the property for pairs $(x,y)$ such that $\values_1(x) \ne y$ and $\values_2(x) = y$. In fact, there is at most one variable whose points-to target changes, therefore there is at most one such pair $(x,y)$, and it may only exist when $\aninstr$ is an assignment of depth $k \geq 1$ (\emph{i.e.}, a pointer assignment), which we now suppose.

We make a case disjunction on the shape of $\aninstr$. The cases where the right-hand side is not properly initialized and where the value on the left-hand side is a dereference of an uninitialized pointer will not yield valid steps. We now assume that all relevant values are initialized in $\config_1$. The six cases treated below are very similar to the six cases from \autoref{proof-pta-soundness-simplified}, except that $\newpt{\exec_1}$ and $\newpt{\exec_2}$ are used instead of $\pto(\anode_1)$ and $\pto(\anode_2)$.
\begin{itemize}
\item If $\aninstr$ is $\instrstyle{x \deff y}$ then $\values_2(x) = \values_1(y)$. By induction hypothesis, $y$ points to $\values_1(y)$ in $\newpt{\exec_1}$, therefore also in $\newpt{\exec_2}$. In $\procpto(\anode_2)$, by construction of $\procpto$, $x$ and $y$ are in the same vertex. In $\newpt{\exec_2}$, the same must hold since $\procpto(\anode_2) \subptg \newpt{\exec_2}$, so that $x$ points to the same variables as $y$ and in particular $x \pt{\newpt{\exec_2}} \values_1(y)$.  
\item If $\aninstr$ is $\instrstyle{x \deff \deref{y}}$ then $\values_2(x) = \values_1(\values_1(y))$. In $\newpt{\exec_1}$, $y$ points to $\values_1(y)$ and $\values_1(y)$ points to $\values_1(\values_1(y))$. In $\procpto(\anode_2)$, by construction, $y$ points to $x$. 
In $\newpt{\exec_2}$, since $\newpt{\exec_1} \subptg \newpt{\exec_2}$ and $\procpto(\anode_2) \subptg \newpt{\exec_2}$, this implies that $y \pt{\newpt{\exec_2}} \values_1(y)$, $\values_1(y) \pt{\newpt{\exec_2}} \values_1(\values_1(y))$ and $y \pt{\newpt{\exec_2}} x$.
Thanks to the diamond completion rule (DCR) applied to $(\vertexof{y},\vertexof{\values_1(y)},\vertexof{x},\vertexof{\values_1(\values_1(y))})$, $x \pt{\newpt{\exec_2}} \values_1(\values_1(y))$.
\item If $\aninstr$ is $\instrstyle{\deref{x} \deff y}$ then, by letting $p_x \deff \values_1(x)$ and $p_y \deff \values_1(y)$, we have $\values_2(p_x) = p_y$. By construction, $x$ points to $y$ in $\procpto(\anode_2)$. In $\newpt{\exec_1}$ therefore in $\newpt{\exec_2}$, $x$ points to $p_x$ and $y$ points to $p_y$. By DCR applied to $(\vertexof{x}, \vertexof{y}, \vertexof{p_x}, \allowbreak \vertexof{p_y})$, $p_x \pt{\newpt{\exec_2}}p_y$.
\item If $\aninstr$ is $\instrstyle{x \deff \addr{y}}$ then $\values_2(x) = y$. By construction, $x$ points to $y$ in $\procpto(\anode_2)$. Therefore, it also does in $\newpt{\exec_2}$.
\item If $\aninstr$ is $\instrstyle{\deref{x} \deff \addr{y}}$, then, by letting $p_x \deff \values_1(x)$, $\values_2(p_x) = y$. By construction, there is an edge from $\vertexof{x}$ to $\vertexof{\dummyderef{x}{\athread}}$ and an edge from $\vertexof{\dummyderef{x}{\athread}}$ to $\vertexof{y}$ in $\procpto(\anode_2)$. The same must therefore hold in $\newpt{\exec_2}$. By DCR applied to $(\vertexof{x}, \vertexof{\dummyderef{x}{\athread}}, \vertexof{p_x}, \vertexof{y})$, $p_x$ points to $y$ in $\newpt{\exec_2}$. See \autoref{soundness-deref-address} for an illustration. 
\item If $\aninstr$ is $\instrstyle{\deref{x} \deff \deref{y}}$ then, by letting $p_x \deff \values_1(x)$ and $p_y \deff \values_1(y)$, we have $\values_2(p_x) = \values_1(p_y)$. In $\newpt{\exec_1}$ therefore in $\newpt{\exec_2}$, $x$ points to $p_x$, $y$ points to $p_y$ and $p_y$ points to $\values_1(p_y)$. In $\procpto(\anode_2)$, by construction, $\vertexof{x}$ and $\vertexof{y}$ share a child $v$ (the vertex with variables $\dummyderef{x}{\athread}$ and $\dummyderef{y}{\athread}$). This implies that $\vertexof{x}$ and $\vertexof{y}$ share a child $\avertex$ in $\newpt{\exec_2}$. By DCR applied to $(\vertexof{y}, \vertexof{p_y}, \avertex, \vertexof{\values_1(p_y)})$, we have that $\avertex$ has an edge to $\vertexof{\values_1(p_y)}$ in $\newpt{\exec_2}$. By DCR applied to $(\vertexof{x}, \avertex, \vertexof{p_x}, \vertexof{\values_1(p_y)})$, $\vertexof{p_x}$ has an edge to $\vertexof{\values_1(p_y)}$ in $\newpt{\exec_2}$, which proves that $p_x \pt{\newpt{\exec_2}} \values_1(p_y)$. See \autoref{soundness-double-deref} for an illustration. 
\end{itemize}
It must be noted that, with the new induction hypothesis, the case where $\aninstr$ is a thread creation is trivial: the induction hypothesis is expressed at the level of the entire execution, so the presence of a new thread has no impact on the truth status of the induction hypothesis. 
\end{proof} 

We now conclude the proof of \autoref{concurrent-alias-soundness}. Let $\exec: \initialconfig \step{*} \config = \tuple{\stateFunction, \values, \held}$, let $\athread$ be a thread, let $\anode$ such that $\stateFunction(\athread) = \anode$ and let $x, y \in \vars$ such that $\values(x) = y$. 
Thanks to \autoref{lemma-newpt-larger-values}, $x \pt{\newpt{\exec}}$. By \autoref{lemma-newpt-smaller-pto}, $\newpt{\exec} \subptg \pto(\anode)$. Therefore $x \pt{\pto(\anode)} y$, concluding the proof of \autoref{concurrent-alias-soundness}.

\section{Automatic Mutex Placement}\label{sec:automatic-placement}
We introduce a technique that, from a program $\prog$, adds mutexes to obtain a new program $\newprog$ with no data race. \emph{Data races} are common concurrency bugs which occur when two threads simultaneously access the same memory location. To do that, the technique identifies locations of potential data races (\autoref{subsec:data-race-detection}) and protects them using mutexes (\autoref{subsec:data-race-protection}). In addition, we extend this technique to detect potential atomicity violation of user-indicated critical sections (\autoref{subsec:atomicity-detection}) and automatically protect these critical sections (\autoref{subsec:atomicity-protection}). The goal is to simplify development of safe concurrent programs by, from an input program (in the form of its source code),  automatically inserting locks to obtain formal guarantees about the absence of data races and atomicity violations.

A crucial concern is that the transformation must not create deadlocks. A deadlock corresponds to a set of threads mutually blocking one another: all these threads become unable to progress. For our purposes, the only blocking operation is locking. Therefore, a deadlock here corresponds to a set of threads each waiting upon some mutex held by another thread in the set.
More precisely, deadlocks may only occur in the presence of a \emph{cyclic dependency}, \emph{i.e.}, a sequence of mutexes $\amutex_1, \dots, \amutex_n, \amutex_{n+1} = \amutex_1$ (with $n \geq 2$) such that, for each $i \leq n$, a thread tries to lock $\amutex_{i+1}$ while holding $\amutex_i$ (see, \emph{e.g.}, \cite{LK14}). A classic strategy for deadlock avoidance is to impose a \emph{lock ordering}, \emph{i.e.}, a total order $\lockordering$ on the set of mutexes such that a mutex smaller for $\lockordering$ must be acquired earlier, \emph{i.e.}, for every $\amutex \lockordering \amutex'$ with $\amutex \ne \amutex'$, a thread can never try to perform a $\lock{\amutex}$ instruction while holding $\amutex'$. In order to remain as general as possible, we do not assume any lock ordering for $\prog$. Instead, we impose a lock ordering on newly-introduced mutexes; we will prove in \autoref{first-transfo-no-deadlock} that no deadlock is introduced this way. 

Below, we consider a fixed input program $\prog$. 
The transformation introduces, for each variable $x$, a new mutex $\newmut{x}$ in charge of protecting $x$. We let $\newmutexes \deff \set{\newmut{x} \mid x \in \vars}$; the full set of mutexes becomes $\mutexes \disjointunion \newmutexes$. To prevent deadlocks, we consider an arbitrary ordering $\lockordering$ on $\newmutexes$, which we consider fixed throughout the remainder of this section.  

\subsection{Data Race Detection}
\label{subsec:data-race-detection}

In our model, the only instructions that access variables are assignments and conditional guards. We call \emph{access instructions} such instructions, and \emph{access transitions} the transitions whose instructions are access instructions. 

\begin{defi}
\label{defi-written-read}
Given an access transition $\atrans = (\anode, \aninstr,\anode')$ and a valuation $\values: \vars \to \rels \cup \vars \cup \set{\uninit}$, we let $\writeacc{\atrans}{\values} \subseteq \vars$ (resp. $\readacc{\atrans}{\values} \subseteq \vars$) denote the set of variables defined as follows: 
\begin{itemize}
\item if $\aninstr$ is a conditional guard $\trueguard{e}, \falseguard{e}$, then $\writeacc{\atrans}{\values}$ is empty and $\readacc{\atrans}{\values}$ contains all variables accessed to evaluate $e$, \emph{i.e.}, any $x$ syntactically appearing in $e$ as well as $\values(x)$ for any $\deref{x}$ appearing in $e$ (address-of terms $\addr{x}$ are not considered, they cannot be involved in data races nor atomicity violations);
\item if $\aninstr$ is an assignment $x \deff e$, then $\writeacc{\atrans}{\values} = \set{x}$ and $\readacc{\atrans}{\values}$ contains all variables accessed to evaluate $e$, as defined in the first case above, as well as $x$;
\item if $\aninstr$ is an assignment $\deref{x} \deff e$, then $\writeacc{\atrans}{\values} = \set{\values(x)}$ and $\readacc{\atrans}{\values}$ is as above except that it contains both $x$ and $\values(x)$.
\end{itemize}
\end{defi}
In words, $\writeacc{\atrans}{\values}$ (resp. $\readacc{\atrans}{\values}$) denotes the set of variables that are written to (resp. read) in $\aninstr$, assuming that the current valuation is $\values$. 
Note that $\writeacc{\values}{\atrans} \subseteq \readacc{\values}{\atrans}$.
Given $\atrans = (\anode, \aninstr, \anode')$, we overapproximate the set of variables read (resp. written to) by $\atrans$ thanks to the points-to analysis $\pto(\anode)$ from \autoref{sec:points-to-analysis}. 
We define its may-read set $\mayread{\atrans}$ (resp. may-write set $\maywrite{\atrans}$) similarly to $\readacc{\atrans}{\values}$ (resp. $\writeacc{\atrans}{\values}$) except that for $\deref{x}$, instead of $\values(x)$, we add all $y$ such that $x \pt{\pto(\anode)} y$. Formally:
\begin{defi}
\label{def-mr_mw}
Given an access transition $\atrans = (\anode, \aninstr,\anode')$, we let $\maywrite{\atrans} \subseteq \vars$ (resp. $\mayread{\atrans} \subseteq \vars$) denote the set of variables defined as follows: 
\begin{itemize}
\item if $\aninstr$ is a conditional guard $\trueguard{e}, \falseguard{e}$, then $\maywrite{\atrans}$ is empty and $\mayread{\atrans}$ contains any $x \in \vars$ syntactically appearing in $e$ as well as, for each $x \in \vars$ such that $\deref{x}$ appears in $e$, any $y \in \vars$ such that $x \pt{\pto(\anode)} y$;
\item if $\aninstr$ is an assignment $x \deff e$, then $\maywrite{\atrans} = \set{x}$ and $\mayread{\atrans}$ contains all variables potentially accessed to evaluate $e$, as defined in the first case above, as well as $x$;
\item if $\aninstr$ is an assignment $\deref{x} \deff e$, then $\maywrite{\atrans} = \set{y \mid x \pt{\pto(\anode) y}}$ and $\mayread{\atrans}$ is as above except that it contains both $x$ and any $y$ such that $x \pt{\pto(\anode)} y$.
\end{itemize}
\end{defi}

By soundness of the concurrent points-to analysis (\autoref{concurrent-alias-soundness}), this gives over-approximations of the set of variables read and written:
\begin{prop}
\label{prop:mw-mr-soundness}
Let $\config = \tuple{\stateFunction, \values, \held}$ be a reachable configuration and $\atrans$, $\config'$ such that $\config \step{\atrans} \config'$. We have $\writeacc{\atrans}{\values} \subseteq \maywrite{\atrans}$ and $\readacc{\atrans}{\values} \subseteq \mayread{\atrans}$. 
\end{prop}
\begin{proof}
Let $\atrans \ffed (\anode, \aninstr, \anode')$ and $\athread = \threadof{\atrans}$.
Since $\atrans$ can be applied from $\config$ , we have $\stateFunction(\athread) = \anode$. 
By \autoref{concurrent-alias-soundness}, if $\values(x) = y \in \vars$ then $x \pt{\pto(\anode)} y$. 
Suppose by contradiction that there is $y \in \writeacc{\atrans}{\values} \setminus \maywrite{\atrans}$.
Variable $y$ does not syntactically appear in $\aninstr$ as $y \notin \maywrite{\atrans}$. By a direct case disjunction, we have $y \in \values(x)$ where $\deref{x}$ appears in $\aninstr$. But this implies that $x \pt{\pto(\anode)} y$ therefore $y$ is added to $\maywrite{\atrans}$, a contradiction. This proves that $\writeacc{\atrans}{\values} \subseteq \maywrite{\atrans}$. The proof that $\readacc{\atrans}{\values} \subseteq \mayread{\atrans}$ is similar.  
\end{proof}

In our model, a program $\prog$ has a data race when there exists a reachable configuration from which two transitions of different threads are enabled which access the same variable, with at least one of the transitions writing to that variable. A transition $\atrans$ is \emph{enabled} from a configuration $\config$ if there is $\config'$ such that $\config \step{\atrans} \config'$. 
\begin{defi}
\label{def-data-race}
Given two access transitions $\atrans_1$ and $\atrans_2$ such that $\threadof{\atrans_1} \ne \threadof{\atrans_2}$ and a variable $x$, $\prog$ has \emph{a data race between $\atrans_1$ and $\atrans_2$ on $x$} when there is a reachable configuration $\config=\tuple{\stateFunction, \values, \held}$ from which $\atrans_1$ and $\atrans_2$ are enabled and such that \[x \in (\writeacc{\atrans_1}{\values} \cap \readacc{\atrans_2}{\values}) \cup (\readacc{\atrans_1}{\values} \cap \writeacc{\atrans_2}{\values}).\] 
\end{defi}
The condition $x \in (\writeacc{\atrans_1}{\values} \cap \readacc{\atrans_2}{\values}) \cup (\readacc{\atrans_1}{\values} \cap \writeacc{\atrans_2}{\values})$ expresses that at least one of the transitions writes to $x$. 
Observe that $\writeacc{\atrans}{\values} \subseteq \readacc{\atrans}{\values}$ so this also includes the case where both of them write to $x$. Since the semantics represents concurrency by interleaving, simultaneous enabledness of two conflicting accesses represents the possibility that these accesses execute concurrently in the actual concurrent program.

We define a relation $\MHP$ that overapproximates when two nodes or transitions may happen in parallel. 

\begin{multline*}
\anode_1 \MHP \anode_2 \quad \iff \quad \threadof{\anode_1} \ne \threadof{\anode_2} \quad \land \quad \athread_1 \notin \uncreated{\anode_2} \cup \joined{\anode_2} \quad \land \quad \athread_2 \notin \uncreated{\anode_1} \cup \joined{\anode_1}  \\\land \quad \lockset{\anode_1} \cap \lockset{\anode_2} = \emptyset. \end{multline*} 
We extend relation $\MHP$ to access transitions by letting $\atrans_1 \MHP \atrans_2$ whenever, by letting $\atrans_1= (\anode_1, \aninstr_1, \anode_1')$ and $\atrans_2 = (\anode_2, \aninstr_2, \anode_2')$,  we have $\anode_1 \MHP \anode_2$. 

\begin{lem}
\label{soundness-MHP}
For every reachable configuration $\config = \tuple{\stateFunction,\values, \held}$, for every $\athread \ne \athread'$, for every $\anode, \anode'$ such that $\stateFunction(\athread) = \anode$ and $\stateFunction(\athread') = \anode'$, we have $\anode \MHP \anode'$.
\end{lem}
\begin{proof}
By \autoref{soundness-joined} and \autoref{soundness-uncreated}, $\athread \notin \uncreated{\anode'} \cup \joined{\anode'}$ and $\athread' \notin \uncreated{\anode} \cup \joined{\anode}$. Also, there cannot be $\amutex \in \lockset{\anode} \cap \lockset{\anode'}$ thanks to \autoref{soundness-lockset}. This proves that $\anode \MHP \anode'$.
\end{proof}

For every variable $x$, we compute a \emph{may-race relation} $\mayrace{x}$ on transitions as follows.  
\begin{multline*}
\forall x \in \vars, \quad
\atrans_1 \mayrace{x} \atrans_2 \; \iff \; \atrans_1 \MHP \atrans_2 \; \land \\ x \in (\maywrite{\atrans_1} \cap \mayread{\atrans_2}) \cup  (\mayread{\atrans_1} \cap \maywrite{\atrans_2}).
\end{multline*}
In words, we consider that $\atrans_1$ may race with $\atrans_2$ on $x$ when they may happen in parallel and, based on their respective points-to overapproximations, there is a risk that they concurrently access $x$ with at least one of them writing to $x$. Recall that $\maywrite{\atrans} \subseteq \mayread{\atrans}$, so this also includes concurrent writes. 

\begin{lem}
\label{soundness-data-race-analysis}
If $\prog$ has a data race between two access transitions $\atrans_1$ and $\atrans_2$ on a variable $x$, then $\atrans_1 \mayrace{x} \atrans_2$.
\end{lem}
\begin{proof}
Let $\config = \tuple{\stateFunction,\values,\held}$ be a reachable configuration from which $\atrans_1 = (\anode_1, \aninstr_1, \anode_1')$ and $\atrans_2 = (\anode_2, \aninstr_2, \anode_2')$ are enabled such that $x \in (\writeacc{\atrans_1}{\values} \cap \readacc{\atrans_2}{\values}) \cup (\readacc{\atrans_1}{\values} \cap \writeacc{\atrans_2}{\values})$.
In $\config$, $\athread_1 \deff \threadof{\atrans_1}$ is in $\anode_1$ since $\athread_1$ is enabled and $\athread_2 \deff \threadof{\atrans_2}$ is in $\anode_2$ since $\athread_2$ is enabled. 
By \autoref{soundness-MHP}, $\atrans_1 \MHP \atrans_2$.
Also, by \autoref{prop:mw-mr-soundness}, for all $i \in \set{1,2}$, $\writeacc{\atrans_i}{\values} \subseteq \maywrite{\atrans_i}$ and $\readacc{\atrans_i}{\values} \subseteq \mayread{\atrans_i}$.
Therefore,
$x \in (\maywrite{\atrans_1} \cap \mayread{\atrans_2}) \cup  (\mayread{\atrans_1} \cap \maywrite{\atrans_2})$, so that $\atrans_1 \mayrace{x} \atrans_2$.
 \end{proof}

\subsection{Data Race Protection} 
\label{subsec:data-race-protection}
We insert mutexes and locking instructions in order to transform program $\prog$ into a new program $\newprog$ guaranteed free of data races. This program $\newprog$ is constructed by inserting, around some transitions of $\prog$, locking and unlocking instructions of relevant mutexes. The aim is to ensure that, at a transition $\atrans$ in $\newprog$, the thread must hold all mutexes $\newmut{x}$ for all variables $x$ potentially involved in a data race at $\atrans$. 

Formally, $\prog$ and $\newprog$ share the same sets of threads and variables; it has all mutexes in $\mutexes$ and additional fresh mutexes from $\newmutexes$.
In the control-flow graph of every thread in $\threads$, its set of nodes in $\newprog$ is equal to its set of nodes of $\prog$ plus some additional intermediate nodes. We identify copies of nodes of $\prog$ by giving them the same name in $\newprog$. Moreover, to every intermediate node $\newof{\anode}$ added to $\newprog$, we associate a node $\oldnodeof{\newof{\anode}}$ of $\prog$, defined below. 

Given a transition $\atrans= (\anode_a, \aninstr, \anode_b)$ in $\prog$ and a sequence $\sigma= \aninstr_1, \dots, \aninstr_k$ of instructions, we call \emph{inserting $\sigma$ before $\atrans$} the operation that consists in adding intermediate nodes immediately after $\anode_a$ so that all instructions in $\sigma$ are performed before $\aninstr$. We formalize this by expressing how $\newprog$ differs from $\prog$. $\newprog$ also has nodes $\anode_a$ and $\anode_b$; but instead of being connected by $\atrans$, they are connected using a sequence of intermediate states and instructions. $\newprog$ has fresh nodes $\newof{\anode_2}$ to $\newof{\anode_{k+1}}$ which do not correspond to nodes of $\prog$; by letting $\newof{\anode_1} \deff \anode_a$, there are transitions $(\newof{\anode_1}, \aninstr_1, \newof{\anode_2})$, \dots, $(\newof{\anode_{k}}, \aninstr_k, \newof{\anode_{k+1}})$ (corresponding to all instructions in $\sigma$) and a transition $(\newof{\anode_{k+1}}, \aninstr, \anode_b)$ (corresponding to $\atrans$). Overall, this guarantees that, before $\atrans$ is performed, all instructions in $\sigma$ are performed. All intermediate nodes $\newof{\anode_2}$ to $\newof{\anode_{k+1}}$ are mapped to $\anode_a$: for every $i \in \nset{2}{k+1}$, $\oldnodeof{\newof{\anode_i}} = \anode_a$. 
We define similarly \emph{inserting $\sigma$ after $\atrans$}, inserting a sequence of transitions after $\aninstr$ that ends in $\anode_b$. In this case, all newly inserted nodes are mapped to $\anode_b$.
Overall, if $\sigma_1$ is inserted before $\atrans$ and $\sigma_2$ is inserted after $\atrans$, there is in $\newprog$ a sequence of transitions from $\anode_a$ to $\anode_b$ whose sequence of instructions is $\sigma_1 \cdot \aninstr \cdot \sigma_2$. 

To obtain $\newprog$ from $\prog$, we do the following for every access transition $\atrans$. Let $M_\atrans \deff \set{\newmut{x} \mid x \in \vars, \, \exists \atrans': \, \atrans \mayrace{x} \atrans'}$; $M_\atrans$ contains all mutexes of variables on which $\atrans$ may be involved in a data race. Let $\lockseq$ be the sequence containing all $\lock{\newmut{x}}$ for $\newmut{x} \in M_\atrans$, in the order given by $\lockordering$. Similarly, let $\unlockseq$ be the sequence containing all $\unlock{\newmut{x}}$ for $\newmut{x} \in M_\atrans$ (the order is irrelevant for unlocking). The transformation inserts $\lockseq$ before $\atrans$ and $\unlockseq$ after $\atrans$, guaranteeing that the new corresponding transition can only be performed by a thread that holds all mutexes in $M_\atrans$. 
The transformation introduces a mutex for every variable, with the goal to yield lower contention than simpler approaches such as coarse-grained locking. 

Given a transition $\atrans = (\anode_a, \aninstr, \anode_b)$ of $\prog$, we define the sequence of transitions $\newseqof{\atrans}$ as the sequence of transitions of $\newprog$ corresponding to $\atrans$, defined as follows:
\begin{itemize}
\item if $\atrans$ is not modified by the transformation, then $\newseqof{\atrans} = [\atrans]$ is the sequence containing only $\atrans$.
\item otherwise, $\atrans$ is an access transition such that sequences $\lockseq$ and $\unlockseq$ were introduced before and after it. In this case, $\newseqof{\atrans}$ contains the sequence of transitions in $\newprog$ from $\anode_a$ to $\anode_b$  corresponding to $\atrans$, with sequence of instructions $\lockseq \cdot \aninstr \cdot \unlockseq$. 
\end{itemize}

By immediate induction, we can build an execution of $\newprog$ from an execution of $\prog$ by making each thread perform $\newseqof{\atrans}$ instead of $\atrans$. 
Conversely, an execution of $\newprog$ can be turned into an execution of $\prog$ by removing all lock and unlock instructions of mutexes in $\newmutexes$:
\begin{prop}
\label{old-exec-from-new}
For every reachable configuration $\newof{\config} = \tuple{\newof{\stateFunction}, \newof{\values}, \newof{\held}}$ of $\newprog$, there exists a reachable configuration $\config = \tuple{\stateFunction, \values, \held}$ of $\prog$ such that, for every $\athread \in \threads$:
\begin{itemize}
\item if $\newof{\stateFunction}(\athread) \notin \set{\stateUncreated, \stateJoined}$ then $\stateFunction(\athread) = \oldnodeof{\newof{\stateFunction}(\athread)}$;
\item if $\newof{\stateFunction}(\athread) \in \set{\stateUncreated, \stateJoined}$ then $\newof{\stateFunction}(\athread) = \stateFunction(\athread)$;
\item for every $x \in \vars$, $\newof{\values}(x) = \values(x)$;
\item for every $\amutex \in \mutexes$, $\newof{\held}(\amutex) = \held(\amutex)$. 
\end{itemize} 
\end{prop}
\begin{proof}
The proof proceeds by induction on the length of the execution $\newof{\exec}: \newof{\initialconfig} \step{*} \newof{\config}$. It trivially holds for execution of length $0$. Let $\newof{\exec}: \newof{\initialconfig} \step{*} \newof{\config_m} \step{\newof{\atrans}} \config$, and suppose that the property holds for the prefix up until $\newof{\config_m}$, so that there is a configuration $\config_m$ that satisfies the conditions above with respect to $\newof{\config_m}$. If $\newof{\atrans}$ corresponds to a transition $\atrans$ that was unaffected by the transformation, it suffices to apply $\atrans$ from $\config_m$. If $\newof{\atrans}$ is a lock or unlock transition introduced, then it can be ignored since the source and target nodes of $\atrans$ are mapped to the same node of $\prog$ through $\oldnodeofsymb$. 
Note that this condition on $\config$ does not speak about mutexes in $\newmutexes$. 
If $\newof{\atrans}$ is the access transition corresponding to a transition $\atrans$, but lies in $\newprog$ in the middle of a sequence $\newseqof{\atrans}$ of transitions, it suffices to make the thread perform $\atrans$ in $\prog$ from $\config_m$; the destination nodes of $\newof{\atrans}$ and of $\atrans$ correspond for $\oldnodeofsymb$. 
\end{proof}

\begin{exa}
\label{data-race-example}
Consider again the example from \autoref{fig:example-program}; the program obtained is depicted in \autoref{fig:mutex-insertion}. As highlighted in \autoref{example-reachable-config}, there is an execution with a data race on $k$ between transition $\atrans_{\ref{p2_access-out}}$ at line~\ref{p2_access-out} (with instruction \proglike{out = k}) and transition $\atrans_{\ref{p2_access-q}}$ at line~\ref{p2_access-q} (with instruction \proglike{\deref{q} = 2}). Let $\anode_{\ref{p2_access-q}}$ and $\anode_{\ref{p2_access-out}}$ denote their source nodes.
As displayed in \autoref{fig:dcr-needed}, $q \pt{\pto(\ref{p2_access-q})} k$, therefore $k \in \maywrite{\atrans_{\ref{p2_access-q}}}$. Trivially, $k \in \mayread{\atrans_{\ref{p2_access-out}}}$ since $k$ syntactically appears in $\atrans_{\ref{p2_access-out}}$. Moreover, $\atrans_{\ref{p2_access-q}} \MHP \atrans_{\ref{p2_access-out}}$; indeed, $\uncreated{\anode_{\ref{p2_access-q}}} = \joined{\anode_{\ref{p2_access-q}}} = \uncreated{\anode_{\ref{p2_access-out}}} = \joined{\anode_{\ref{p2_access-out}}} = \emptyset$. Also, $\lockset{\anode_{\ref{p2_access-q}}} = \set{\texttt{m}}$, $\lockset{\anode_{\ref{p2_access-out}}} = \emptyset$ 
so that $\lockset{\anode_{\ref{p2_access-q}}} \cap  \lockset{\anode_{\ref{p2_access-out}}} = \emptyset$.
This implies that $\atrans_{\ref{p2_access-q}} \mayrace{k} \atrans_{\ref{p2_access-out}}$, illustrating soundness from \autoref{soundness-data-race-analysis} on this example. In fact, this is the only may-race relation in this program. Therefore, the program $\newprog$ has a single new mutex, $\newmut{k}$. Two $\lock{\newmut{k}}$ instructions are inserted, as displayed in \autoref{fig:mutex-insertion}: one at line \ref{p2_lock1} and one at \ref{p2_lock2}. Corresponding $\unlock{\newmut{k}}$ instructions are inserted immediately after the instruction to protect. 
Had there been a need for a mutex $\newmut{\texttt{out}}$ protecting $\texttt{out}$ at line~\ref{p2_access-out}, the transformation would have introduced an arbitrary ordering $\lockordering$ on $\set{\newmut{k}, \newmut{\texttt{out}}}$, and inserted instructions $\lock{\newmut{k}}$ and $\lock{\newmut{\texttt{out}}}$ in the order given by $\lockordering$. 
\end{exa}

\begin{figure}
\centering
\begin{lstlisting}[language=C, basicstyle=\ttfamily\scriptsize, escapeinside={(*@}{@*)}, label=input, multicols = 4]
int** dbl_ptr;
int* p;
int* q;
int k;
int j;
int out;


main() {
  dbl_ptr = &p;
  q = &j;
  k=0, j=0;
  create(t1);
  create(t2); (*@\label{p2_create-t2}@*)
  lock(mu_k); (*@\label{p2_lock1}@*)
  out = k; (*@\label{p2_access-out}@*) 
  unlock(mu_k); (*@\label{p2_unlock1}@*)
} (*@\label{end-main}@*) 
t1() {
  lock(m); (*@\label{p2_t1-line1}@*) 
  dbl_ptr = &q; (*@\label{p2_t1-line2}@*) 
  p = &k; 
  unlock(m); 
  return;
}

t2() {
  lock(m);
  *dbl_ptr = p; (*@\label{p2_dblptr-to-p}@*) 
  lock(mu_k); (*@\label{p2_lock2}@*)
  *q = 2; (*@\label{p2_access-q}@*) 
  unlock(mu_k); (*@\label{p2_unlock2}@*)
  unlock(m); (*@\label{t2-unlock-m}@*)
  return;
}
\end{lstlisting}
\caption{The example from \autoref{fig:example-program} with the mutexes added.}
\label{fig:mutex-insertion}
\end{figure}
 The obtained program is free of data races:

 \begin{thm}
    \label{no-data-race-transfo}
There is no data race in $\newprog$.
\end{thm}
\begin{proof}
Suppose by contradiction that there is a data race in $\newprog$. 
Let $\newof{\exec}: \newof{\initialconfig} \step{*} \newof{\config} = \tuple{\newof{\stateFunction}, \newof{\values}, \newof{\held}}$ and two access transitions $\newof{\atrans_1}$ and $\newof{\atrans_2}$ enabled from $\newof{\config}$ such that, by letting $\athread_1 \deff \threadof{\newof{\atrans_1}}$ and $\athread_2 \deff \threadof{\newof{\atrans_2}}$, $\athread_1 \ne \athread_2$ and $x \in (\writeacc{\atrans_1}{\newof{\values}} \cap \readacc{\atrans_2}{\newof{\values}}) \cup (\readacc{\atrans_1}{\newof{\values}} \cap \writeacc{\atrans_2}{\newof{\values}})$. By \autoref{old-exec-from-new}, $\newof{\config}$ projects to reachable configuration $\config = \tuple{\stateFunction,\values,\held}$ of $\prog$ with the same valuation. 
Since the transformation introduces no access transitions, each $\newof{\atrans_i}$ corresponds to a unique access transition $\atrans_i = (\anode_i, \aninstr_i, \anode_i')$ in $\prog$. 
Because $\newof{\atrans_i}$ is enabled in $\newof{\config}$, we have $\oldnodeof{\newof{\stateFunction}(\athread_i)} = \anode_i$ hence $\stateFunction(\athread_i) = \anode_i$. Transitions $\newof{\atrans_i}$ and $\atrans_i$ have the same enabling conditions in both programs, therefore $\atrans_i$ is enabled in $\config$. Moreover, $x \in (\writeacc{\atrans_1}{{\values}} \cap \readacc{\atrans_2}{{\values}}) \cup (\readacc{\atrans_1}{{\values}} \cap \writeacc{\atrans_2}{{\values}})$. 
Therefore, there is a data race between $\atrans_1$ and $\atrans_2$ on $x$ in $\prog$. 
This implies that $\atrans_1 \mayrace{x} \atrans_2$ by \autoref{soundness-data-race-analysis}. By construction of $\newprog$, we have $\newmut{x} \in M_{\atrans_1} \cap M_{\atrans_2}$, \emph{i.e.}, $\newmut{x}$ is a protecting mutex for both $\atrans_1$ and $\atrans_2$ in $\newprog$. Hence, in $\newof{\exec}$, $\newof{\atrans_1}$ and $\newof{\atrans_2}$ are both surrounded by a $\lock{\newmut{x}} - \unlock{\newmut{x}}$ pair. In $\newof{\config}$, since $\newof{\atrans_1}$ and $\newof{\atrans_2}$ are enabled, $\athread_1$ and $\athread_2$ must both hold mutex $\newmut{x}$, a contradiction. 
 \end{proof}

We finally argue that the transformation introduces no deadlocks. 

\begin{prop}
\label{first-transfo-no-deadlock}
The transformation does not introduce new deadlocks, \emph{i.e.}, no cyclic lock dependency involves newly introduced mutexes. 
\end{prop}
\begin{proof}
More precisely, we prove that no cyclic lock dependency $\amutex_1, \dots, \amutex_n, \amutex_{n+1} = \amutex_1$ in $\newprog$ ($n \geq 2$) involves a mutex in $\newmutexes$. By contradiction and w.l.o.g., suppose $\amutex_1 = \newmut{x_1} \in \newmutexes$. 
There is an execution where some $\lock{\amutex_{2}}$ is performed while $\amutex_1$ is held by the thread. 
By construction, mutex $\newmut{x_1}$ is only held in the protected block around transitions identified as potentially involved in data races on $x$. 
By hypothesis, some $\lock{\amutex_2}$ can be performed inside a block where $\newmut{x_1}$ is locked. Consider this $\lock{\newmut{x_1}} - \unlock{\newmut{x_1}}$ block, let $\atrans$ be the access transition protected. 
While $\newmut{x}$ is held, by construction, the only \locksymb instructions that can be performed are those of the form $\lock{\newmut{y}}$ with $y$ s.t. $\newmut{x_1} \lockordering \newmut{y}$ (since $\atrans$ itself is an access transition, it cannot be a \locksymb instruction). 
Thus, $\amutex_{2} = \newmut{x_{2}}$ for some $x_{2} \ne x_1$ such that $\newmut{x_1} \lockordering \newmut{x_{2}}$. By iterating this argument, for all $j$, $\amutex_j = \newmut{x_j}$ with $x_j \ne x_{j+1}$ and $\newmut{x_j} \lockordering \newmut{x_{j+1}}$. 
We obtain by transitivity that $\newmut{x_1} \lockordering \newmut{x_n}$. Moreover, we obtain $\newmut{x_n} \lockordering \newmut{x_1}$ since $\amutex_{n+1} = \amutex_1$. We obtain $\newmut{x_1} = \newmut{x_n}$ by antisymmetry, a contradiction since all mutexes in a cyclic lock are assumed distinct. 
 \end{proof}

We now extend this technique to allow for the automatic protection of critical sections.

\subsection{Atomicity Violation Detection}
\label{subsec:atomicity-detection}
We extend the technique above to provide atomicity guarantees for user-annotated critical sections\footnote{In this context, such sections are often called \emph{atomic sections}. We here prefer the term \emph{critical sections} given that 1. atomicity is not guaranteed by default in the semantics, and must be enforced by the transformation, and 2. this is not true atomicity, but only absence of interference.}. In the rest of this paper, we consider again the program $\prog$ given in input, \emph{i.e.}, without the automatic data race protection from the previous section. Indeed, the protection against data races will depend on whether the instruction is in a critical section or not. 

A critical section of a thread $\athread$ is a portion $\criticalsection$ of code meant to run atomically from the perspective of $\athread$, \emph{i.e.}, free from interference of other threads. True atomicity is here not required: for performance reasons, other threads may still run as long as they do not interfere with $\athread$. 
Contrarily to data races, atomicity violations depend on the developer's intent: whether a given execution constitutes an atomicity violation depends on the intended behavior specified by the developer. 
Concurrent systems, by nature, do not always guarantee atomicity. For this reason, we assume that the original program $\prog$ comes with annotations indicating critical sections. At the level of Control Flow Graphs (CFGs), we represent critical sections as the set of all nodes between a first and a last node. 
\begin{defi}
\label{def:critical-section}
A \emph{critical section} of a thread $\athread$ is a set of nodes $\criticalsection \subseteq \nodesof{\athread}$ defined by the existence of two special nodes $\csfirst{\criticalsection},\cslast{\criticalsection} \in \criticalsection$ such that:
\begin{itemize}
\item $\csfirst{\criticalsection}$ has a single predecessor, which is not in $\criticalsection$;
\item $\cslast{\criticalsection}$ has a single successor, which is not in $\criticalsection$;
\item $\criticalsection \setminus \cslast{\criticalsection}$ is exactly the set of nodes $\anode \in \nodes$ such that there is a path from $\csfirst{\criticalsection}$ to $\anode$ that does not visit $\cslast{\criticalsection}$;
\item for every $\anode \in \criticalsection \setminus \cslast{\criticalsection}$, all successors of $\anode$ are in $\criticalsection$;
\item for every $\anode \in \criticalsection \setminus \csfirst{\criticalsection}$, all predecessors of $\anode$ are in $\criticalsection$. 
\end{itemize}
\end{defi}

The definition above is burdensome because it is expressed at the level of the CFG, which can take quite general shapes. We encourage readers to understand critical sections from the perspective of code: it is a section of code that can only be entered through the entry point $\csfirst{\criticalsection}$ and can only be exited through the exit point $\cslast{\criticalsection}$. 
\begin{prop}
\label{prop:critical-executions}
Let $\criticalsection$ be a critical section of a thread $\athread$ and let $\exec: \initialconfig \step{*} \config$ be an execution. 
Whenever $\exec$ enters $\criticalsection$, it must do so through $\csfirst{\criticalsection}$, and once $\csfirst{\criticalsection}$ has been visited, the execution remains in $\criticalsection$ until either $\cslast{\criticalsection}$ is visited or the execution terminates.
\end{prop}

From now on, we suppose that $\prog$ comes with a set of disjoint critical sections. 
A transition $(\anode, \aninstr, \anode')$ is said to be \emph{in} a critical section $\criticalsection$ when $\anode, \anode' \in \criticalsection$. Similarly, a path is said to be in $\criticalsection$ when all its transitions are in $\criticalsection$. For every critical section $\criticalsection$, we add the restriction that there is no transition in $\criticalsection$ with a $\proglike{lock}$ or $\texttt{unlock}$ instruction. This can be obtained by pushing $\proglike{lock}$ (resp. $\texttt{unlock}$) instructions to the start (resp. end) of the critical section when needed, although it must be noted that this modification is \emph{not} semantics-preserving in general. 

Our objective is to \emph{protect} critical sections by preventing any atomicity violation. Informally, a critical section $\criticalsection$ of a thread $\athread$ is protected if variables observed by $\athread$ in $\criticalsection$ cannot change due to interference. The formal definition is burdensome, because it requires to express the value of a variable $x$ from the perspective of $\athread$.

\begin{defi}
\label{def-protected}
A critical section $\criticalsection$ of a thread $\athread$ is \emph{protected} when $\values_1'(x) = \values_2(x)$ for every  $x \in \vars$, $\config_1, \config_1' = \tuple{\stateFunction_1', \values_1', \held_1'}$, $\config_2 = \tuple{\stateFunction_2, \values_2, \held_2},\config_2' \in \configs$,  $\atrans_1, \atrans_2 \in \transitionsof{\athread}$, $s \in \transitions^*$ and $\exec: \config_1 \step{\atrans_1} \config_1' \step{s} \config_2 \step{\atrans_2} \config_2'$ such that:
\begin{itemize}
\item $\config_1$ is reachable from $\initialconfig$;
\item in all configurations along $\exec$, thread $\athread$ is in $\criticalsection$;
\item $x \in \readacc{\atrans_1}{\values_1} \cup \readacc{\atrans_2}{\values_2}$ and $\athread$ does not access $x$ in $\config_1' \step{s} \config_2$, \emph{i.e.}, for every step $\tuple{\stateFunction,\values,\held} \step{\atrans} \tuple{\stateFunction',\values',\held'}$ in $\config_1' \step{s} \config_2$ such that $\atrans \in \transitionsof{\athread}$, we have $x \notin \readacc{\atrans}{\values}$.
\end{itemize}
\end{defi}

We start by identifying transitions that may be involved in atomicity violations. To do so, we define the following two sets. 
\begin{defi}
\label{def-nodesx-interferencetransitions}  
Let $\criticalsection$ be a critical section of thread $\athread$, and let $x \in \vars$.
\begin{itemize}
\item We let $\nodes_x(\criticalsection) \subseteq \criticalsection$ be the set of all $\anode \in \criticalsection$ for which there are two access transitions in $\criticalsection$ $\atrans_1= (\anode_1, \aninstr_1, \anode_1')$ and $\atrans_2 = (\anode_2, \aninstr_2, \anode_2)$ such that $x \in \mayread{\atrans_1} \cap \mayread{\atrans_2}$ and such that there is a path in $\criticalsection$ from $\anode_1'$ to $\anode_2$ that visits $\anode$; 
\item We let $\interferencetransitions{x}{\criticalsection} \subseteq \transitions$ be the set of all $\atrans_w = (\anode_w, \aninstr_w, \anode_w')$ such that $\threadof{\atrans_w} \ne \athread$, such that $x \in \maywrite{\atrans_w}$ and such that $\anode_w \MHP \anode_x$ for some $\anode_x \in \nodes_x(\criticalsection)$. 
\end{itemize}
\end{defi}

Intuitively, $\nodes_x(\criticalsection) \subseteq \criticalsection$ contains all nodes $\anode$ in $\criticalsection$ such that $x$ can be accessed both before and after $\anode$ in $\criticalsection$, therefore a write to $x$ by another thread while $\threadof{\anode}$ is in node $\anode$ could lead to an atomicity violation. The set $\interferencetransitions{x}{\criticalsection}$ contains all transitions $\atrans$ of other threads such that $\atrans$ may write to $x$ and cause an atomicity violation for $\criticalsection$. Whenever $\interferencetransitions{x}{\criticalsection} \ne \emptyset$, we shall consider that $\criticalsection$ has a risk of atomicity violation on $x$. 
Based on the two definitions, we identify, for every critical section, the set of transitions that need protection with respect to a particular variable. 

\begin{defi}
\label{prot-trans-vars}
\begin{itemize}
Let $\criticalsection$ be a critical section.
\item For every $x \in \vars$, we let $\prottrans{x}{\criticalsection}$ be the set of transitions $\atrans$ in $\criticalsection$ that satisfy one of the following conditions:
\begin{itemize}
\item $x \in \mayread{\atrans}$ and $\interferencetransitions{x}{\criticalsection} \ne \emptyset$,
\item there is a critical section $\criticalsection'$ such that $\atrans \in \interferencetransitions{x}{\criticalsection'}$, or
\item there is $\atrans' \in \transitions$ such that $\atrans \mayrace{x} \atrans'$.
\end{itemize} 
\item 
We let $\protvars{\criticalsection} \deff \set{x \in \vars \mid \prottrans{x}{\criticalsection} \ne \emptyset}$.
\end{itemize}
\end{defi}

The set $\prottrans{x}{\criticalsection}$ contains transitions of $\criticalsection$ on which variable $x$ must be protected. There are three reasons why we may want to protect $x$ at a transition $\atrans$: that $\atrans$ is vulnerable to interference on $x$ that may lead to an atomicity violation, that $\atrans$ may itself interfere on $x$ with another critical section, and simply that $\atrans$ may be involved in a data race on $x$. 
Moreover, $\protvars{\criticalsection}$ is the set of variables that should be protected in $\criticalsection$. The goal is now to ensure that, for every variable $x \in \protvars{\criticalsection}$, mutex $\newmut{x}$ is held at every transition in $\prottrans{x}{\criticalsection}$. This is the topic of the next section. 

\subsection{Protection of Critical Section}
\label{subsec:atomicity-protection}
We now introduce the transformation that protects critical sections. As in \autoref{subsec:data-race-protection}, it relies on a mutex $\newmut{x}$ for every variable $x$. It requires that, whenever there is a risk of atomicity violation for a critical section $\criticalsection$ on a variable $x$, mutex $\newmut{x}$ must be held in $\criticalsection$ as well as at any transition that may interfere with $\criticalsection$ on $x$.  
We proceed in two steps. First, it places lock instructions around interfering transitions, but only those outside of critical sections. Second, it takes care of each critical section $\criticalsection$, both to protect $\criticalsection$ from interference and to treat transitions in $\criticalsection$ that themselves may interfere with other critical sections.

The first step handles potential interference transitions that are themselves \emph{not} in critical sections (transitions in critical sections are treated differently in the second step). Let $\atrans$ be a transition such that $\atrans \notin \criticalsection$ for every $\criticalsection$. 
Let $\interfvars{\atrans}$ be the set of all variables $x$ for which $\atrans \in \interferencetransitions{x}{\criticalsection}$ for some critical section $\criticalsection$; moreover, let $\racevars{\atrans}$ be the set of variables $x$ such that, for some $\atrans'$, $\atrans \mayrace{x} \atrans'$. 
We insert around $\atrans$ a $\lock{\newmut{x}}-\unlock{\newmut{x}}$ pair for every $x \in \racevars{\atrans} \cup \interfvars{\atrans}$ following the same approach as in \autoref{subsec:data-race-protection}, and again respecting the lock ordering given by $\lockordering$.  

The second step iterates over every critical section $\criticalsection$. Inside $\criticalsection$, for every $x$ that must be protected, we rely on a weak two-phase locking to place a single $\lock{\newmut{x}}-\unlock{\newmut{x}}$ pair covering all relevant transitions. 
Acquiring all mutexes at the start of the critical section and releasing them at the end of the critical section would be reasonable, but creates unnecessary contention: instead, we use (weak) two-phase locking approach to place \locksymb instructions as late as possible and unlock instructions as early as possible. 

In general, placing the locks (resp. unlocks) ``as late'' (resp. ``as early'') as possible is hard, because the CFG of $\criticalsection$ can take arbitrary shapes. For this reason, we only perform this optimization in the linear initial (resp. final) portion of the CFG of $\criticalsection$. Variables that must be protected are those in the set $\protvars{\criticalsection}$ defined in \autoref{prot-trans-vars}. 
We consider variables in $\protvars{\criticalsection}$ in decreasing order for $\lockordering$, \emph{i.e.}, starting with variables whose mutex $\newmut{x}$ is greatest for $\lockordering$ (such mutexes must be locked last). In order to minimize contention, we insert the $\lock{\newmut{x}}$ (resp. $\unlock{\newmut{x}}$) transition as late (resp. as early) in the critical section as possible while protecting all relevant transitions (and respecting the lock ordering). This approach is often called \emph{(weak) two-phase locking}. Crucially, mutex $\newmut{x}$ must be held at every transition in $\criticalsection$ that requires protection for variable $x$; these transitions are those in set $\prottrans{x}{\criticalsection}$, as defined in \autoref{prot-trans-vars}. 

We start by explaining how $\proglike{lock}$ instructions are inserted.
Let $\anode_0, \anode_1, \dots, \anode_k$ be the longest path in $\criticalsection$ starting from $\anode_0= \csfirst{{\criticalsection}}$ such that $\anode_{i+1}$ is the only successor of $\anode_i$ and $\anode_i$ the only predecessor of $\anode_{i+1}$ for all $i \leq k-1$. This sequence represents the longest portion of $\criticalsection$ including $\csfirst{\criticalsection}$ where the CFG is linear. Note that $\anode_k = \cslast{\criticalsection}$ means that the CFG in $\criticalsection$ is a line, while $k=0$ means that $\csfirst{\criticalsection}$ already has several successors. 
 Given a variable $x$ such that $\prottrans{x}{\criticalsection} \ne \emptyset$, we let $f(x) \in \nset{0}{k}$ be the largest index such that, for every $i \in \nset{1}{f(x)}$, the transition from $\anode_{i-1}$ to $\anode_i$ is not in $\prottrans{x}{\criticalsection}$. 
With this definition, we know that the lock of mutex $\newmut{x}$ must be inserted before index $f(x)$.
 
For every $i \in \nset{0}{k}$, we construct the set $M_i$ of mutexes whose lock is inserted immediately after $\anode_i$. Let $M \deff \set{\newmut{x} \mid x \in \protvars{\criticalsection}}$ be the set of mutexes that must be held in $\criticalsection$. We let $M_0$ be equal to all $\newmut{x} \in M$ such that $f(x)=0$, along with all mutexes smaller for $\lockordering$ than those mutexes, \emph{i.e.}, all $\amutex \in M$ for which there is $x$ such that $f(x)=0$ and $m \lockordering \newmut{x}$. For $M_1$, we proceed similarly but excluding mutexes already in $M_0$, \emph{i.e.}, we include all mutexes $\newmut{x} \in M \setminus M_0$ such that $f(x) =1$ along with all $\amutex \in M \setminus M_0$ such that, for some $\newmut{x} \in M_1$, $m \lockordering M_1$. We proceed similarly for $M_2$ but excluding mutexes in $M_0 \cup M_1$, and so on. For every $i \in \nset{0}{k}$, we order mutexes in $M_i$ by $\lockordering$ into a sequence $m_0, \dots, m_p$, and we insert immediately after $\anode_i$ the sequence of instructions $\lock{m_0}, \dots, \lock{m_p}$. 

For \unlocksymb instructions, we proceed similarly, except that there is no need to consider the lock ordering $\lockordering$. We let $\anode_k, \dots, \anode_1, \anode_0$ be the longest path in $\criticalsection$ ending in $\anode_0 = \cslast{\criticalsection}$ such that, for all $i \leq k-1$, $\anode_{i+1}$ is the only successor of $\anode_i$ and $\anode_i$ is the only predecessor of $\anode_{i+1}$. We let $f(x)$ be the function that maps to every $x \in \protvars{\criticalsection}$ the largest index in $\nset{0}{k}$ such that, for every $i \in \nset{1}{k}$, the transition from $\anode_i$ to $\anode_{i-1}$ is not in $\prottrans{x}{\criticalsection}$. This time, we simply let $M_i \deff \set{\newmut{x} \mid f(x) = i}$ because $\lockordering$ is irrelevant for $\unlocksymb$ instructions. We insert immediately before $\anode_i$  $\unlock{m}$ instructions for all $m \in M_i$.

\begin{exa}
The example from \autoref{fig:code-2pl} displays the obtained program when applying the transformation to a program with a critical section.  
The original program includes a critical section $\criticalsection$ in task $\proglike{t1}$.
All \locksymb and \unlocksymb instructions in the transformed program depicted in \cref{fig:code-2pl} were inserted by the transformation. 
Variables $x$ and $y$ must be protected, because \proglike{main} may cause an atomicity violation on either of those, therefore $\newmut{x}$ and $\newmut{y}$ are used to protect the critical section. 
Here, we assume that $\newmut{x} \lockordering \newmut{y}$, hence $\newmut{x}$ is locked first even though the first instruction accessing $x$  in $\criticalsection$ comes later. 
Both \locksymb instructions come after the instruction 
$\proglike{z = 0;}$ in $\criticalsection$, allowing to minimize unnecessary contention. 
For the same reason, the $\unlock{\newmut{y}}$ 
is inserted before the $\proglike{print(x)}$ 
instruction; it is not inserted before the 
\proglike{if} block because the CFG is no longer linear there.
\end{exa}

\begin{figure}
\centering
\begin{lstlisting}[language=C, basicstyle=\ttfamily\footnotesize, label=input, breaklines = true, multicols = 2]
int x;
int y;
int z;


main() {
    create(t1);
    lock(mu_x);
    x=1;
    unlock(mu_x);
    lock(mu_y);
    y=2;
    unlock(mu_y);
}


t1() {
    critical{
        z = 0;
        lock(mu_x);
        lock(mu_y);
        y=0;
        x=1;
        print(y);
        if (cond) {
            x=2;
        }
        unlock(mu_y);
        print(x);
        unlock(mu_x);
    }
}
\end{lstlisting}
\caption{An example program after the transformation that protects the critical section using two-phase locking.
}
\label{fig:code-2pl}
\end{figure}

Performing the above operation for every critical section concludes the transformation. Let $\newprog$ denote the program obtained. By construction, for every critical section $\criticalsection$ of thread $\athread$, for every variable $x$:
\begin{itemize}
\item whenever $\athread$ enters $\criticalsection$, it does not hold $\newmut{x}$;
\item whenever $\athread$ performs a transition in $\prottrans{x}{\criticalsection}$, it holds $\newmut{x}$;
\item whenever $\athread$ leaves $\criticalsection$, it does not hold $\newmut{x}$.
\end{itemize} 
Moreover, if $\prottrans{x}{\criticalsection} = \emptyset$ then $\athread$ never holds $\newmut{x}$ in $\criticalsection$.
There is no hope in arguing that $\newprog$ has the same behaviors as $\prog$ (atomicity violation protection necessarily alters behaviors). We can at least observe that $\newprog$ is obtained from $\prog$ by introducing $\lock{\amutex}$ and $\unlock{\amutex}$ instruction for mutexes $\amutex \in \newmutexes$, and that locking - unlocking patterns are well-formed in the sense that, when $(\anode, \lock{\newmut{x}}, \anode')$ is a transition of $\newprog$, node $\anode$ cannot be visited by a thread that already holds $\newmut{x}$.

We must now argue that $\newprog$ has no data race and that critical sections are protected.

\begin{prop}
The transformed program $\newprog$ has no data race, and all its critical sections are protected. Moreover, the transformation from $\prog$ to $\newprog$ does not introduce new deadlocks, \emph{i.e.}, no cyclic lock dependency involves newly introduced mutexes. 
\label{transformation-safe-with-cs}
\end{prop}
\begin{proof}
As in \autoref{no-data-race-transfo}, every execution prefix of $\newprog$ projects, after removing transitions on newly introduced mutexes and collapsing intermediate nodes, to an execution prefix of $\prog$ with the same valuation and corresponding original nodes. We use this projection implicitly in the proof below. 

Suppose that $\newprog$ has a data race between $\atrans$ and $\atrans'$ on $x$. By a projection argument, this implies the existence of a data race in $\prog$ between the corresponding transitions $\atrans$ and $\atrans'$ on $x$. Thanks to \autoref{soundness-data-race-analysis}, this implies $\atrans \mayrace{x} \atrans'$. To obtain a contradiction, it suffices to prove that $\atrans$ and $\atrans'$ can only be performed in $\newprog$ by threads holding mutex $\newmut{x}$. 
Let $\atrans$ be such a transition, let $\athread$ be the corresponding thread. 
If $\atrans$ is outside of atomic sections, then $x \in \racevars{\atrans}$ therefore a $\lock{\newmut{x}} - \unlock{\newmut{x}}$ was inserted around $\atrans$, so that $\athread$ can only perform $\atrans$ when holding $\newmut{x}$. 
If $\atrans$ is in some critical section $\criticalsection$, by definition of $\prottrans{x}{\criticalsection}$, $\atrans \in \prottrans{x}{\criticalsection}$ therefore $\athread$ must hold $\newmut{x}$ while performing $\atrans$. 

We now prove that all critical sections in $\newprog$ are protected. Suppose by contradiction that there is a critical section $\criticalsection$ that is not protected in $\newprog$. By \autoref{def-protected}, this implies the existence of  $\values_1'(x) \ne \values_2(x)$ for some $x \in \vars$, $\config_1, \config_1' = \tuple{\stateFunction_1', \values_1', \held_1'}$, $\config_2 = \tuple{\stateFunction_2, \values_2, \held_2},\config_2' \in \configs$,  $\atrans_1, \atrans_2 \in \transitionsof{\athread}$, $s \in \transitions^*$ and $\exec: \config_1 \step{\atrans_1} \config_1' \step{s} \config_2 \step{\atrans_2} \config_2'$ such that:
\begin{itemize}
\item $\config_1$ is reachable from $\initialconfig$;
\item in all configurations along $\exec$, $\athread$ is in $\criticalsection$;
\item $x \in \readacc{\atrans_1}{\values_1} \cup \readacc{\atrans_2}{\values_2}$ and $\athread$ does not access $x$ in $\config_1' \step{s} \config_2$, \emph{i.e.}, for every step $\tuple{\stateFunction,\values,\held} \step{\atrans} \tuple{\stateFunction',\values',\held'}$ in $\config_1' \step{s} \config_2$ such that $\atrans \in \transitionsof{\athread}$, $x \notin \readacc{\atrans}{\values}$.
\end{itemize}
Since $\values_1'(x) \ne \values_2(x)$, there is at least one step in $\config_1' \step{s} \config_2$ that modifies the value of $x$. Let $\config_w \step{\atrans_w} \config_w'$ be such a step, let $(\anode_w, \aninstr_w, \anode_w') \deff \atrans_w$ and let $\athread_w$ be the corresponding thread. By hypothesis, $\athread$ does not access $x$ in $\config_1' \step{s} \config_2$, which implies that $\athread_w \ne \athread$. Thanks to \autoref{prop:mw-mr-soundness}, $x \in \maywrite{\atrans}$. Moreover, again by \autoref{prop:mw-mr-soundness}, $x \in \mayread{\atrans_1} \cap \mayread{\atrans_2}$. As witnessed by $\exec$, there is a path from the target of $\atrans_1$ to the source of $\atrans_2$. By \autoref{def-nodesx-interferencetransitions}, every node visited by $\athread$ in $\config_1' \step{s} \config_2$ is in $\nodes_x(\criticalsection)$. 
Therefore, by letting $\anode$ be the node of $\athread$ in $\config_w$, we have $\anode \in \nodes_x(\criticalsection)$; also, $\anode \MHP \anode_w$ by \autoref{soundness-MHP}. By \autoref{def-nodesx-interferencetransitions}, $\atrans_w \in \interferencetransitions{x}{\criticalsection}$. By construction, $\athread_w$ must hold $\newmut{x}$ whenever it performs $\atrans_w$. 

In particular, this implies that $\interferencetransitions{x}{\criticalsection} \ne \emptyset$. Since $x \in \mayread{\atrans_1} \cap \mayread{\atrans_2}$, we obtain that $\atrans_1, \atrans_2 \in \prottrans{x}{\criticalsection}$. Therefore, the $\lock{\newmut{x}}$ instruction was inserted before $\atrans_1$ in $\criticalsection$, and the $\unlock{\newmut{x}}$ instruction was inserted after $\atrans_2$. This implies that $\athread$ must hold $\newmut{x}$ in the entirety of $\exec$. However, as proven above, $\athread_w \ne \athread$ must hold $\newmut{x}$ to perform $\atrans_w$, a contradiction. 

A last remaining point is that the transformation introduces no deadlock. As in \autoref{first-transfo-no-deadlock}, what we mean by that is that no cyclic lock dependency involves mutexes in $\newmutexes$. As in the proof of \autoref{first-transfo-no-deadlock}, this relies on two facts: 1. that no thread can try to acquire a mutex in $\mutexes$ while holding a mutex in $\newmutexes$, and 2. that mutexes in $\newmutexes$ are acquired in the order given by $\lockordering$. For 1., it is clear outside of critical sections because mutexes in $\newmutexes$ are only held during access transitions. In a critical section $\criticalsection$, mutexes in $\newmutexes$ can be held longer, but there is no \locksymb of mutexes in $\mutexes$ in $\criticalsection$ by hypothesis, and mutexes in $\newmutexes$ are released before the thread leaves $\criticalsection$. Regarding 2., the order $\lockordering$ is respected between mutexes in $\newmutexes$: it is trivial outside of critical sections, and holds in critical sections by construction as any needed mutex smaller for $\lockordering$ is inserted earlier if necessary. 
 \end{proof}

\subsection{Deleting Redundant Locking}
\label{subsec:redundant-locking}
As a final step, we remove any introduced mutex identified as redundant in $\newprog$. For each variable $x$ whose mutex $\newmut{x}$ appears in $\newprog$, the transformation identified a set of transitions $\transitions_x$ that must be protected by $\newmut{x}$. More precisely, this set $\transitions_x$ is composed of:
\begin{itemize}
\item the union of all $\prottrans{x}{\criticalsection}$ for all $\criticalsection$ (transitions vulnerable to interference on $x$),
\item the union of all $\interferencetransitions{x}{\criticalsection}$ for all $\criticalsection$ (transitions that can interfere on $x$ and risk violating atomicity), and
\item all $\atrans$ such that $\atrans \mayrace{x} \atrans'$ for some $\atrans'$.
\end{itemize} 
If, for some $x$, there is another mutex that is held whenever $x$ should be protected, then $\newmut{x}$ is redundant and can safely be removed. More formally, mutex $\newmut{x}$ can be removed if there is $\newmut{y} \ne \newmut{x}$ guaranteed held at every transition in $\transitions_x$. This test can be made easier by maintaining relevant information along the transformation. Outside of critical sections, $\newmut{x}$ is only held at individual access transitions, and it suffices to check that $y$ must also be protected at every such transition. In a critical section $\criticalsection$, we simply must check that the portion of the atomic section where $\newmut{y}$ is held includes the portion where $\newmut{x}$ is held.
When $\newmut{x}$ is redundant with $\newmut{y}$, it suffices to delete any transition locking or unlocking $\newmut{x}$ and to merge its source and target nodes. Note that there can be cases where two mutexes are mutually redundant, in which case only one of them should be removed. 

As a side note, it is not useful to check whether a newly introduced mutex $\newmut{x}$ in $\newmutexes$ is redundant with respect to a mutex $\amutex \in \mutexes$ of the original program. Indeed, if we could detect such a redundancy, then we would have had $\amutex \in \lockset{\anode}$ for every $(\anode, \aninstr, \anode') \in \transitions_x$. But then, the transformation would have detected that mutex $\amutex$ already prevented any data race or atomicity violation on variable $x$ in $\prog$, so $\newmut{x}$ would not have been introduced in $\newprog$ in the first place. 

\section{Conclusion}\label{sec:Conclusion}
We have presented a method to automatically place locks in a concurrent program so as to prevent any data race and atomicity violation.
Our approach is self-contained and formally proven. It includes a formal model for C-like POSIX concurrent programs, a preliminary analysis of concurrency and a must-lockset analysis. One of the key contributions is a concurrent points-to analysis derived from a procedural, Steensgaard-like points-to analysis such as the one implemented in the \texttt{alias} plugin of Frama-C \cite{le_gall_alias_2024}. This is achieved using a unification technique called Diamond Completion Rule. 
Our method uses the result of the analysis to automatically place locks and respects a lock ordering to prevent deadlocks.

\bibliographystyle{alphaurl}
\bibliography{ctask-publication.bib}
\end{document}